\documentclass[11pt]{article}

\usepackage[preprint]{acl}  

\usepackage{times}
\usepackage{latexsym}
\usepackage[T1]{fontenc}
\usepackage[utf8]{inputenc}
\usepackage[protrusion=true,expansion=true,final,babel]{microtype}
\usepackage{inconsolata}

\usepackage{amssymb}
\usepackage{amsmath}
\usepackage{bm}

\usepackage{graphicx}
\usepackage{booktabs}
\usepackage{multirow}
\usepackage{threeparttable}
\usepackage{adjustbox}
\usepackage{float}

\usepackage{tikz}
\usetikzlibrary{positioning, arrows.meta, shapes.geometric, calc, fit, backgrounds, decorations.pathreplacing}

\usepackage{xcolor}
\usepackage{color}
\definecolor{revred}{RGB}{200,0,0}

\usepackage{enumitem}
\usepackage{xspace}
\usepackage{listings}
\usepackage{newfloat}
\usepackage[most]{tcolorbox}

\newenvironment{smalldisplay}{%
  \par
  \begingroup
  \small
  \setlength{\abovedisplayskip}{8pt plus 2pt minus 2pt}%
  \setlength{\belowdisplayskip}{8pt plus 2pt minus 2pt}%
  \setlength{\abovedisplayshortskip}{4pt plus 2pt minus 1pt}%
  \setlength{\belowdisplayshortskip}{6pt plus 2pt minus 2pt}%
  \setlength{\jot}{2pt}%
}{%
  \par
  \endgroup
  \noindent\ignorespacesafterend
}
\newenvironment{scriptsizedisplay}{%
  \par
  \begingroup
  \scriptsize
  \setlength{\abovedisplayskip}{7pt plus 2pt minus 2pt}%
  \setlength{\belowdisplayskip}{7pt plus 2pt minus 2pt}%
  \setlength{\abovedisplayshortskip}{3pt plus 2pt minus 1pt}%
  \setlength{\belowdisplayshortskip}{5pt plus 2pt minus 2pt}%
  \setlength{\jot}{2pt}%
}{%
  \par
  \endgroup
  \noindent\ignorespacesafterend
}

\definecolor{macaroongreen}{RGB}{255, 220, 220}

\newcommand{\dsnew}[1]{#1}

\lstdefinestyle{prompt}{
  basicstyle=\ttfamily\footnotesize,
  breaklines=true,
  breakatwhitespace=false,
  breakindent=0pt,
  postbreak=\mbox{\textcolor{gray}{$\hookrightarrow$}\space},
  columns=fullflexible,
  keepspaces=true,
  showstringspaces=false,
  upquote=true,
  xleftmargin=4pt,
  xrightmargin=4pt,
  framexleftmargin=2pt,
  framexrightmargin=2pt,
  frame=single,
  rulecolor=\color{black!30},
  belowcaptionskip=2pt,
  aboveskip=4pt,
  belowskip=4pt,
}
\definecolor{promptnavy}{RGB}{29, 47, 113}
\definecolor{promptbody}{RGB}{238, 240, 252}
\definecolor{promptcodebg}{RGB}{249, 249, 252}
\definecolor{promptcodeframe}{RGB}{205, 205, 215}

\newtcolorbox{promptbox}[1]{
  enhanced,
  colback=promptbody,
  colframe=promptnavy,
  colbacktitle=promptnavy,
  coltitle=white,
  fonttitle=\bfseries,
  title=#1,
  boxrule=0.5pt,
  arc=1.5pt,
  left=8pt, right=8pt, top=5pt, bottom=5pt,
  fontupper=\small,
}

\definecolor{rubricamber}{RGB}{166, 119, 24}
\definecolor{rubricbody}{RGB}{254, 247, 218}

\newtcolorbox{rubricbox}[1]{
  enhanced,
  colback=rubricbody,
  colframe=rubricamber,
  colbacktitle=rubricamber,
  coltitle=white,
  fonttitle=\bfseries,
  title=#1,
  boxrule=0.5pt,
  arc=1.5pt,
  left=8pt, right=8pt, top=5pt, bottom=5pt,
  fontupper=\small,
}

\tcbuselibrary{listings}
\newtcblisting{promptcode}{
  enhanced,
  colback=promptcodebg,
  colframe=promptcodeframe,
  listing only,
  listing options={
    basicstyle=\ttfamily\footnotesize,
    breaklines=true,
    breakatwhitespace=false,
    columns=fullflexible,
    keepspaces=true,
    aboveskip=0pt,
    belowskip=0pt,
  },
  boxrule=0.4pt,
  arc=1pt,
  left=4pt, right=4pt, top=3pt, bottom=3pt,
}

\newtcblisting{promptcodesmall}{
  enhanced,
  colback=promptcodebg,
  colframe=promptcodeframe,
  listing only,
  listing options={
    basicstyle=\ttfamily\scriptsize,
    breaklines=true,
    breakatwhitespace=false,
    columns=fullflexible,
    keepspaces=true,
    aboveskip=0pt,
    belowskip=0pt,
  },
  boxrule=0.4pt,
  arc=1pt,
  left=3pt, right=3pt, top=2pt, bottom=2pt,
}

\renewcommand{\thefootnote}{\fnsymbol{footnote}}

\title{PersonalPlan: Planning Multi-Agent Systems for Personalized Programming Learning}

\author{
  \textbf{Zhiyuan Wen}\textsuperscript{1}\thanks{\,Corresponding authors.} \quad
  \textbf{Jiannong Cao}\textsuperscript{1} \quad
  \textbf{Peng Gao}\textsuperscript{1} \quad
  \textbf{Haochen Shi}\textsuperscript{1} \\[2pt]
  \textbf{Wengpan Kuan}\textsuperscript{1} \quad
  \textbf{Bo Yuan}\textsuperscript{2}\footnotemark[1] \quad
  \textbf{Xiuxiu Qi}\textsuperscript{1,\,3} \\[4pt]
  \textsuperscript{1}Department of Computing, The Hong Kong Polytechnic University, Hong Kong SAR, China \\
  \textsuperscript{2}JIUTIAN Team, China Mobile Research Institute, Beijing, China \\
  \textsuperscript{3}School of Artificial Intelligence, Nankai University, Tianjin, China \\[3pt]
  \texttt{\{zhiyuan.wen,\,jiannong.cao,\,penggao\}@polyu.edu.hk} \\
  \texttt{\{haochen8.shi,\,wengpan.kuan\}@connect.polyu.hk} \\
  \texttt{yuanboyjy@chinamobile.com} \quad
  \texttt{xiuxqihm@gmail.com}
}

\begin{document}
\maketitle
\renewcommand{\thefootnote}{\arabic{footnote}}
\setcounter{footnote}{0}

\begin{abstract}
Effective programming education requires personalized instruction adapted to diverse learner backgrounds. However, while LLM-based multi-agent systems (MAS) excel at complex planning, existing planners often lack profile-grounding and pedagogical scaffolding, thereby undermining personalized programming learning. To fill in the gap, we first introduce \textbf{MAP-PPL} (\textbf{M}ulti-\textbf{A}gent \textbf{P}lans for \textbf{P}ersonalized \textbf{P}rogramming \textbf{L}earning), a profile-conditioned multi-agent planning dataset with 3{,}043 query--profile--plan instances from 1{,}730 Stack Overflow question groups and 2{,}738 learner profiles. Each plan specifies agents, subtasks, executable steps, and prerequisite dependencies. Then, we propose \textbf{PersonalPlan}, a two-stage MAS planner that first performs hierarchical SFT with separate LoRA adapters for profile-aware task decomposition and step dependency planning, then applies a Reward-Adaptive GRPO to encourage the model to generate executable, personalized, and pedagogically scaffolded plans. Extensive experiments on MAP-PPL comparing PersonalPlan against frontier LLMs, generic MAS frameworks, and agentic planners demonstrate its superiority. With only 8B and 32B variants, PersonalPlan achieves state-of-the-art plan executability, personalization, and pedagogical quality, effectively orchestrating MAS for agent-student interactions.
\end{abstract}

\section{Introduction}
\label{introduction}
Programming literacy is increasingly useful beyond traditional computer science classrooms. With AI assistants, learners from diverse backgrounds ask models to automate analyses, inspect code, connect APIs, debug workflows, or turn informal goals into executable programs \citep{denny2024computing,hsu2025from}. Programming learning often requires resource retrieval, prerequisite explanation, code demonstration, execution, testing, debugging, and reflection \citep{barua2014developers,ponzanelli2013seahawk}. This motivates a planning view of programming learning, in which a learner query should be mapped to coordinated instructional roles, subtasks, tools, and dependencies.

Recent code tutors and LLM-based systems provide concept explanation, Socratic debugging, and multi-role tutoring \citep{kazemitabaar2024codeaid,liffiton2024codehelp,kargupta2024treeinstruct,zhao2025codeedu,david2026intellicode}. However, they rarely produce the specialized plans required by personalized programming MAS, which must satisfy three core requirements. First, plans should be \emph{personalized in structure}, adapting agent roles, subtask granularity, and prerequisite paths to diverse learner profiles \citep{nabizadeh2021learning,chen2008intelligent,jiang2022data}. Second, they must be \emph{executable}, avoiding latent errors like cyclic dependencies or agent--tool mismatches often hidden in fluent surface plans \citep{kim2024llmcompiler,wei2025beyondreact,xiong2025mpo}. Third, they must encode \emph{pedagogical scaffolding} as a checkable structure, explicitly defining instructional sequences (e.g., concept-before-application) and feedback checkpoints before execution \citep{hmelosilver2007scaffolding,sentance2019primm}.

In this paper, we propose \textbf{PersonalPlan}, a profile-conditioned multi-agent planning framework for personalized programming-learning MAS. A full tutoring plan couples two decisions: the high-level instructional scaffold and the low-level executable workflow. PersonalPlan therefore uses hierarchical SFT with two LoRA adapters: \emph{Profile-Aware Decomposition} (PAD) selects learner-conditioned subtasks and pedagogical agents, while \emph{Step Dependency Planning} (SDP) grounds this scaffold into step actions, prerequisite edges, agent--step assignments, and tool bindings. A lightweight joint-alignment stage then exposes SDP to PAD-produced scaffolds and gives PAD a downstream-compatible scaffold signal, reducing the hierarchical exposure-bias mismatch between gold scaffolds used during SFT and generated scaffolds seen at inference. Finally, the reward-adaptive GRPO optimizes complete plans with verifiable composable rewards for structural validity, profile grounding, and pedagogical phase coverage, while a hard gate penalizes schema violations, cycles, and invalid tool bindings.

To facilitate PersonalPlan and broader research on profile-conditioned MAS planning, we also construct \textbf{MAP-PPL} (\textbf{M}ulti-\textbf{A}gent \textbf{P}lans for \textbf{P}ersonalized \textbf{P}rogramming \textbf{L}earning), a dataset of 3{,}043 query--profile--plan instances from 1{,}730 Stack Overflow question groups and 2{,}738 learner profiles across seven learning intents \citep{beyer2018automatically}. Each instance pairs a learner profile and correctness reference with a strict JSON MAS plan generated by an advanced LLM (Claude Sonnet 4.6 \cite{anthropic2026claude46}) that specifies agents, subtasks, executable steps, tool bindings, execution order, and dependency edges. All released plans pass deterministic schema and DAG checks plus an execution-effectiveness gate, and the corpus records profile-grounded structure and pedagogical scaffolding for evaluating whether planners adapt agent roles, subtask decomposition, dependencies, and teaching phases to learner background.

We evaluate PersonalPlan on MAP-PPL against frontier LLM, generic MAS, and agentic workflow planners on both static plan-quality metrics and dynamic MAS execution with tutor--learner interactions. With only 8B and 32B variants, PersonalPlan achieves state-of-the-art plan executability, personalization, and pedagogical quality, effectively orchestrating MAS for agent-student interactions. Ablations show that the two-stage SFT and reward-guided refinement are both critical: single-stage SFT with only decomposition or dependency supervision fails to produce executable plans, while joint alignment and GRPO each contribute to improving tool binding and pedagogical quality. To summarize, our contributions are threefold:
\begin{enumerate}
    \item We introduce \textbf{PersonalPlan}, a profile-conditioned multi-agent planning framework for personalized programming-learning MAS that combines hierarchical SFT and reward-adaptive GRPO with verifiable rewards for executable structure, profile grounding, and pedagogy.
    
    \item We construct \textbf{MAP-PPL}, a dataset of 3{,}043 query--profile--plan instances from 1{,}730 Stack Overflow question groups and 2{,}738 learner profiles across seven learning intents to facilitate related research. All plans are executable and contain profile-grounded structure and pedagogical scaffolding for evaluating personalized programming-learning MAS planners.
    
    \item Experiments show that PersonalPlan improves the three target properties of personalized programming-learning MAS: executability better tool-bound plans and compact executed traces; personalization through stronger profile fit and profile-sensitive variation; and pedagogy through better instructional quality and realized tutoring behavior\footnote{Our code and data are at \url{https://github.com/preke/PersonalPlan}}. 
\end{enumerate}

\section{Related Work}
\label{related_work}
\paragraph{LLM-based multi-agent planning.}
LLM-based multi-agent systems decompose tasks across specialized agents that communicate, call tools, and coordinate through conversations or workflow graphs. General frameworks such as AutoGen, CAMEL, AgentVerse, MetaGPT, ChatDev, and AutoAgents instantiate agent teams for broad task solving \citep{autogen,camel,chen2024agentverse,hong2024metagpt,qian2024chatdev,chen2024autoagents}, while AIPOM, AFlow, AOP, and WorFBench generate explicit agent/task graphs \citep{kim2025aipom,zhang2025aflow,li2025aop,qiao2025worfbench}. Planner-training methods further improve high-level plans with trajectories, hierarchical supervision, and post-training rewards \citep{yin2024lumos,zhao2024epo,xiong2025mpo,parmar2025plantuning}. These works motivate our MAS and workflow-planning baselines, but they mainly optimize generic completion or orchestration; learner profiles rarely determine the agent roster, prerequisite graph, tool bindings, or instructional phases.

\paragraph{Personalized programming education.}
Personalized tutoring adapts hints, resources, problem sequences, or learning paths to learner knowledge and goals \citep{graesser2004autotutor,piech2015dkt,nabizadeh2021learning,chen2008intelligent}. Recent LLM educational agents add lesson planning, learner simulation, Socratic teaching, and pedagogy-aware RL \citep{zhang2025eduplanner,liu2024personalitysim,peng2025kele,dinucu2025aligningpedagogy}. For programming education, CodeAid, CodeHelp, TreeInstruct, AdaCoder, CodeEdu, and IntelliCode provide guardrailed coding help, Socratic debugging, and multi-role tutoring \citep{kazemitabaar2024codeaid,liffiton2024codehelp,kargupta2024treeinstruct,zhu2025adacoder,zhao2025codeedu,david2026intellicode}. However, their adaptation is usually a dialogue policy, hint choice, recommendation, or session outline rather than an inspectable MAS execution plan, making it difficult to validate whether agent roles, tool bindings, dependencies, and teaching phases form an executable personalized tutoring workflow. Therefore, PersonalPlan fills this gap by generating learner-conditioned MAS plans optimized for executability, profile grounding, and pedagogy.

\section{PersonalPlan}
\label{methodology}
\subsection{Problem Formulation}
\label{sec:prelim}
Let $I_q$ denote a query raised by a learner and $I_p$ denote the learner profile containing skills, knowledge, prior experience, and background context. The goal is to generate a structured multi-agent plan
$\mathcal{P}=(\mathcal{A},\mathcal{T},\mathcal{S}, G)$ for a tutoring-oriented MAS that executes a personalized instructional workflow, where $\mathcal{A}=\{a_1,\dots,a_N\}$ is a set of specialized agents, $\mathcal{T}=\{\tau_1,\dots,\tau_M\}$ is a set of subtasks, and $\mathcal{S}=\{s_1,\dots,s_K\}$ denotes the executable steps. Additionally, the prerequisite relations among steps define a dependency graph $G=(\mathcal{S},\mathcal{E})$, where $\mathcal{E}=\{(s_i,s_j): s_i \in \operatorname{dep}(s_j)\}$. Our objective is to learn a planner $\mathcal{F}_\theta:(I_q,I_p)\mapsto\mathcal{P}$ under three requirements: executable MAS structure with schema-valid references and an acyclic $G$; personalization to $I_p$ in agents, subtasks, and instructions; and pedagogical scaffolding \citep{hmelosilver2007scaffolding} through concept-before-application ordering \citep{sentance2019primm} and explicit solution checks.

\begin{figure*}[t]
 \centering
 \includegraphics[width=0.99\linewidth,height=0.31\textheight]{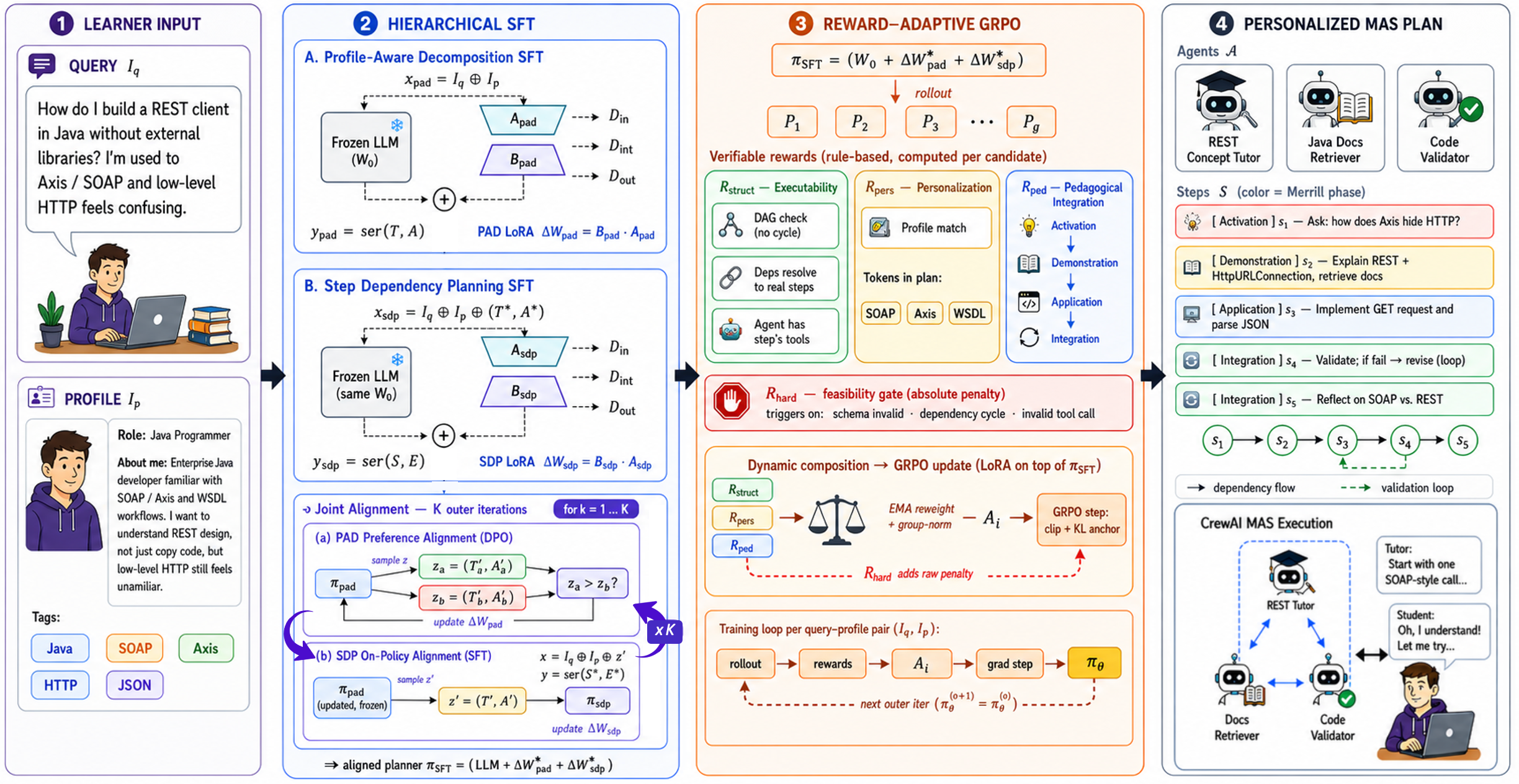}
 \caption{Overview of PersonalPlan. Given a query--profile pair, hierarchical profile-aware SFT first learns the plan scaffold through Profile-Aware Decomposition (PAD) and Step Dependency Planning (SDP) on separate LoRA adapters with a lightweight joint alignment; Reward-Adaptive GRPO then refines full-plan generation under three verifiable rewards: structural validity, personalization, and pedagogy, together with a hard feasibility gate.}
 \label{fig:personalplan-overview}
\end{figure*}

To solve this problem, PersonalPlan uses a two-stage training pipeline: hierarchical profile-aware SFT first learns the profile-conditioned scaffold and step-dependency structure, and GRPO then refines full-plan generation with verifiable rewards for structural validity, learner grounding, and pedagogical coverage, as shown in Figure~\ref{fig:personalplan-overview}.

\subsection{Hierarchical SFT}

An MAS plan orchestrating a tutoring MAS for personalized programming learning combines two decisions: constructing the instructional scaffold, including agents and subtasks, and grounding it into executable steps, dependencies, and tool use. Because one-stage generation entangles these signals and fixes execution details before the pedagogy is stable, we factorize the planning into two LoRA~\citep{hu2022lora} stages: Profile-Aware Decomposition for scaffold generation and Step Dependency Planning for executable workflow construction.

\paragraph{Profile-Aware Decomposition (PAD).}

PAD predicts the high-level scaffold $(\mathcal{T},\mathcal{A})$ from the query--profile pair $(I_q, I_p)$. 
For each training instance, the input is
$x_{\mathrm{pad}}=I_q\oplus I_p$,
and the target sequence is
$y_{\mathrm{pad}}=\operatorname{ser}(\mathcal{T}^{\star},\mathcal{A}^{\star})$,
where $\operatorname{ser}(\cdot)$ serializes the gold subtasks $\mathcal{T}^{\star}$ and pedagogical agents $\mathcal{A}^{\star}$ into the target text format. 
The PAD supervision signal is constructed from the high-level components of the gold plan only, excluding step-level workflow details. 
PAD is trained by minimizing the sequence-level autoregressive SFT objective:
\begin{smalldisplay}
\begin{equation}
\mathcal{L}_{\mathrm{pad}}
=
-\sum_{(x_{\mathrm{pad}},y_{\mathrm{pad}})\in\mathcal{D}_{\mathrm{pad}}}
\log P_{W_{\mathrm{pad}}}
\left(y_{\mathrm{pad}}\mid x_{\mathrm{pad}}\right).
\end{equation}
\end{smalldisplay}
Here,
$W_{\mathrm{pad}} = W_0 + \Delta W_{\mathrm{pad}}$,
where $W_0$ denotes the frozen backbone parameters and $\Delta W_{\mathrm{pad}}$ is the trainable low-rank LoRA update.

\paragraph{Step Dependency Planning (SDP).}

SDP grounds the high-level scaffold into an executable workflow by predicting the step set and dependency graph $G=(\mathcal{S},\mathcal{E})$ from $I_q$, $I_p$, and the gold scaffold $(\mathcal{T}^{\star},\mathcal{A}^{\star})$.
For each training instance, the input is
$x_{\mathrm{sdp}}=I_q\oplus I_p\oplus \mathcal{T}^{\star}\oplus\mathcal{A}^{\star}$,
and the target sequence is
$y_{\mathrm{sdp}}=\operatorname{ser}(\mathcal{S}^{\star},\mathcal{E}^{\star})$.
Thus, the target contains only step-level gold components, while the gold scaffold is used as context to isolate SDP from early PAD errors and learn dependencies, agent--step assignments, tool bindings, and execution order under a clean scaffold.
SDP is trained by minimizing the sequence-level autoregressive SFT objective:
\begin{smalldisplay}
\begin{equation}
\mathcal{L}_{\mathrm{sdp}}
=
-\sum_{(x_{\mathrm{sdp}},y_{\mathrm{sdp}})\in\mathcal{D}_{\mathrm{sdp}}}
\log P_{W_{\mathrm{sdp}}}
\left(y_{\mathrm{sdp}}\mid x_{\mathrm{sdp}}\right).
\end{equation}
\end{smalldisplay}
Here,
$W_{\mathrm{sdp}} = W_0 + \Delta W_{\mathrm{sdp}}$,
where $\Delta W_{\mathrm{sdp}}$ is the trainable low-rank LoRA update for step-dependency planning.

\paragraph{Joint Alignment.}

Although PAD and SDP are trained independently, directly composing them creates a distribution mismatch: SDP is trained on the gold scaffold $z^{\star}=(\mathcal{T}^{\star},\mathcal{A}^{\star})$ but receives the PAD-generated scaffold $z'=(\mathcal{T}',\mathcal{A}')$ at inference. This hierarchical exposure bias~\citep{bengio2015scheduled,ranzato2016sequence} leaves SDP with unseen contexts and gives PAD no downstream grounding signal. We therefore add a lightweight joint alignment stage following on-policy hierarchical alignment and distillation work \citep{yin2024lumos,erdogan2025planandact,agarwal2024onpolicy}: over $K$ batched outer iterations, PAD is preference-aligned toward gold-like scaffolds, and SDP is then adapted to PAD-produced scaffolds. Appendix~\ref{app:joint-alignment-details} gives the candidate scoring rules and losses.

\subsection{Reward-Adaptive GRPO}
\label{sec:grpo}

Hierarchical SFT provides token-level imitation, but plan-level properties such as acyclicity, valid bindings, profile grounding, and pedagogical coverage must be optimized over complete trajectories. We therefore use \emph{Reward-Adaptive GRPO} \citep{shao2024deepseekmath}: for each $(I_q,I_p)$, the model samples plans $\{\mathcal{P}_i\}_{i=1}^{g}$ and scores each parsed plan with three verifiable soft rewards plus a hard feasibility gate. Rewards depend only on $\mathcal{P}_i$ and $I_p$, not a gold plan or LLM judge, and soft rewards are normalized within the rollout group while non-recoverable feasibility failures are handled separately. We next introduce the reward components and GRPO training procedure\footnote{Appendix~\ref{app:grpo-reward-details} gives the exact predicates, token sets, and phase detectors used for all reward subcomponents.}.

\paragraph{Structural Reward.}
An MAS plan $\mathcal{P}_i$ is structurally executable only if its steps admit an acyclic execution order, all declared step prerequisites resolve to valid steps, and each step is assigned to an agent whose tools and capabilities can support its execution. To make executability a verifiable training signal, we combine three rule-based predicates into an equally weighted score:
\begin{smalldisplay}
\begin{equation}
\label{eq:r-struct}
R^{\mathrm{struct}}_i
=
\tfrac{1}{3}\bigl(
\mathrm{DAG}(\mathcal{P}_i) + \mathrm{DC}(\mathcal{P}_i) + \mathrm{ATR}(\mathcal{P}_i)
\bigr).
\end{equation}
\end{smalldisplay}
$\mathrm{DAG}(\mathcal{P}_i)\in\{0,1\}$ indicates whether the induced dependency graph is acyclic, $\mathrm{DC}(\mathcal{P}_i)\in[0,1]$ is the fraction of declared prerequisite edges that resolve to existing steps, and $\mathrm{ATR}(\mathcal{P}_i)\in[0,1]$ combines a step-level agent--tool validity check with a plan-level capability-coverage check. 

\paragraph{Personalization Reward.}
For personalized programming tutoring, an MAS plan should reflect concrete learner-profile signals $\mathcal{C}(I_p)$ (\textit{e.g.}, keywords from self-description that indicate skills, knowledge, prior experience, and background context) in its agent roster, subtask design, and step execution. To reward this profile grounding, we define the personalization reward as the fraction of profile signals covered by the plan content:
\begin{smalldisplay}
\begin{equation}
\label{eq:r-pers}
R^{\mathrm{pers}}_i
=
\frac{
\left|\{c\in\mathcal{C}(I_p): c\text{ appears in }\operatorname{Text}(\mathcal{P}_i)\}\right|
}{
\max(1,|\mathcal{C}(I_p)|)
},
\end{equation}
\end{smalldisplay}
where $\operatorname{Text}(\mathcal{P}_i)$ denotes the concatenated agent, subtask, and step text of the generated plan (\textit{e.g.}, agent name, subtask instruction, expected output).

\paragraph{Pedagogy Reward.}
To make the MAS plans function as tutoring strategies, we reward coverage of a complete instructional cycle: probing prior knowledge, demonstrating the relevant concept or API, guiding application/implementation, and checking the solution. These stages align with Merrill's first principles of instruction \citep{merrill2002first}: \textsc{Activation}, \textsc{Demonstration}, \textsc{Application}, and \textsc{Integration}. Let $\mathcal{M}$ denote this phase set, and let $\operatorname{phase}(\mathcal{P}_i)\subseteq\mathcal{M}$ denote the subset detected in the generated plan $\mathcal{P}_i$. The pedagogy reward is the fraction of required phases covered by the plan:
\begin{smalldisplay}
\begin{equation}
\label{eq:r-ped}
R^{\mathrm{ped}}_i
=
\frac{|\operatorname{phase}(\mathcal{P}_i)|}{|\mathcal{M}|}.
\end{equation}
\end{smalldisplay}
In particular, to encourage interactive plans, we require $\textsc{Activation}\in\operatorname{phase}(\mathcal{P}_i)$ only when an \textsc{Activation}-matched step also requires explicit learner input.

\paragraph{Hard Feasibility Penalty.}
Although the soft rewards encourage structurally valid, personalized, and pedagogically complete plans, they are preference signals rather than feasibility constraints. Following constraint-style multi-objective RL \citep{huang2025mogrpo}, we apply a hard penalty to plans that would cause the downstream MAS to crash before execution because of schema parsing errors, dependency cycles, or invalid tool assignments:
\begin{smalldisplay}
\begin{equation}
\begin{aligned}
R^{\mathrm{hard}}_i
=
-\lambda_{\mathrm{hard}}\,\mathbb{I}\bigl[
&\mathrm{SchemaInvalid}(\mathcal{P}_i)
\vee \mathrm{HasCycle}(G_i)\\
&\vee \mathrm{InvalidToolCall}(\mathcal{P}_i)
\bigr],
\end{aligned}
\end{equation}
\end{smalldisplay}
where $G_i$ is the dependency graph induced by $\mathcal{P}_i$.

\paragraph{Dynamic Composition.}
Optimizing structure, personalization, and pedagogy jointly requires combining rewards with different scales and learning speeds. Following dynamic multi-objective reward balancing \citep{huang2025mogrpo}, we design a two-step composition. First, for each soft reward axis, we convert raw rewards into relative z-scores within the rollout group for the same query--profile pair. Concretely, for each $k\in\{\mathrm{struct},\mathrm{pers},\mathrm{ped}\}$,
\begin{smalldisplay}
\begin{equation}
\bar{R}^k_i = \frac{R^k_i - \mu_g(R^k)}{\sigma_g(R^k) + \varepsilon},
\end{equation}
\end{smalldisplay}
where $\mu_g(R^k)$ and $\sigma_g(R^k)$ are the mean and standard deviation over candidates in that rollout group, and $\varepsilon$ is a small numerical stabilizer. Second, we maintain normalized soft-reward weights $w_k$ over the three axes. During training, an exponential moving average (EMA) tracks each axis's recent raw reward, and a multiplicative update increases $w_k$ for axes that lag behind the target level. The composed per-rollout reward is
\begin{smalldisplay}
\begin{equation}
R_i
=
\sum_{k\in\{\mathrm{struct},\mathrm{pers},\mathrm{ped}\}}
w_k\bar{R}^k_i
+
R^{\mathrm{hard}}_i.
\end{equation}
\end{smalldisplay}
Only the soft rewards are normalized and dynamically reweighted; $R^{\mathrm{hard}}_i$ remains an absolute penalty, so invalid plans cannot compensate with high personalization or pedagogy scores.

\paragraph{GRPO Training.}
When training with GRPO on the composed rewards, we increase the likelihood of higher-reward plans while keeping the updated policy close to the SFT reference model. For each input $I=(I_q,I_p)$, GRPO converts the composed rewards within its rollout group into group-relative advantages
$A_i=(R_i-\mu_R)/(\sigma_R+\varepsilon)$, where $\mu_R$ and $\sigma_R$ are computed over the $g$ sampled plans for the same $I$.
Let $o_i$ denote the token sequence of the $i$-th sampled plan and $o_{i,t}$ its $t$-th token. The token-level probability ratio is
\begin{smalldisplay}
\begin{equation}
 r_{i,t}(\theta)=\frac{\pi_\theta(o_{i,t}\mid I,o_{i,<t})}{\pi_{\theta_{\mathrm{old}}}(o_{i,t}\mid I,o_{i,<t})},
\end{equation}
\end{smalldisplay}
where $\pi_{\theta_{\mathrm{old}}}$ is the policy before the current update. We then optimize
\begin{scriptsizedisplay}
\begin{equation}
\begin{aligned}
\mathcal{L}_{\mathrm{GRPO}}(\theta) ={}&{-}\mathbb{E}_{I,\{o_i\}}\!\Bigl[\tfrac{1}{g}\!\sum_i\!\tfrac{1}{|o_i|}\!\sum_t\!\min\!\bigl(r_{i,t}(\theta)A_i, \\
&\quad\operatorname{clip}(r_{i,t}(\theta),1{-}\epsilon,1{+}\epsilon)A_i\bigr)\Bigr] + \beta\,\mathbb{D}_{\mathrm{KL}}\!\left[\pi_\theta\Vert\pi_{\mathrm{SFT}}\right].
\end{aligned}
\end{equation}
\end{scriptsizedisplay}
The clipping parameter $\epsilon$ limits the size of each policy update, and the fixed KL term weighted by $\beta$ keeps the reward-optimized policy close to the model behavior learned during Hierarchical SFT.

\section{MAP-PPL}
\label{maple_dataset}

\subsection{Dataset Overview}
\label{sec:maple-motivation}
\label{sec:maple-collection}
\label{sec:maple-construction}

To support PersonalPlan and related research on multi-agent planning for personalized programming learning, we construct \textbf{MAP-PPL} (\textbf{M}ulti-\textbf{A}gent \textbf{P}lans for \textbf{P}ersonalized \textbf{P}rogramming \textbf{L}earning). It contains $3{,}043$ query--profile--plan instances $\{I_q,I_p,\mathcal{P}\}$ derived from $1{,}730$ Stack Overflow question groups, their accepted or high-vote answers, and $2{,}738$ unique learner profiles. Each plan $\mathcal{P}{=}(\mathcal{A},\mathcal{T},\mathcal{S},G)$ specifies agents, subtasks, executable steps, and a step-dependency graph. Stack Overflow is a natural source because its duplicate-question groups pair the same technical problem with users from different backgrounds, yielding a one-to-many structure where the problem stays fixed while the plan adapts to the learner profile \citep{barua2014developers,beyer2018automatically}.

\begin{figure}[t]
 \centering
 \includegraphics[width=\columnwidth]{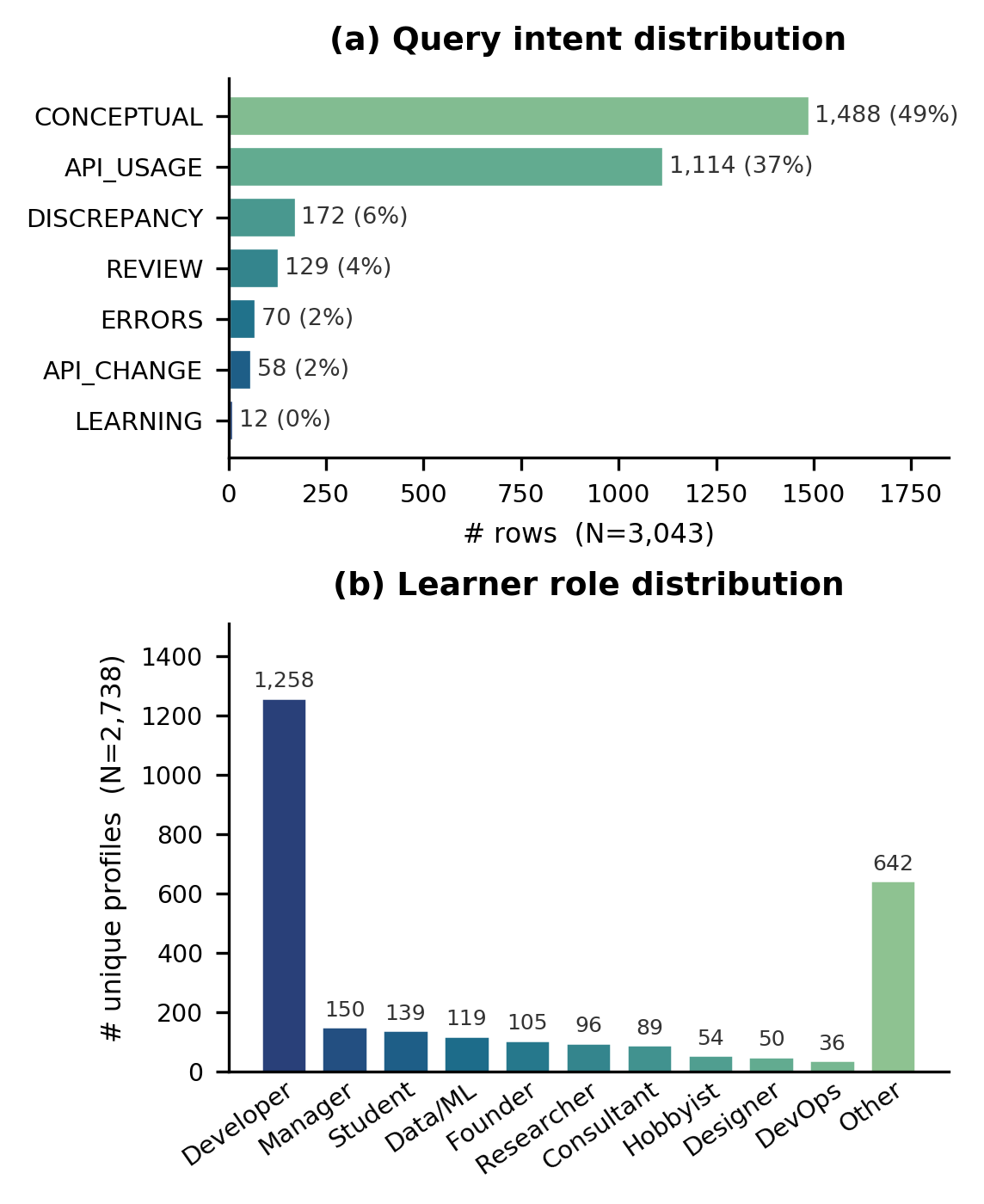}
 \caption{Input-side distributions in \textbf{MAP-PPL}. (a) Primary query intent across the 3{,}043 records in the Stack Overflow-derived source corpus. (b) Learner role across the 2{,}738 unique profiles (deduplicated by \texttt{(about\_me, top\_tags)}).}
 \label{fig:mapppl_inputs}
\end{figure}

\paragraph{Data Collection.} \begingroup\sloppy
We construct Stack Overflow question groups by tracing duplicate links in the \texttt{original\_questions} field, retaining each group's cleaned question, learner profile, and accepted or strongly upvoted answer. We filter out one-line lookups, weak answers, and profiles without concrete personalization signals such as role, stack, experience level, or learning interest. From a raw collection of tens of thousands of pages, $1{,}730$ question groups survive, yielding $3{,}043$ final plans after validation. The source corpus is dominated by conceptual explanation ($1{,}488$, $48.9\%$) and API-usage ($1{,}114$, $36.6\%$) queries, with discrepancy, review, error, API-change, and learning-oriented questions forming the long tail (Figure~\ref{fig:mapppl_inputs}a). The profiles also contain role context beyond tags: $45.9\%$ are developer/engineer profiles, while managers, students, data/ML practitioners, founders, researchers, consultants, hobbyists, designers, and DevOps roles form a diverse tail (Figure~\ref{fig:mapppl_inputs}b). Table~\ref{tab:maple_headline} summarizes the corresponding plan-scale statistics; extended input distributions are in Appendix~\ref{appendix:dataset-overview}.
\endgroup

\paragraph{Plan Generation.} Given a surviving $\langle$question, profile, answer$\rangle$ triple, Claude Sonnet~4.6 receives a plan-generation prompt that uses the answer as the correctness reference, infers the learner's starting point from profile text and tags, chooses a personalization strategy, decomposes the learning gap into observable subtasks, and emits strict JSON with \texttt{agents}, \texttt{subtasks}, \texttt{steps}, \texttt{execution\_order}, and explicit \texttt{depends\_on} edges. We retain only plans that pass deterministic schema checks and an execution-effectiveness gate; all released plans are executable and complete successfully under the MAS executor. The abridged prompt and the detailed validation/audit protocol are reported in Appendix~\ref{appendix:plan-generation-prompt} and Appendix~\ref{appendix:dataset-quality}.

\subsection{Dataset Characteristics}
\label{sec:maple-characteristics}

\begin{table}[t]
 \centering
 \small
 \setlength{\tabcolsep}{4pt}
 \begin{adjustbox}{max width=\columnwidth}
 \begin{tabular}{lr}
 \toprule
 \textbf{Statistic} & \textbf{Value} \\
 \midrule
 \multicolumn{2}{l}{\emph{Scale and one-to-many coverage}}\\
 Query--profile--plan instances       & $3{,}043$ \\
 Unique question groups               & $1{,}730$ \\
 Unique learner profiles              & $2{,}738$ \\
 Questions with $\ge\!2$ profiles     & $971$ ($56.1\%$) \\
 \midrule
 \multicolumn{2}{l}{\emph{Plan structure}}\\
 Agent declarations / unique roles    & $8{,}849$ / $4{,}380$ \\
 Agents per plan (mean / max)         & $2.91$ / $4$ \\
 Subtasks per plan (mean / max)       & $3.98$ / $6$ \\
 Steps total / per plan (mean / max)  & $31{,}580$ / $10.38$ / $23$ \\
 Declared / used tool types           & $8$ / $5$ \\
 \midrule
 \multicolumn{2}{l}{\emph{Executability}}\\
 Executable / DAG-valid plans         & $3{,}043$ ($100.0\%$) \\
 Critical path / layer width (mean)   & $5.45$ / $3.58$ \\
 Inter-agent dependency edges         & $58.5\%$ \\
	 \midrule
	 \multicolumn{2}{l}{\emph{Pedagogical scaffold}}\\
	 Tutor / retriever / validator families & $98.1\%$ of agents  \\
	 Merrill phase coverage (A/D/Ap/I)    & $100.0/98.1/99.7/100.0\%$ \\
	 \quad Phases per plan (4 / 3 / $\leq$2) & $2{,}974$ / $69$ / $0$ \\
	 \bottomrule
 \end{tabular}
 \end{adjustbox}
 \caption{Headline statistics for \textbf{MAP-PPL}. The main text emphasizes scale, one-to-many profile coverage, multi-agent plan structure, executability, and pedagogical scaffolding; extended distributions and action-level pedagogy diagnostics are in Appendix~\ref{appendix:dataset}.}
 \label{tab:maple_headline}
\end{table}

\begin{figure}[t]
 \centering
 \includegraphics[width=\columnwidth]{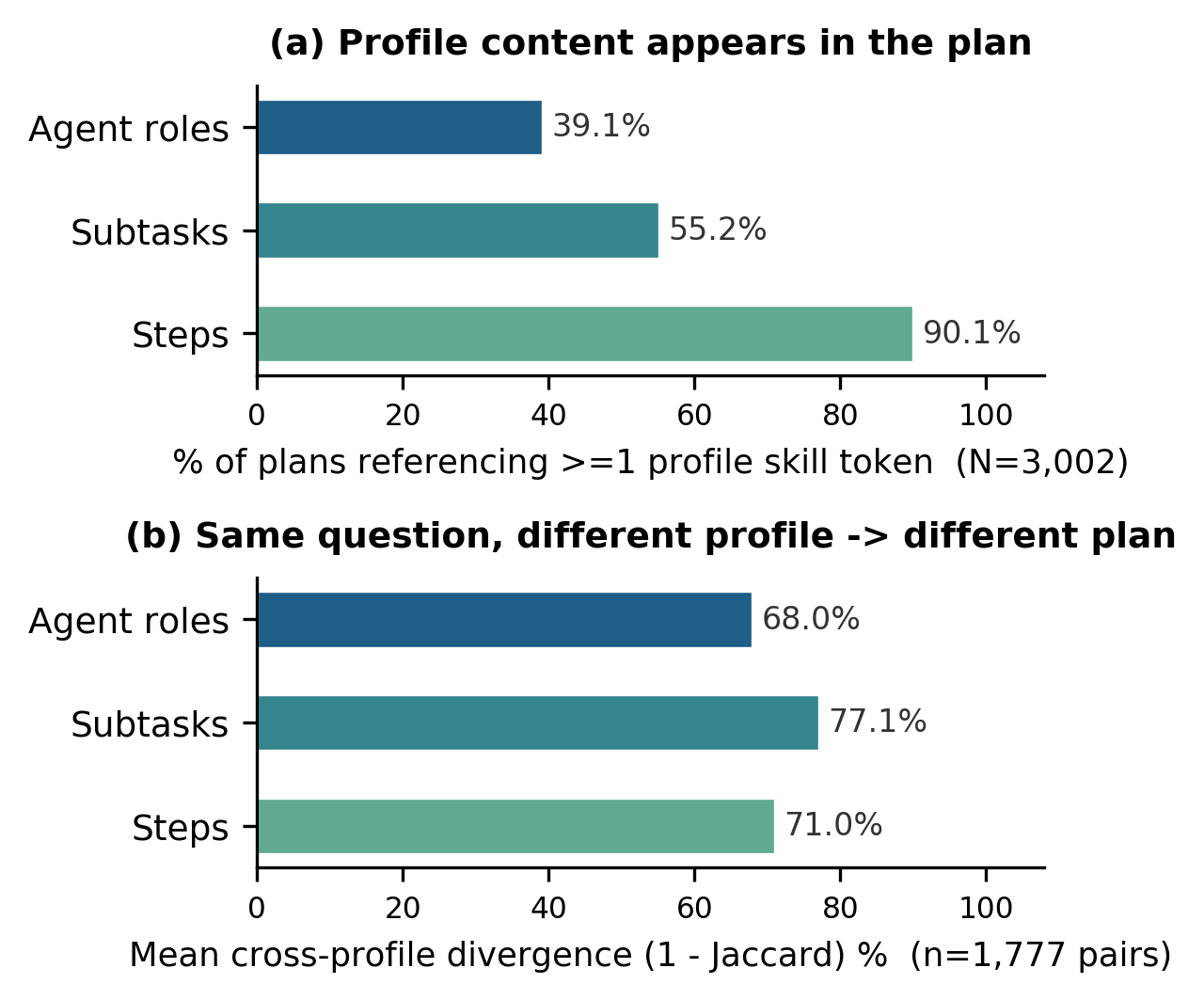}
 \caption{Profile grounding and profile-conditioning effects in \textbf{MAP-PPL}. The figure summarizes how learner-profile signals appear in plan layers and how those layers change under same-question profile swaps.}
 \label{fig:mapppl_personalization_showcase}
\end{figure}

\paragraph{Personalization.}
Personalization in MAP-PPL is both grounded in and conditioned on the learner profile. Figure~\ref{fig:mapppl_personalization_showcase} first measures whether concrete profile skills appear in different plan layers: $39.1\%$ of plans mention profile skills in agent-role names, $55.2\%$ in subtask wording, and $90.1\%$ in step-level instructions. It then holds the programming question fixed and compares $1{,}777$ same-question profile pairs. Swapping only the learner profile produces large rewrites across all layers, with $68.0\%$ agent-role divergence, $77.1\%$ subtask divergence, and $71.0\%$ step-text divergence. Together, these views show that profile signals are not merely copied into the prompt; they re-instantiate the multi-agent teaching plan.

\paragraph{Pedagogy.}
Plans in MAP-PPL align with Merrill's first principles \citep{merrill2002first}: they activate prior knowledge, demonstrate the target concept, ask the learner to apply it, and integrate it through validation or transfer. Agents in MAP-PPL are categorized into tutor roles that explain concepts, retriever/docs roles that find materials, and validator/checker roles that verify learner attempts. These three families of agents account for $98.1\%$ of agent declarations and co-occur in $2{,}534$ plans ($83.3\%$), forming an explain--ground--validate scaffold. At the plan level, the four Merrill phases cover $100.0\%$/$98.1\%$/$99.7\%$/$100.0\%$ of plans, with $2{,}974/3{,}043$ plans ($97.7\%$) containing all four phases and no plan covering two or fewer. Besides, most plans contain concrete pedagogical moves, including Socratic probing, practice, validation, documentation grounding, worked examples, feedback, consolidation, and analogy. Phase order is intentionally not treated as a rigid template; the detailed order diagnostic is reported in Appendix~\ref{appendix:dataset-pedagogy} and Figure~\ref{fig:maple_merrill_order}.

\paragraph{Executability.}
MAP-PPL is not a collection of flat teaching checklists; it contains executable multi-agent workflows with explicit handoffs, tool calls, and dependency constraints. Executability is evidenced by both admission checks and graph structure: all released plans are schema-valid, DAG-valid, and executable under the CrewAI-style MAS executor, with non-trivial prerequisite depth, schedulable parallelism, and cross-agent coordination. The consolidated executability audit is reported in Appendix~\ref{appendix:dataset-executability}.

\section{Experiments}
\label{experiments}
To validate whether PersonalPlan can generate executable and personalized multi-agent tutoring plans, we conduct extensive experiments on MAP-PPL against state-of-the-art MAS planning methods.

\subsection{Baselines and Implementation}
\label{sec:baselines}
We compare PersonalPlan with three groups of planner baselines: frontier LLM planners (GPT-5.4~\citep{openai2025gpt5}, Claude Opus~4.6~\citep{anthropic2026claude46}, and Qwen3-Max~\citep{qwen2025qwen3max}), generic MAS framework planners (AutoGen~\citep{autogen} and AutoAgents~\citep{chen2024autoagents}), and agentic workflow planners (AIPOM~\citep{kim2025aipom} and AFlow~\citep{zhang2025aflow}). These groups cover direct LLM planning, general-purpose agent orchestration, and explicit workflow construction; the selection rationale and excluded tutoring systems are detailed in Appendix~\ref{app:implementation-details}.

For fair comparison, all open or framework-based baselines and PersonalPlan use Qwen3-32B-Instruct as the backbone, share the same tool pool, and are converted to the same MAP-PPL plan schema. Detailed implementation settings, the shared baseline plan-generation prompt, and the evaluation metrics are provided in Appendix~\ref{app:implementation-details} and Appendix~\ref{sec:eval-metrics}. 

\subsection{Data Splits}
\label{sec:splits}
We split MAP-PPL by \texttt{question\_id}, so all learner profiles attached to the same programming question stay in the same partition. This prevents profile variants of one question from leaking across training and testing. The split is stratified by the number of profiles per question, preserving the one-to-many structure that is central to profile-conditioned planning. The held-out test set contains $305$ query--profile--plan instances from $173$ question groups, and all final comparisons are reported on this split. The remaining $2{,}738$ instances (the $3{,}043$ total minus the $305$ held-out test instances), drawn from the other $1{,}557$ question groups, are used to train PersonalPlan.

\subsection{Evaluation Method}
\label{sec:evaluation-protocol}
We evaluate PersonalPlan and all baselines in two categories. 

\emph{Static Plan Quality} measures the plan properties that can be inspected before running a real MAS. It covers three aspects: \emph{(1) Executability}, which checks the average retry rate (\textbf{Atps}) for generating all executable plans, residual format repair rate after three reruns (\textbf{RR}), tool binding quality (\textbf{TBQ}) measuring
whether tool assignments are semantically appropriate for
the agent role and the current step, and dependency-graph similarity to the gold plan (\textbf{TS}); \emph{(2) Personalization}, which measures LLM-judged profile fit (\textbf{Pers.}) and two profile-counterfactual probes: profile-induced structural variation (\textbf{PVS}) and personalization advantage of target profile over random profiles (\textbf{PNG}); and \emph{(3) Pedagogy}, which measures whether the plan teaches rather than
simply delivers the solution, via a composite \textbf{Ped.} score assessed by LLM judges and rule-based checks from: prerequisite-respecting progression,
no-direct-answer guidance, Merrill-phase coverage, and
profile-appropriate instructional style.

\emph{Plan Execution Quality} measures outcomes after the generated plan is executed by an MAS and produces an agents--learner trace. For all methods, we instantiate the same CrewAI runtime: GPT-4o agents execute the planned roles and interact with GPT-4o-mini as a simulated learner conditioned on the learner profile. We report deterministic structural compactness (\textbf{SCS}) of the MAS execution trace (fewer steps, fewer trivial outputs, and declared tools actually invoked), post-execution pedagogical quality (\textbf{PQS}) computed from the teacher--learner transcript, and post-dialog problem solve rate based on final-code correctness (\textbf{$r_{\text{sol}}$}).  We additionally report profile-conditioned pairwise preference rates under the \textbf{Satisfaction} (Sati.) protocol in Fig.~\ref{fig:execution-pairwise-preferences}.  All metrics are defined in Appendix~\ref{sec:eval-metrics}, and all judge-based metrics are supported by rubrics in Appendix~\ref{app:judge-rubrics} and anti-hacking safeguards in Appendix~\ref{app:anti-hacking}.

\section{Result Analysis}
\label{result_analysis}
\begin{table*}[!t]
\centering
\caption{Evaluation results. Arrows indicate whether lower ($\downarrow$) or higher ($\uparrow$) values are better.}
\label{tab:main_results}
\renewcommand{\arraystretch}{1.08}
\setlength{\tabcolsep}{3pt}
\small
\resizebox{\textwidth}{!}{%
\begin{tabular}{l|cccc|ccc|c|ccc}
\toprule
\textbf{Method} &
\multicolumn{8}{c|}{\textbf{Static Plan Quality}} &
\multicolumn{3}{c}{\textbf{Plan Execution Quality}} \\
&
\multicolumn{4}{c|}{\textbf{Executability}} &
\multicolumn{3}{c|}{\textbf{Personalization}} &
\textbf{Pedagogy} &
\multicolumn{3}{c}{\textbf{}} \\
& Atps$\downarrow$ & RR$\downarrow$ & TBQ$\uparrow$ & TS$\uparrow$ & Pers.$\uparrow$ & PVS$\uparrow$ & PNG$\uparrow$ & Ped.$\uparrow$ & SCS$\uparrow$ & PQS$\uparrow$ & $r_{\text{sol}}\uparrow$ \\
\midrule
\multicolumn{12}{l}{\textit{Frontier LLM planners}}\\
GPT-5.4~\citep{openai2025gpt5}                & \underline{1.00} & \underline{0.00} & 0.55 & 0.43 & 0.53 & 0.47 & 0.10 & 0.62 & 0.35 & 0.33 & 0.90 \\
Claude Opus 4.6~\citep{anthropic2026claude46} & 1.09 & 0.04 & 0.64 & 0.59 & 0.67 & 0.39 & 0.24 & 0.55 & 0.26 & 0.39 & 0.81 \\
Qwen3-Max~\citep{qwen2025qwen3max}            & 1.06 & 0.00 & 0.61 & 0.75 & 0.49 & 0.29 & 0.22 & 0.47 & 0.46 & 0.20 & 0.81 \\
\midrule
\multicolumn{12}{l}{\textit{Generic MAS framework planners}}\\
AutoGen~\citep{autogen}                       & 1.04 & 0.01 & 0.61 & 0.76 & 0.29 & 0.37 & 0.08 & 0.31 & 0.84 & 0.38 & 0.83 \\
AutoAgents~\citep{chen2024autoagents}         & 2.12 & 0.03 & 0.63 & 0.71 & 0.28 & 0.34 & 0.03 & 0.33 & 0.75 & 0.35 & \underline{0.92} \\
\midrule
\multicolumn{12}{l}{\textit{Agentic workflow planners}}\\
AIPOM~\citep{kim2025aipom}                    & 2.03 & 0.01 & 0.58 & 0.79 & 0.31 & 0.33 & 0.02 & 0.35 & 0.55 & 0.40 & 0.81 \\
AFlow~\citep{zhang2025aflow}                  & 2.06 & 0.01 & 0.56 & 0.79 & 0.26 & 0.40 & 0.06 & 0.33 & 0.36 & 0.42 & 0.89 \\
\midrule
\textbf{PersonalPlan (Qwen3-8B-Instruct)}                  & 1.03 & \underline{0.00} & 0.69 & \textbf{0.82} & 0.70 & 0.48 & \textbf{0.24} & 0.61 & \textbf{0.88} & 0.36 & 0.77 \\
\textbf{PersonalPlan (Qwen3-32B-Instruct)}                 & \bf{1.00} & \textbf{0.00}& \textbf{0.74}	& 0.81 & \textbf{0.76} & \textbf{0.57} & 0.22 & \textbf{0.65} & 0.82 & \textbf{0.56} & \textbf{0.92} \\
\bottomrule
\end{tabular}%
}
\end{table*}

\subsection{Static Plan Quality}
\label{sec:main-results}

Table~\ref{tab:main_results} shows that \textbf{PersonalPlan} attains the most consistently balanced static plan quality before execution, and the two model sizes split the gains in a telling way. Going from 8B to 32B mainly improves profile-conditioning and pedagogy. The 32B variant leads tool binding (TBQ), profile fit (Pers.), profile-induced structural variation (PVS), and pedagogy (Ped.), while the smaller 8B variant stays best on raw dependency-graph similarity (TS) and ties for the best profile-sensitivity score (PNG). Both remain the most admissible plans, with the best Atps and RR. The clearest evidence of personalization is the profile-counterfactual pair. PVS asks whether a plan restructures when only the learner profile changes, and PNG asks whether the intended profile receives a better-matched plan than a swapped one. PersonalPlan leads PVS outright and stays competitive on PNG, which is consistent with conditioning plan \emph{structure} on the learner instead of emitting one generically strong plan. Frontier LLMs are competitive on individual cells but do not reproduce this balance, and the generic MAS and workflow planners make the contrast sharpest. They keep competitive dependency structure (TS) yet collapse on the profile-swap probe (PNG $0.02$--$0.08$ versus $0.22$--$0.24$ for PersonalPlan) and on pedagogy, producing valid-looking workflows whose structure barely responds to the learner.

\subsection{Plan Execution Results}
\label{sec:execution-results}
The post-execution block of Table~\ref{tab:main_results} evaluates whether static plans can be faithfully executed in a tutoring MAS, measuring compactness with SCS, realized pedagogy with PQS, and final-code correctness with $r_{\text{sol}}$.

\begin{figure}[t]
\centering
\includegraphics[width=0.92\columnwidth]{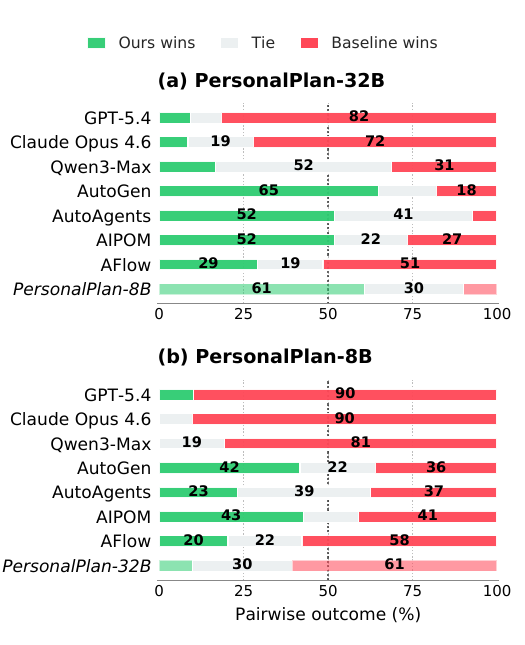}
\caption{\textbf{Profile-conditioned pairwise preferences over executed interaction traces.} GPT-5.4, Claude Opus 4.6, and Gemini 3.1 Pro compare PersonalPlan against each baseline over $305$ MAP-PPL test set instances after execution.}
\label{fig:execution-pairwise-preferences}
\end{figure}

PersonalPlan separates these axes instead of trading one for another. The 8B model produces the most compact execution traces (best SCS), while the 32B model sustains the richest realized tutoring (best PQS) and ties for the best post-dialog solve rate ($r_{\text{sol}}$). The profile-conditioned \textbf{Satisfaction} (Sati.) preferences in Figure~\ref{fig:execution-pairwise-preferences} agree at the holistic level, but with an honest ceiling. PersonalPlan-32B is clearly preferred over AutoGen ($65\%$ wins), more narrowly over AutoAgents and AIPOM ($52\%$ wins each) and over its own 8B variant ($61\%$ wins), and it roughly ties Qwen3-Max ($52\%$ \emph{ties}), but it still trails GPT-5.4 and Claude Opus~4.6. We read this gap as informative rather than contradictory. PersonalPlan's measured advantage is concentrated in plan \emph{structure}, namely personalization, tool binding, and pedagogical coverage. The residual preference against the strongest frontier planners plausibly reflects axes that our holistic protocol does not isolate, such as surface dialogue quality at frontier scale, rather than weaker planning, and we leave a human-rated decomposition of it to future work. Structurally faithful planning therefore closes much of the gap to far larger frontier planners at a fraction of their scale, while a holistic-preference gap to the strongest two remains.

\subsection{Ablation Study}
\label{sec:ablation}
The ablation analysis focuses on \emph{static plan quality}: executable structure, personalization, and pedagogy. This isolates the training pipeline itself, where PAD/SDP build the skeleton, joint alignment stabilizes it, and GRPO refines tool binding and topology.

\begin{figure}[t]
\centering
\includegraphics[width=\columnwidth]{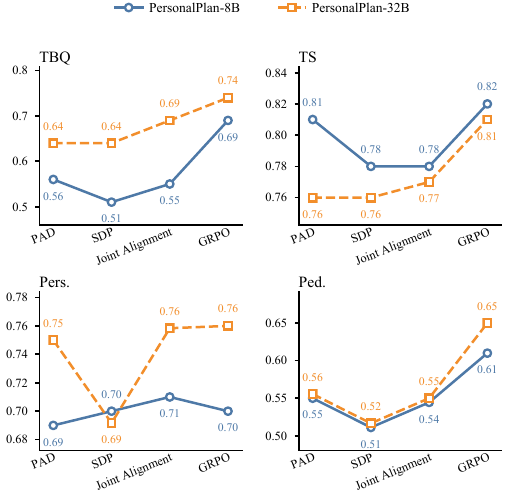}
\caption{\textbf{Metric-wise ablation trends.} Static-quality scores for 8B and 32B variants across PAD-SFT, SDP-SFT, joint alignment, and GRPO. }
\label{fig:ablation-metric-fork}
\end{figure}

Figure~\ref{fig:ablation-metric-fork} decomposes the pipeline into imitation,
alignment, and reward optimization. Two-stage SFT (PAD then SDP) establishes a
usable planning scaffold, and joint alignment makes it more coherent across the
two stages, plausibly by narrowing the train/inference scaffold mismatch. The
largest gains, however, come from GRPO, which explicitly targets plan-level
properties that token-level SFT objectives do not score directly, namely tool
binding, dependency topology, and pedagogical phase coverage. The clearest jumps
appear at the joint-alignment$\rightarrow$GRPO step (for instance, 8B tool
binding rises from $0.55$ to $0.69$), which indicates that these harder, global
properties are best acquired under verifiable plan-level reward rather than
imitation alone. The two scales stay complementary across stages, so the choice
between 8B and 32B is best read as a compactness-versus-richness trade-off rather
than a strict ordering.

\section{Conclusion and Future Work}
\label{conclusions}
In this paper, we introduced \textbf{PersonalPlan}, a profile-conditioned
multi-agent planning framework for personalized programming-learning MAS that
combines hierarchical SFT, joint alignment, and reward-adaptive GRPO with
verifiable rewards for executable structure, profile grounding, and pedagogy. We
also constructed \textbf{MAP-PPL}, a 3{,}043-instance query--profile--plan
dataset with executable plans that expose profile-grounded structure and
pedagogical scaffolding. Experiments show that PersonalPlan improves
executability, personalization, and pedagogy across static plan-quality metrics,
MAS execution, and profile-conditioned pairwise preferences over executed interaction traces.
Several directions remain open. One is to let learner profiles evolve over
time. Our profiles are fixed when a plan is generated, but a deployed tutor
follows the same student across many sessions, so updating the profile from
observed progress would turn one-shot personalization into a longitudinal
process. A second direction is to test how far profile-conditioned planning
generalizes beyond programming. MAP-PPL is built from programming questions, and
we do not yet know whether the same approach transfers to neighboring technical
subjects such as data analysis or mathematics. A third direction follows from
the residual preference gap in Section~\ref{sec:execution-results}, where
PersonalPlan wins on plan structure but still trails the strongest frontier
planners on holistic satisfaction. Pairing our plans with a stronger executor,
and checking the LLM-judged scores against real human raters, would reveal
whether that gap comes from execution quality rather than from the plan itself.

\clearpage
\section*{Limitations}
Currently, the system generates and executes plans without live input from the learner. Although our online simulation offers a controlled proxy, it cannot fully capture the unpredictability of real student behavior. Moving forward, the most logical extension is to ground the system in real-world interactions, such as treating per-step solve-rate gains as verifiable rewards to continuously align the pedagogy \citep{dinucu2025aligningpedagogy}. This would also allow us to test the system's adaptability to learner feedback and evolving needs, which is crucial for personalized education.

\section*{Ethical Considerations}
MAP-PPL is constructed from public Stack Overflow content and profile text. Before release, profile fields should be filtered for personally identifying details, normalized to coarse skill and background attributes, and distributed under licenses compatible with the source platform. PersonalPlan is designed as a planning assistant for educators or learners, not as an automated replacement for instruction; generated plans can encode incorrect assumptions about a learner's prior knowledge, so deployments should keep teacher or learner review in the loop. Because profile-conditioned generation may amplify stereotypes if profiles contain sensitive attributes, the released benchmark and model card should document filtering rules, supported profile fields, and intended educational use.

\bibliography{sample}

\appendix
\section{Appendix}
\label{Appendix}

\subsection{Method Details}
\label{app:method-details}

This appendix details the method components of Section~\ref{methodology}: the joint-alignment losses and the exact reward computation used by Reward-Adaptive GRPO.

\subsubsection{Joint Alignment Details}
\label{app:joint-alignment-details}

This appendix expands the lightweight joint alignment stage introduced in Section~\ref{methodology}. Each outer iteration alternates between a PAD preference update and an SDP on-policy update.

\paragraph{PAD Preference Alignment.}
Let $\pi_{\mathrm{pad}}$ denote the distribution induced by the current PAD adapter over $z=(\mathcal{T},\mathcal{A})$. For each query--profile pair, we sample two candidates
$z_a=(\mathcal{T}^{\prime}_a,\mathcal{A}^{\prime}_a)$ and
$z_b=(\mathcal{T}^{\prime}_b,\mathcal{A}^{\prime}_b)$ from
$\pi_{\mathrm{pad}}(\cdot\mid I_q,I_p)$. Each candidate is scored against the gold $z^{\star}$ with a rule-based structural similarity:

\begin{smalldisplay}
\begin{equation}
s(z)
=
\tfrac{1}{2}\operatorname{Jaccard}(\mathcal{T}^{\prime},\mathcal{T}^{\star})
+
\tfrac{1}{2}\operatorname{Jaccard}(\mathcal{A}^{\prime},\mathcal{A}^{\star}),
\end{equation}
\end{smalldisplay}
where both subtasks and agent rosters are compared as sets. The higher-scoring candidate is treated as the preferred output $z^{+}$ and the lower-scoring one as the rejected output $z^{-}$. We update the PAD adapter $\Delta W_{\mathrm{pad}}$ by minimizing
\begin{smalldisplay}
\begin{equation}
\begin{aligned}
&r_{\mathrm{pad}}(z)
=
\log\pi_{\mathrm{pad}}(z\mid I_q,I_p)
-
\log\pi_{\mathrm{pad}}^{\mathrm{ref}}(z\mid I_q,I_p), \\
&\mathcal{L}^{\mathrm{ja}}_{\mathrm{pad}}
=
-\mathbb{E}_{(z^{+},z^{-})}
\![\log\sigma(\beta(r_{\mathrm{pad}}(z^{+})\!-r_{\mathrm{pad}}(z^{-})))].
\end{aligned}
\end{equation}
\end{smalldisplay}
Here, $\pi_{\mathrm{pad}}^{\mathrm{ref}}$ is the reference distribution before alignment, and $\beta$ is the temperature parameter.

\paragraph{SDP On-Policy Alignment.}
After PAD Preference Alignment, we freeze the updated PAD adapter $\Delta W_{\mathrm{pad}}$ and adapt SDP to PAD-produced scaffolds. For each query--profile pair, we sample
$z'=(\mathcal{T}^{\prime},\mathcal{A}^{\prime})\sim\pi_{\mathrm{pad}}(\cdot\mid I_q,I_p)$
and construct
$x^{\mathrm{ja}}_{\mathrm{sdp}}=I_q\oplus I_p\oplus z'$, while the target remains the gold step-dependency serialization
$y_{\mathrm{sdp}}^{\star}=\operatorname{ser}(\mathcal{S}^{\star},\mathcal{E}^{\star})$.
We fine-tune the existing SDP LoRA $\Delta W_{\mathrm{sdp}}$ by minimizing
\begin{smalldisplay}
\begin{equation}
\mathcal{L}^{\mathrm{ja}}_{\mathrm{sdp}}
=
-\mathbb{E}_{z'\sim\pi_{\mathrm{pad}}(\cdot\mid I_q,I_p)}
\log P_{W_{\mathrm{sdp}}}
\left(y_{\mathrm{sdp}}^{\star}\mid I_q,I_p,z'\right).
\end{equation}
\end{smalldisplay}
This update adapts SDP to the PAD distribution observed at inference time.

\subsubsection{GRPO Reward Computation Details}
\label{app:grpo-reward-details}

This appendix specifies the four reward components and the composition rule introduced in Section~\ref{sec:grpo}. All rewards are deterministic functions of the candidate plan and learner profile; none reads a reference plan or calls a learned model at training time.

\paragraph{Structural Reward.}
\label{app:reward-struct}

Let $\mathcal{P}_i$ denote a candidate plan and $G_i=(\mathcal{S}_i,\mathcal{E}_i)$ its step-dependency graph induced from \texttt{depends\_on} fields. Following Eq.~\ref{eq:r-struct}, the structural reward is
\[
R^{\mathrm{struct}}_i
=
\tfrac{1}{3}\bigl(
\mathrm{DAG}(\mathcal{P}_i)
+ \mathrm{DC}(\mathcal{P}_i)
+ \mathrm{ATR}(\mathcal{P}_i)
\bigr).
\]
This appendix specifies each of the three plan-level predicates.

\paragraph{Acyclicity (DAG).}
A linear-time cycle detector based on topological-sort traversal reports whether $G_i$ contains any directed cycle:
\begin{smalldisplay}
\begin{equation}
\mathrm{DAG}(\mathcal{P}_i) = \mathbb{I}[G_i \text{ admits a topological order}] \in \{0,1\}.
\end{equation}
\end{smalldisplay}

\paragraph{Dependency Completeness (DC).}
A declared edge $(u,v)\in\mathcal{E}_i$ is valid when its source resolves to a known step in $\mathcal{S}_i$:
\begin{smalldisplay}
\begin{equation}
\mathrm{DC}(\mathcal{P}_i) =
\begin{cases}
1 & \text{if } \mathcal{E}_i = \emptyset, \\[2pt]
\dfrac{\bigl|\{(u,v)\in\mathcal{E}_i : u \in \mathcal{S}_i\}\bigr|}{|\mathcal{E}_i|} & \text{otherwise.}
\end{cases}
\end{equation}
\end{smalldisplay}

\paragraph{Agent--Tool Relevance (ATR).}
ATR averages a local step-level validity check (subscore A) and a plan-level capability coverage check (subscore B):
\begin{smalldisplay}
\begin{equation}
\mathrm{ATR}(\mathcal{P}_i) = \tfrac{1}{2}\bigl(\mathrm{ATR}_{\mathrm{A}}(\mathcal{P}_i) + \mathrm{ATR}_{\mathrm{B}}(\mathcal{P}_i)\bigr).
\end{equation}
\end{smalldisplay}
Writing $\mathrm{tool}(s)$ for the tool declared by step $s$, $\mathrm{agent}(s)$ for the responsible agent, $\mathrm{tools}(a)$ for the agent's declared tool set, $\mathcal{S}_i^{\text{tool}} = \{s : \mathrm{tool}(s) \neq \varnothing\}$, and $v(s)=\mathbb{I}[\mathrm{tool}(s)\in\mathrm{tools}(\mathrm{agent}(s))]$,
\begin{smalldisplay}
\begin{equation}
\mathrm{ATR}_{\mathrm{A}}(\mathcal{P}_i) =
\begin{cases}
1 & \text{if } \mathcal{S}_i^{\text{tool}} = \emptyset, \\[2pt]
\dfrac{1}{|\mathcal{S}_i^{\text{tool}}|}
\sum_{s\in\mathcal{S}_i^{\text{tool}}}
v(s)
& \text{otherwise.}
\end{cases}
\end{equation}
\end{smalldisplay}
Subscore B checks whether the plan as a whole declares the tools implied by the step text. A lexical \emph{intent detector} matches keyword regexes against each step's instruction, objective, and expected-output fields and returns the set $\mathrm{Intent}(s)$ of tools the text implies, e.g., \texttt{CodeInterpreterTool} for \emph{execute / run / verify / compile}, or \texttt{CodeDocsSearchTool} for \emph{retrieve / documentation / specification}. Patterns for frequent tools are mined from gold-step token statistics with recall validated on gold ($99.6\%$ for \texttt{CodeInterpreterTool}, $100\%$ for \texttt{CodeDocsSearchTool}, $98.6\%$ for \texttt{FirecrawlSearchTool}); patterns for sparse tools are derived from tool-name semantics. Letting $\mathcal{I}_i = \bigcup_{s \in \mathcal{S}_i} \mathrm{Intent}(s)$ and $\mathcal{W}_i = \bigcup_{a\in\mathcal{A}_i} \mathrm{tools}(a)$,
\begin{smalldisplay}
\begin{equation}
\mathrm{ATR}_{\mathrm{B}}(\mathcal{P}_i) =
\begin{cases}
1 & \text{if } \mathcal{I}_i = \emptyset, \\[2pt]
\dfrac{|\mathcal{I}_i \cap \mathcal{W}_i|}{|\mathcal{I}_i|} & \text{otherwise.}
\end{cases}
\end{equation}
\end{smalldisplay}
Subscore B is intentionally plan-level rather than step-level: many gold steps contain teaching narratives such as ``ask the learner to run X'' that match a code-execution intent regex but do not require the agent itself to carry the tool (the learner does). Aggregating intents across the plan avoids penalizing such narrative steps. Subscore A retains the strict step-level check as a backstop, so an agent can never call a tool it has not declared.

\paragraph{Profile-Grounded Personalization Reward.}
\label{app:reward-pers}

The main text treats the profile input abstractly through $\mathcal{C}(I_p)$ in Eq.~\ref{eq:r-pers}. In the MAP-PPL implementation, $I_p$ contains a structured tag list $\text{tags}(I_p)$ and a free-text self-description $\text{about}(I_p)$. We instantiate $\mathcal{C}(I_p)=\mathcal{K}_p\cup\mathcal{D}_p$ from the following two sources:
\begin{itemize}\itemsep1pt
  \item $\mathcal{K}_p = \{\mathrm{lower}(t) : t \in \text{tags}(I_p)\}$: lower-cased skill tags.
  \item $\mathcal{D}_p$: domain tokens of length $\ge 4$ extracted from $\text{about}(I_p)$ using the regex \texttt{[A-Za-z][A-Za-z0-9\_+\#\textbackslash-\textbackslash.]+}, then filtered against a stopword list of (a) common English function words and (b) generic profession words such as \emph{developer}, \emph{engineer}, \emph{experience}, \emph{working}. $\mathcal{D}_p$ is further deduplicated against $\mathcal{K}_p$ so that a token already in the tag set is not double-counted.
\end{itemize}
Let $\operatorname{Text}(\mathcal{P}_i)$ denote the lower-cased concatenation of every agent's role, goal, and backstory; every subtask's name and objective; and every step's instruction, objective, and expected output. The matched profile-signal set is
\begin{smalldisplay}
\begin{equation}
\operatorname{match}(I_p,\mathcal{P}_i)
=
\{c \in \mathcal{C}(I_p) : c \text{ appears in } \operatorname{Text}(\mathcal{P}_i)\}.
\end{equation}
\end{smalldisplay}
Substring matches use no word boundaries and no stemming, which is asymmetric in direction: the tag \texttt{json} matches a plan text containing \texttt{json.net}, but the tag \texttt{rails-activerecord} does not match a plan that only mentions \texttt{activerecord}. If $\mathcal{C}(I_p)$ is empty, Eq.~\ref{eq:r-pers} returns zero through its $\max(1,|\mathcal{C}(I_p)|)$ denominator.

\paragraph{Pedagogy Reward.}
\label{app:reward-ped}

We track the four Merrill-aligned teaching phases used in the main reward: \textsc{Activation}, \textsc{Demonstration}, \textsc{Application}, and \textsc{Integration}; let $\mathcal{M}$ denote this phase set. The Problem-centred principle is treated as a dataset-level by-construction condition rather than a per-plan check, since the programming query is given by the user and not produced by the model.

A lexical detector inspects the instruction, objective, and expected-output fields of each step and returns one phase in $\mathcal{M}$ or $\varnothing$ if no signature fires. Aggregating detector outputs over all steps gives $\operatorname{phase}(\mathcal{P}_i)$ in Eq.~\ref{eq:r-ped}. For \textsc{Activation}, lexical evidence alone is insufficient; the phase is counted only when the matched step also requires learner input:
\begin{smalldisplay}
\begin{equation}
\mathbf{1}_{\mathrm{Act}}(\mathcal{P}_i)
=
\mathbb{I}\!\left[
\exists s\in\mathcal{P}_i:
\mathrm{ActSig}(s)\wedge \mathrm{LearnerInput}(s)
\right],
\end{equation}
\end{smalldisplay}
This implements Merrill's requirement that \textsc{Activation} actively elicits learner input, not merely mentions the learner. The reward thus takes values in $\{0,\tfrac{1}{4},\tfrac{1}{2},\tfrac{3}{4},1\}$. The reward does not enforce a fixed global order on phases. Merrill's principles form a cycle that can be iterated within a problem-centred plan, and only $16.6\%$ of MAP-PPL gold plans satisfy the strict canonical prefix from Activation to Demonstration to Application to Integration; an order constraint would therefore mis-penalize a large fraction of pedagogically valid plans. We also do not include a plan-size envelope, since plan size is not a Merrill condition and earlier thresholded versions calibrated against gold percentiles risked rewarding superficial size mimicry.

\paragraph{Hard Feasibility Penalty.}
\label{app:reward-hard}

The hard penalty fires for failures that prevent the plan from being executed at all:
\begin{smalldisplay}
\begin{equation}
\begin{aligned}
R^{\mathrm{hard}}_i
=
-\lambda_{\mathrm{hard}}\,\mathbb{I}\bigl[
&\mathrm{SchemaInvalid}(\mathcal{P}_i)
\vee \mathrm{HasCycle}(G_i)\\
&\vee \mathrm{InvalidToolCall}(\mathcal{P}_i)
\bigr],
\end{aligned}
\end{equation}
\end{smalldisplay}
with $\lambda_{\mathrm{hard}} = 10$. The three predicates cover non-recoverable execution failures:
\begin{itemize}\itemsep1pt
  \item \textbf{Schema invalidity:} the model output is not parsable JSON, lacks required \texttt{agents}/\texttt{subtasks} blocks, or omits required step fields (\texttt{id}, \texttt{agent}, \texttt{instruction}).
  \item \textbf{Cyclic dependency:} the same acyclicity predicate as $\mathrm{DAG}(\mathcal{P}_i)$ in the structural reward, escalated here from a soft signal to a hard gate.
  \item \textbf{Invalid tool call:} a step references a tool outside the fixed palette $\Pi$, or assigns a palette-valid tool to an agent that has not declared it. The palette $\Pi$ contains code execution, documentation search, web search, file I/O, retrieval, and paper-search tools.
\end{itemize}
The value $\lambda_{\mathrm{hard}} = 10$ is calibrated to the soft-reward scale: each soft reward lies in $[0,1]$, so the un-z-scored soft contribution to $R_i$ is at most $\sum_k w_k = 1$; with $\lambda_{\mathrm{hard}}=10$, no combination of soft scores can override an invalid plan, while larger values would inflate the within-group standard deviation enough to compress z-scored soft advantages between valid candidates to near zero. Earlier versions of this gate also penalized plans with zero learner-interaction steps and plans with no code-tool step; both were removed after we observed entire rollout groups failing the same binary check, which collapsed group variance and erased the GRPO advantage signal. The learner-interaction requirement is now expressed softly through the \textsc{Activation} phase of $R^{\mathrm{ped}}$, and the code-tool requirement through $\mathrm{ATR}_{\mathrm{B}}$ of $R^{\mathrm{struct}}$.

\paragraph{Group Normalization and Dynamic Reweighting.}
\label{app:reward-composition}

Within a rollout group $\mathcal{G}_I$ of candidates sharing the same input $I=(I_q,I_p)$, each soft axis $k \in \{\mathrm{struct}, \mathrm{pers}, \mathrm{ped}\}$ is z-scored:
\begin{smalldisplay}
\begin{equation}
\bar{R}^k_i =
\begin{cases}
\dfrac{R^k_i - \mu_{I}(R^k)}{\sigma_{I}(R^k) + \varepsilon} & \text{if } \sigma_{I}(R^k) > \tau, \\[4pt]
0 & \text{otherwise.}
\end{cases}
\end{equation}
\end{smalldisplay}
where $\mu_I$ and $\sigma_I$ are the within-group mean and standard deviation, $\varepsilon$ is a numerical stabilizer, and the $\tau$ threshold prevents zero-variance groups from blowing up the z-score. The hard penalty $R^{\mathrm{hard}}_i$ is never z-scored, so it preserves its absolute magnitude even when an entire rollout group is invalid.

Soft-axis weights $(w_{\mathrm{struct}}, w_{\mathrm{pers}}, w_{\mathrm{ped}})$ are initialized to $(\tfrac{1}{3}, \tfrac{1}{3}, \tfrac{1}{3})$. At each training step, we maintain an exponential moving average (EMA) of the raw batch mean per axis,
\begin{smalldisplay}
\begin{equation}
m^{(t)}_k = (1-\alpha)\,m^{(t-1)}_k + \alpha\,\overline{R^k}_{\text{batch}},
\end{equation}
\end{smalldisplay}
with $\alpha = 0.10$. During a warmup of the first $T_{\text{warm}}=50$ steps the EMAs are tracked but the weights are not yet updated. After warmup, every $T_{\text{refresh}}=100$ steps we apply a clipped multiplicative update against a reference target $\rho$:
\begin{smalldisplay}
\begin{equation}
\begin{aligned}
\mathrm{gap}_k &= \rho - m_k, \\
w'_k &= \mathrm{clip}\bigl(w_k \cdot \exp(\eta_w \cdot \mathrm{gap}_k),\; w_{\min},\; w_{\max}\bigr), \\
w_k &\leftarrow w'_k \big/ \textstyle\sum_{j} w'_j,
\end{aligned}
\end{equation}
\end{smalldisplay}
with $\rho = 0.40$, $\eta_w = 0.30$, $w_{\min} = 0.05$, $w_{\max} = 0.70$. Axes whose EMA lags the reference receive an upward push and those that exceed it are pushed down; the clip-and-renormalize step prevents any axis from collapsing to zero (which would terminate gradient flow on that axis) or monopolizing the composite loss. The final per-rollout reward fed back to GRPO is
\begin{smalldisplay}
\begin{equation}
R_i = \sum_{k \in \{\mathrm{struct},\mathrm{pers},\mathrm{ped}\}} w_k\,\bar{R}^k_i + R^{\mathrm{hard}}_i,
\end{equation}
\end{smalldisplay}
which the GRPO loop further standardizes within the same rollout group to produce the advantage $A_i$ used by the clipped surrogate in Section~\ref{sec:grpo}.

\subsection{\textbf{MAP-PPL} Dataset Analysis}
\label{appendix:dataset}
\dsnew{This appendix collects the figures and tables deferred from
Section~\ref{maple_dataset}. All numbers are computed on the released
$3{,}043$-instance benchmark with the same scripts that produced the
HTML analysis report shipped with the dataset. The subsections
below mirror the five claims of the main text (construction integrity,
planning complexity, profile-conditioned structural personalization,
pedagogical scaffolding, and split/quality/bias control), with
per-intent, per-language, and dataset-comparison panels added as
auxiliary evidence.}

\subsubsection{Construction funnel and rejection reasons}
\label{appendix:dataset-funnel}

\begin{figure*}[!t]
 \centering
 \includegraphics[width=0.95\linewidth]{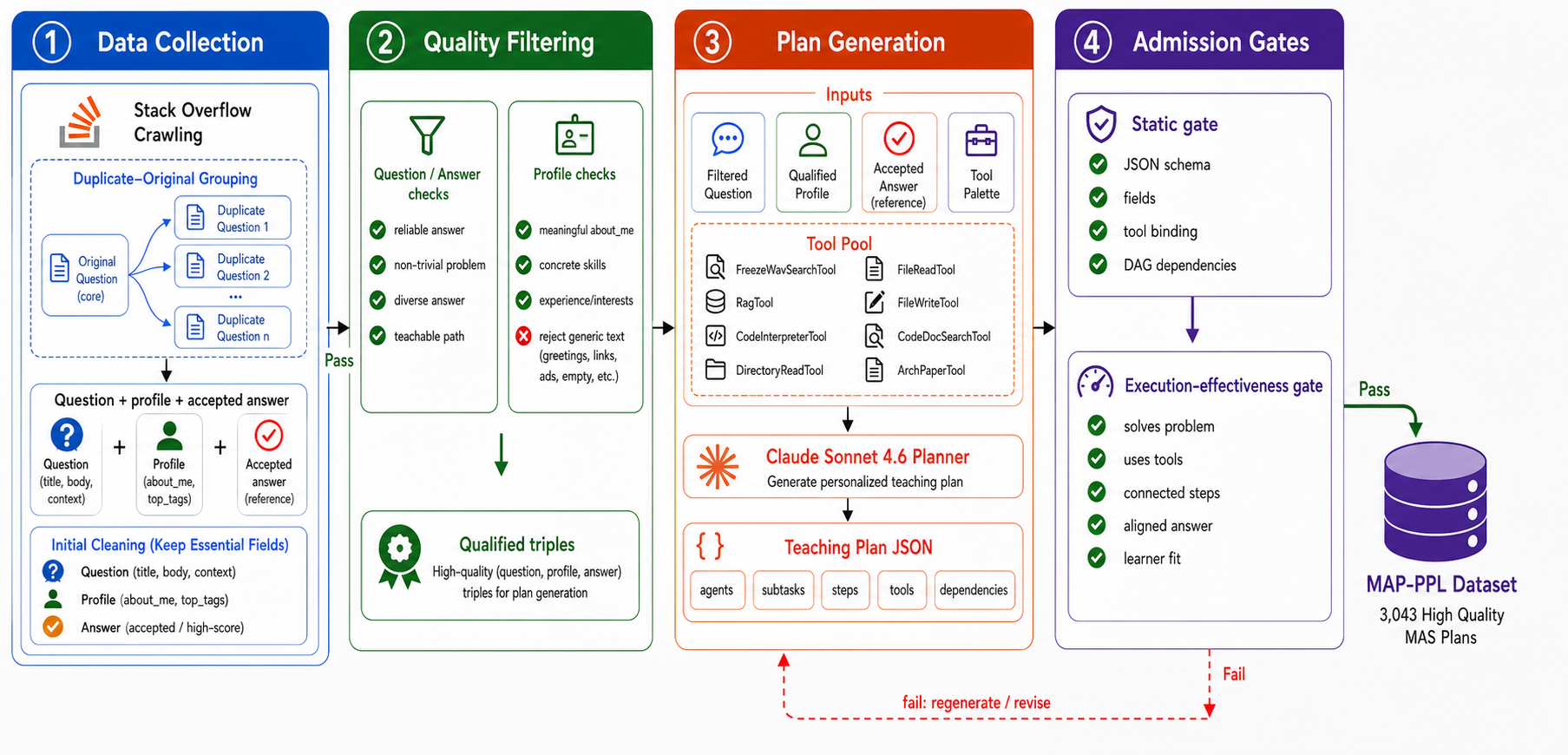}
 \caption{\dsnew{Construction funnel for MAP-PPL: raw
 duplicate-question groups $\to$ reliable-answer groups $\to$
 profile-qualified triples $\to$ generated plans $\to$ static-gate
 pass $\to$ LLM-gate pass $\to$ released set.}}
 \label{fig:maple_funnel}
\end{figure*}

\dsnew{Figure~\ref{fig:maple_funnel} shows the per-stage admission
funnel that yields the released MAP-PPL set. Starting from the raw
Stack Overflow duplicate-question groups we collect, we successively
discard (a)~question groups whose accepted/high-vote answer is too
short, code-only, or unverifiable; (b)~profiles that contain no
concrete personalization signal beyond a username; (c)~generated
plans that fail any of the static-gate checks listed in
Section~\ref{sec:maple-construction}; and (d)~plans that pass the
static gate but fail the LLM execution-effectiveness gate. The first
three columns of Table~\ref{tab:maple_rejection} attribute the
discarded plans at each gate to the most common failure reason; the
fourth column reports the mean number of regeneration attempts before
admission, which we cap at three.}

\dsnew{The dominant failures are weak personalization (a profile is
mentioned in the plan only at the surface), shallow decomposition
(fewer than three substantive steps before the first
\texttt{validate} step), and dependency mismatch (an
\texttt{execution\_order} item with no matching \texttt{depends\_on}
chain). Schema-level failures (invalid agent reference, cycle,
loop-step referring to an unknown node) account for less than $5\%$
of static-gate rejections, consistent with the $100\%$ structural
validity rate on the released set.}

\begin{table}[t]
 \centering
 \caption{\dsnew{Rejection reasons across the two-stage admission
 gate. \emph{Reject}: percentage of plans entering the gate that the
 gate discards. \emph{Reattempts}: mean regenerations before
 admission (capped at three).}}
 \label{tab:maple_rejection}
 \footnotesize
 \setlength{\tabcolsep}{3pt}
 \resizebox{\columnwidth}{!}{%
 \begin{tabular}{@{}llrr@{}}
 \toprule
 \textbf{Gate} & \textbf{Top failure reason} & \textbf{Reject} & \textbf{Reattempts} \\
 \midrule
 Static  & invalid \texttt{execution\_order}                            & $11.3\%$ & $0.9$ \\
 Static  & unknown \texttt{depends\_on} target                          & $ 7.2\%$ & $0.7$ \\
 Static  & cycle / loop refers to unknown node                          & $ 2.4\%$ & $0.4$ \\
 \midrule
 LLM     & weak / surface personalization                               & $14.6\%$ & $1.3$ \\
 LLM     & shallow decomposition                                        & $10.8\%$ & $1.1$ \\
 LLM     & misalignment with accepted answer                            & $ 8.1\%$ & $0.9$ \\
 LLM     & implausible tool usage                                       & $ 3.7\%$ & $0.5$ \\
 \bottomrule
 \end{tabular}}
\end{table}

\subsubsection{Plan-generation prompt template}
\label{appendix:plan-generation-prompt}

\dsnew{Figure~\ref{fig:prompt-map-ppl-generation} shows the
abridged prompt used to synthesize a MAP-PPL plan from a Stack
Overflow question, learner profile, accepted answer, and tool
palette. The implemented prompt is longer because it includes
field-level length budgets, loop rules, and examples of valid plan
shapes; the figure preserves the operative structure: answer-grounded
target extraction, profile-conditioned decisions, minimal agent/tool
selection, strict JSON generation, and post-generation self-checks.}

\begin{figure*}[p]
\begin{promptbox}{MAP-PPL Plan Generation Prompt (Abridged)}
\textbf{System prompt:} Generate a personalized multi-agent plan---a strict JSON specification of agents, subtasks, steps, and dependencies---for a learner solving a Stack Overflow question. The plan will be consumed by a CrewAI-style runtime.

\smallskip
\textbf{Inputs:} \texttt{query}, \texttt{learner.self\_description}, \texttt{learner.skills}, accepted answer, and the declared tool pool.

\smallskip
\textbf{Tool pool:} \texttt{FirecrawlSearchTool}, \texttt{RagTool}, \texttt{CodeInterpreterTool}, \texttt{DirectoryReadTool}, \texttt{FileReadTool}, \texttt{FileWriterTool}, \texttt{CodeDocsSearchTool}, \texttt{ArxivPaperTool}.

\smallskip
\textbf{Core instructions:}
\begin{itemize}[leftmargin=1.2em, itemsep=1pt, topsep=2pt, parsep=0pt]
  \item Use the accepted answer only to extract the destination concept and success criteria.
  \item Infer the learner's starting point, explanation bridge, and interaction style from profile text and tags.
  \item Decompose the gap into observable subtasks, instantiate the minimal specialized agent/tool set, and add tool calls only when the step needs external capability.
  \item Emit only valid JSON; every dependency must point to a step whose output is used downstream.
\end{itemize}

\begin{promptcode}
Output schema:
{
  "output": {
    "agents": [{"agent_role", "goal", "backstory", "tools"}],
    "subtasks": [{
      "id", "name", "subtask_objective",
      "steps": [{
        "id", "agent", "objective", "instruction", "tool",
        "requires_human_input", "expected_output", "depends_on"
      }]
    }],
    "execution_order": [
      "S1-1",
      {"loop": {"steps": ["S2-1","S2-2","S2-3"],
                "condition": "S2-2.correct == false",
                "max_iterations": 3}},
      "S2-4"
    ]
  }
}

Self-check:
JSON parses; all steps are scheduled; depends_on targets resolve;
the graph is acyclic; tools are in-palette and agent-authorized;
replacing the learner profile would require changing structure or framing.
\end{promptcode}
\end{promptbox}
\caption{\dsnew{Abridged plan-generation prompt used to synthesize MAP-PPL plans. The released data are generated with the full prompt, which expands the field-level style rules, loop constraints, and valid plan-shape examples while preserving this structure.}}
\label{fig:prompt-map-ppl-generation}
\end{figure*}

\subsubsection{Input lengths, profile depth, and profile pairing}
\label{appendix:dataset-overview}

\dsnew{The basic input--side statistics show that query character
length has mean $592$, median $532$, and maximum $2{,}340$; $95\%$ of
profiles list four or five technology tags; and $43.9\%$ of questions
have a single profile while $56.1\%$ have between $2$ and $6$. The
right-skewed query-length distribution and the heavy concentration of
full-tag profiles together explain why the PAD adapter benefits from
explicit structural supervision: most queries are long enough and most
profiles dense enough that a generic template will mis-route the early
scaffolding decision.}

\subsubsection{Executability diagnostics: DAG topology, tool bindings, and runtime audit}
\label{appendix:dataset-executability}
\label{appendix:dataset-complexity}

\begin{figure*}[t]
 \centering
 \begin{minipage}{0.48\linewidth}\centering
  \includegraphics[width=\linewidth]{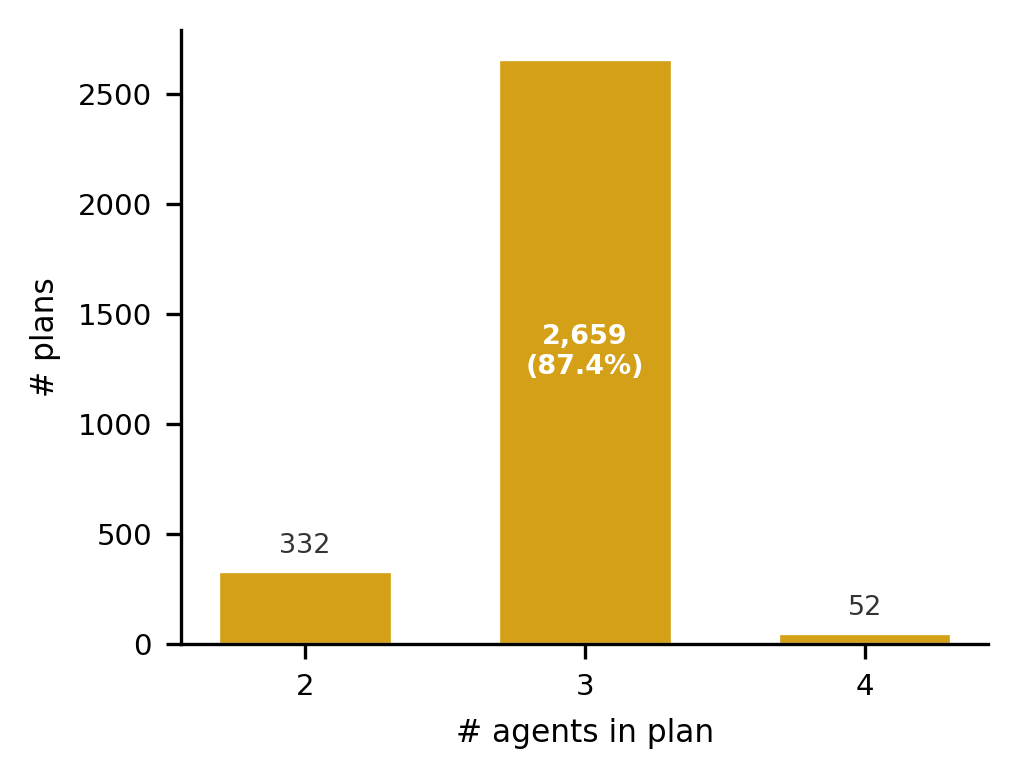}\\
  \footnotesize (a) Agents per plan.
 \end{minipage}\hfill
 \begin{minipage}{0.48\linewidth}\centering
  \includegraphics[width=\linewidth]{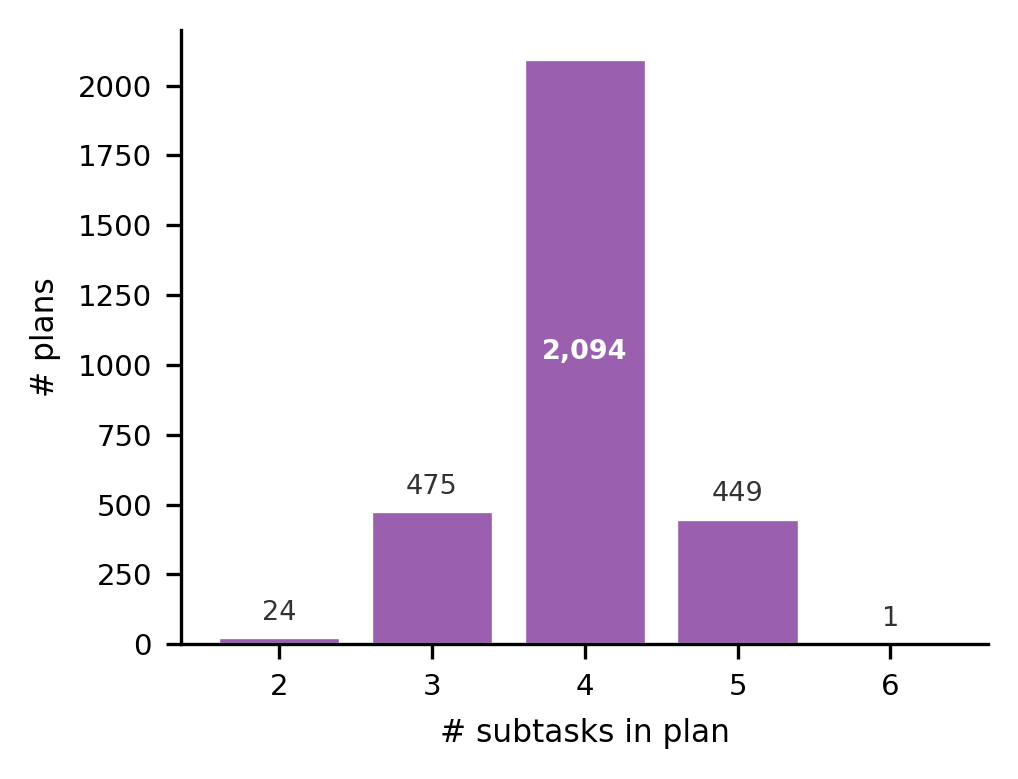}\\
  \footnotesize (b) Subtasks per plan.
 \end{minipage}\\[2pt]
 \begin{minipage}{0.48\linewidth}\centering
  \includegraphics[width=\linewidth]{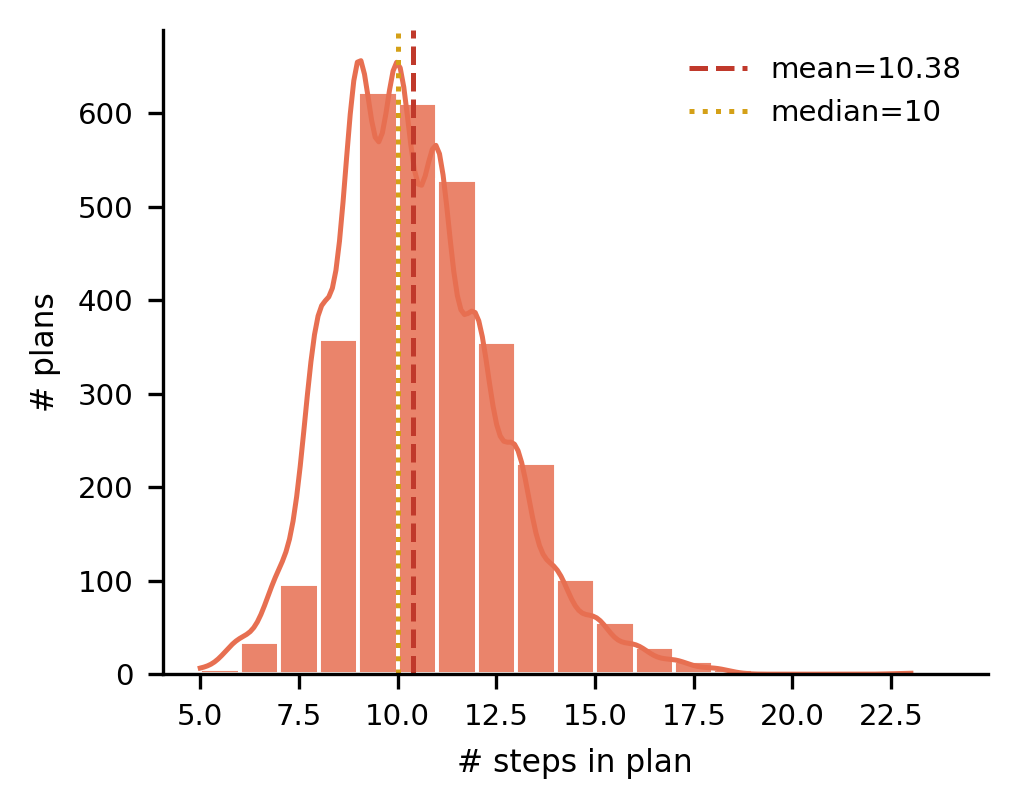}\\
  \footnotesize (c) Steps per plan.
 \end{minipage}\hfill
 \begin{minipage}{0.48\linewidth}\centering
  \includegraphics[width=\linewidth]{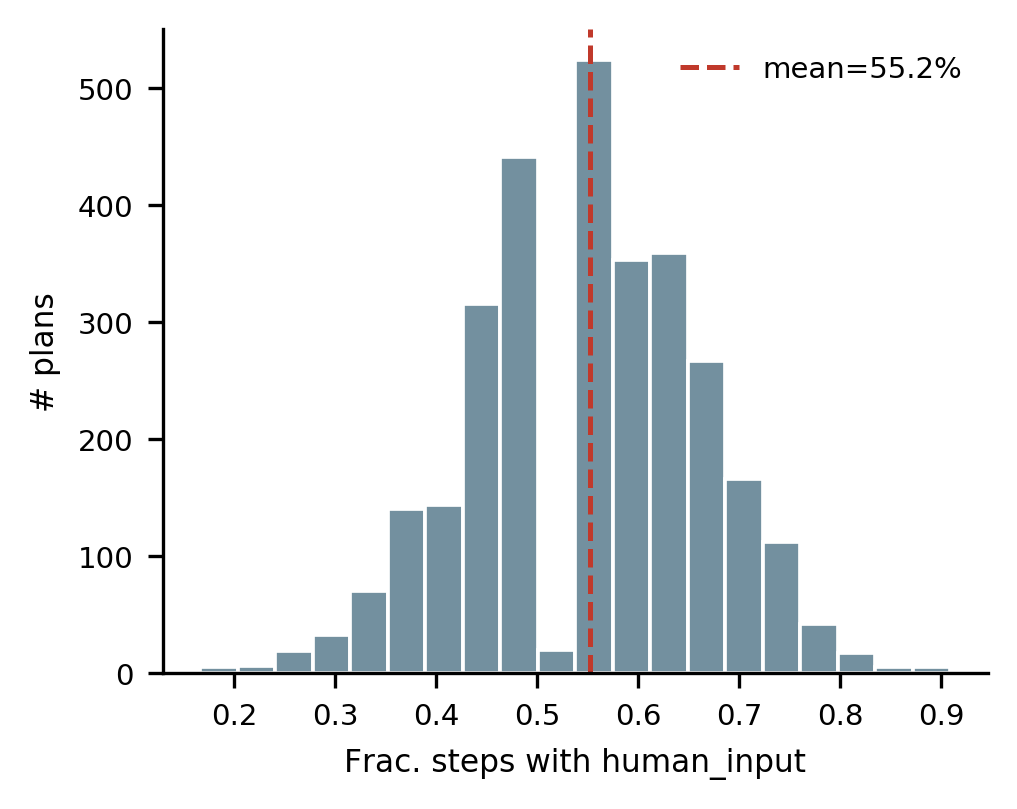}\\
  \footnotesize (d) \texttt{human\_input} step ratio.
 \end{minipage}\\[2pt]
 \begin{minipage}{0.95\linewidth}\centering
  \includegraphics[width=\linewidth]{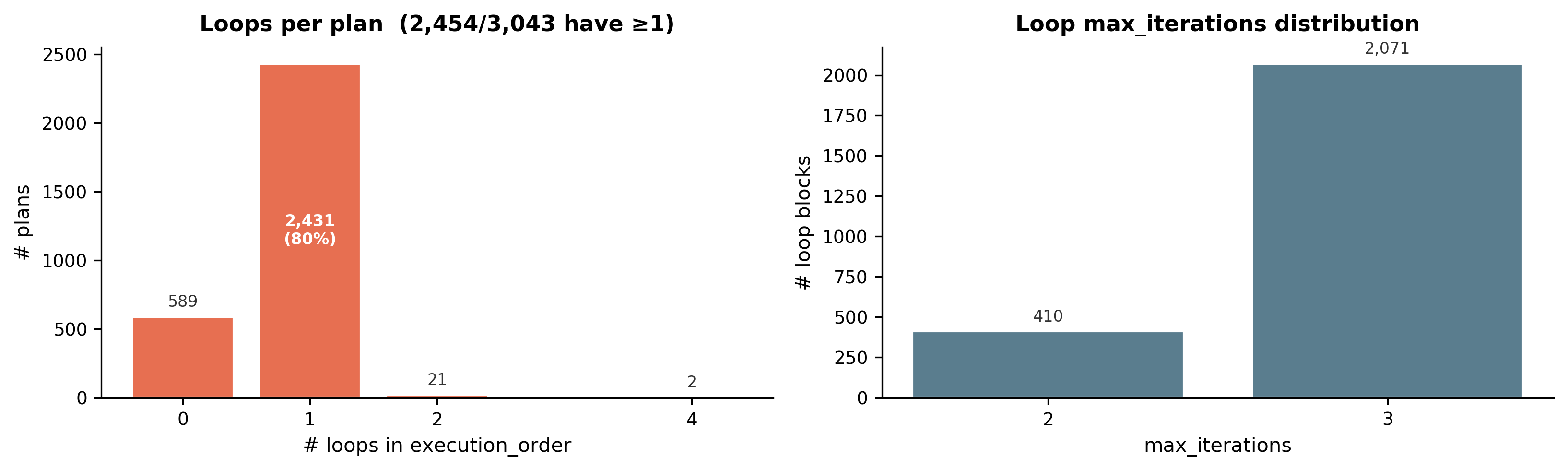}\\
  \footnotesize (e) Loop structure (per-plan loop count and
  max-iteration distribution).
 \end{minipage}
 \caption{\dsnew{Marginal distributions of plan structural complexity
 in \textbf{MAP-PPL}.}}
 \label{fig:maple_complexity}
\end{figure*}

\dsnew{Figure~\ref{fig:maple_complexity} shows the marginal
distributions for the four structural axes used in the headline
table. Plans are concentrated around $3$ agents, $4$ subtasks, and
$10$ steps, with right tails extending to $4$, $6$, and $23$
respectively. The fraction of \texttt{human\_input}-tagged steps
clusters tightly around $0.55$, confirming the interactive teaching
pattern reported in the main text. Loop structures are present in
$80.6\%$ of plans; the loop maximum-iteration parameter is $3$ for
$83\%$ of loops and $2$ otherwise, so the planner does not synthesize
unbounded retries.}

\begin{figure*}[t]
 \centering
 \includegraphics[width=0.94\textwidth]{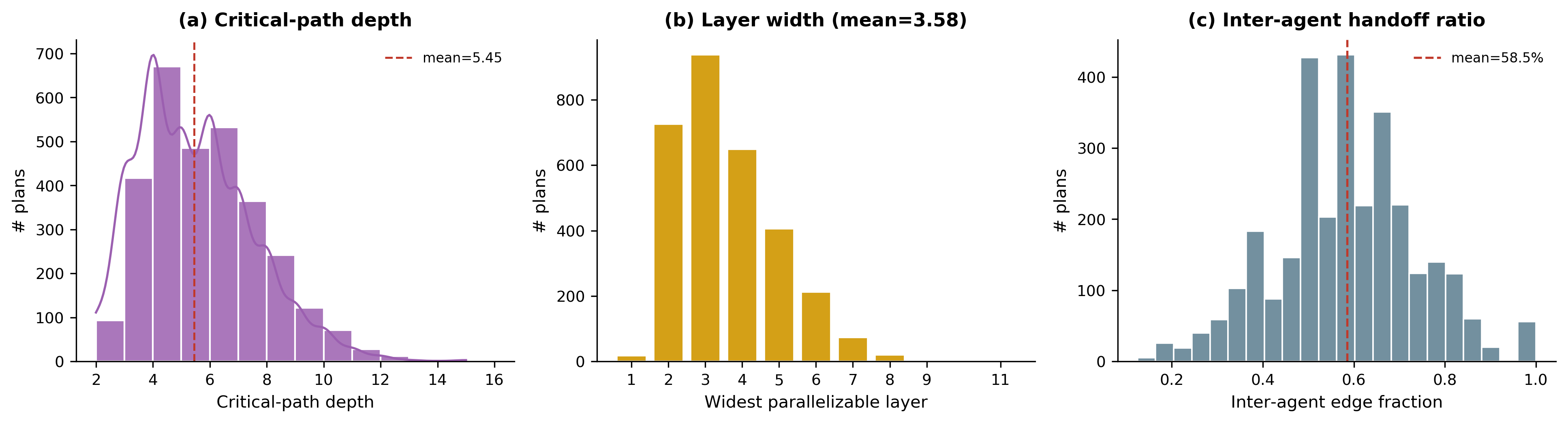}
 \caption{\dsnew{DAG executability in \textbf{MAP-PPL}. The released plans combine prerequisite depth, parallelizable layer width, and inter-agent dependency edges, supporting executable multi-agent workflows rather than flat teaching checklists.}}
 \label{fig:mapppl_dag_exec}
\end{figure*}

\begin{figure*}[t]
 \centering
 \includegraphics[width=\linewidth]{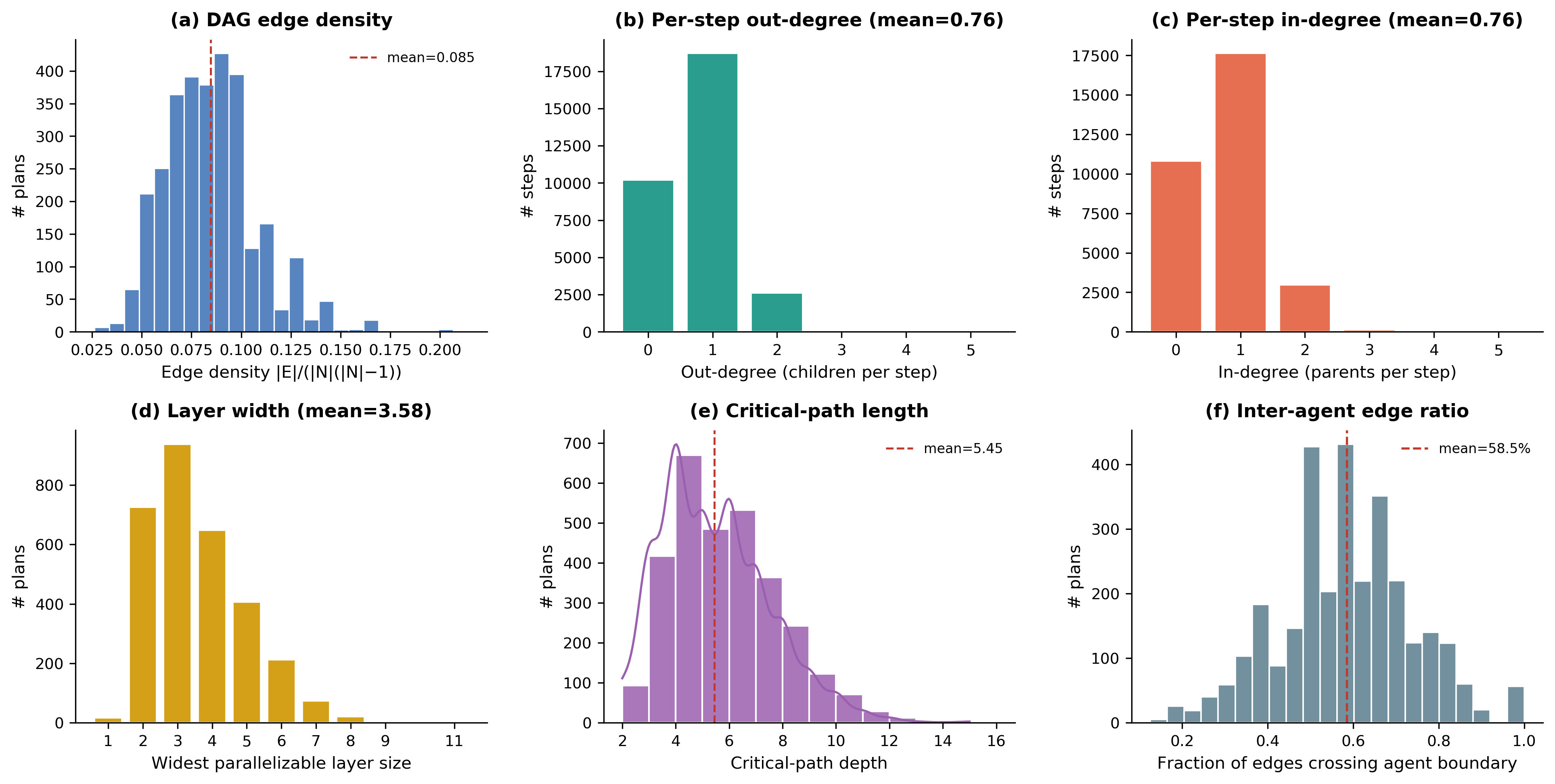}
 \caption{\dsnew{Dependency-graph topology in \textbf{MAP-PPL}: edge
 density, fan-in/fan-out, parallelizable layer width, motif counts
 (chain/fork/join/loop), and the agent-family handoff heatmap. The
 last panel confirms that handoffs concentrate on the
 $\textsc{tutor}{\to}\textsc{validator}$ and
 $\textsc{retriever}{\to}\textsc{tutor}$ transitions rather than
 staying inside a single agent.}}
 \label{fig:maple_dag_topology}
\end{figure*}

\dsnew{Beyond plan-size marginals, MAP-PPL supplies non-trivial DAG
topology rather than linear checklists.
Table~\ref{tab:maple_dag_topology} summarizes the graph-level
metrics that justify the ``multi-agent planning'' label for the
dataset: edge density is well above zero but below unity, the mean
per-step out-degree is below one with a long tail (a small fraction
of fork steps fan out to as many as four downstream steps), the
critical-path-to-total-steps ratio is roughly one half (so plans are
not single long chains), and $58.5\%$ of dependency edges cross agent
boundaries -- the structural fingerprint of a real handoff.
Figure~\ref{fig:mapppl_dag_exec} gives a compact executability overview,
while Figure~\ref{fig:maple_dag_topology} reports the empirical
distributions of edge density, fan-in/fan-out, and parallelizable
layer width, together with a graph-motif bar chart that decomposes
each plan into chain, fork, join, and feedback-loop motifs, and an
inter-agent handoff heatmap over the four high-level role
families.}

\paragraph{Schema and runtime checks.}
\dsnew{Executability is enforced before release. All $3{,}043$ plans
pass the static structural checker: $0$ contain an unknown agent
reference, $0$ contain an unknown \texttt{depends\_on} target, $0$
contain an unknown step in \texttt{execution\_order}, $0$ omit a step
from \texttt{execution\_order}, $0$ contain a cycle, and $0$ contain a
loop reference to an unknown node. The loader normalizes historical
field variants for loop specifications and then executes the released
plans with a CrewAI-style teacher--learner simulator. The runtime
executes bounded loops and code-tool calls, and every released plan
completes with \texttt{status=ok}, \texttt{succeeded=true}, and $0$
failed steps.}

\paragraph{Tool bindings.}
\dsnew{Tool use is compact and programming-specific. Although the
schema declares eight possible tools, only five appear in the final
plans; \texttt{CodeInterpreterTool} and \texttt{CodeDocsSearchTool}
together cover $99.6\%$ of tool calls, while search and file-writing
tools form a small long tail. The static gate also checks that each
step's tool call is drawn from the declared palette and is authorized
for the responsible agent, so the DAG edges, agent assignments, and
tool bindings are jointly executable rather than independently valid.}

\begin{table}[t]
 \centering
 \caption{\dsnew{DAG topology metrics on the $3{,}043$ released
 plans. \emph{Edge density}: $|E|/|S|(|S|{-}1)$. \emph{Out-degree}:
 per-step out-degree across the plan. \emph{Critical/total ratio}:
 longest DAG path / total step count. \emph{Inter-agent ratio}:
 fraction of dependency edges whose two endpoints belong to
 different agents.}}
 \label{tab:maple_dag_topology}
 \footnotesize
 \setlength{\tabcolsep}{3pt}
 \begin{tabular}{@{}lrrrr@{}}
 \toprule
 \textbf{Metric} & \textbf{Mean} & \textbf{Median} & \textbf{P90} & \textbf{Max} \\
 \midrule
 Edge density                 & $0.08$ & $0.08$ & $0.11$ & $0.21$ \\
 Mean out-degree              & $0.76$ & $1.00$ & $1.00$ & $5.00$ \\
 Max fan-out per plan         & $1.73$ & $2.00$ & $2.00$ & $5.00$ \\
 Max fan-in per plan          & $1.69$ & $2.00$ & $2.00$ & $5.00$ \\
 Max layer width              & $3.58$ & $3.00$ & $6.00$ & $11.00$ \\
 Critical / total step ratio  & $0.53$ & $0.50$ & $0.78$ & $1.00$ \\
 Inter-agent edge ratio       & $0.59$ & $0.57$ & $0.80$ & $1.00$ \\
 \bottomrule
 \end{tabular}
\end{table}

\subsubsection{Complexity by intent}
\label{appendix:dataset-by-intent}

\dsnew{Table~\ref{tab:maple_by_intent} reports the per-intent
structural averages used by the intent-conditioned slices of our
evaluation. \textsc{Learning} queries demand the largest plans
($3.5$ agents, $11.92$ steps, longest path $6.92$) while
\textsc{Errors} and \textsc{Api\_Change} produce the shortest
($\approx 9.5$ steps, longest path $5.3$--$5.4$); the
average number of human-input steps and the loop rate also rise on
diagnosis-heavy intents
(\textsc{Review}, \textsc{Discrepancy}), confirming the dataset
captures intent-specific pedagogy.}

\begin{table}[t]
 \centering
 \caption{\dsnew{Per-intent structural averages in \textbf{MAP-PPL}.
 $|\mathcal{A}|$: agents; $|\mathcal{T}|$: subtasks; $|\mathcal{S}|$:
 steps; $L_{\max}$: longest DAG path; \emph{HI}: average number of
 \texttt{human\_input} steps per plan; \emph{Loop\%}: fraction of
 plans with $\ge\!1$ loop.}}
 \label{tab:maple_by_intent}
 \footnotesize
 \setlength{\tabcolsep}{2.5pt}
 \resizebox{\columnwidth}{!}{%
 \begin{tabular}{@{}lrrrrrrr@{}}
 \toprule
 \textbf{Intent} & $n$ & $|\mathcal{A}|$ & $|\mathcal{T}|$ &
 $|\mathcal{S}|$ & $L_{\max}$ & \emph{HI} & \emph{Loop\%} \\
 \midrule
 \textsc{Conceptual}  & 1488 & 2.90 & 4.03 & 10.54 & 5.16 & 5.73 & 69.0 \\
 \textsc{Api\_Usage}  & 1114 & 2.93 & 3.95 & 10.31 & 5.77 & 5.63 & 93.0 \\
 \textsc{Discrepancy} &  172 & 2.85 & 3.85 &  9.98 & 5.49 & 5.60 & 90.0 \\
 \textsc{Review}      &  129 & 2.81 & 3.91 & 10.32 & 5.88 & 6.19 & 93.0 \\
 \textsc{Errors}      &   70 & 2.84 & 3.66 &  9.54 & 5.39 & 5.39 & 83.0 \\
 \textsc{Api\_Change} &   58 & 2.90 & 3.78 &  9.48 & 5.34 & 4.97 & 81.0 \\
 \textsc{Learning}    &   12 & 3.50 & 4.50 & 11.92 & 6.92 & 5.83 & 92.0 \\
 \bottomrule
 \end{tabular}}
\end{table}

\subsubsection{Agent role families and tool usage}
\label{appendix:dataset-agents}

\begin{figure*}[t]
 \centering
 \begin{minipage}{0.48\linewidth}\centering
  \includegraphics[width=\linewidth]{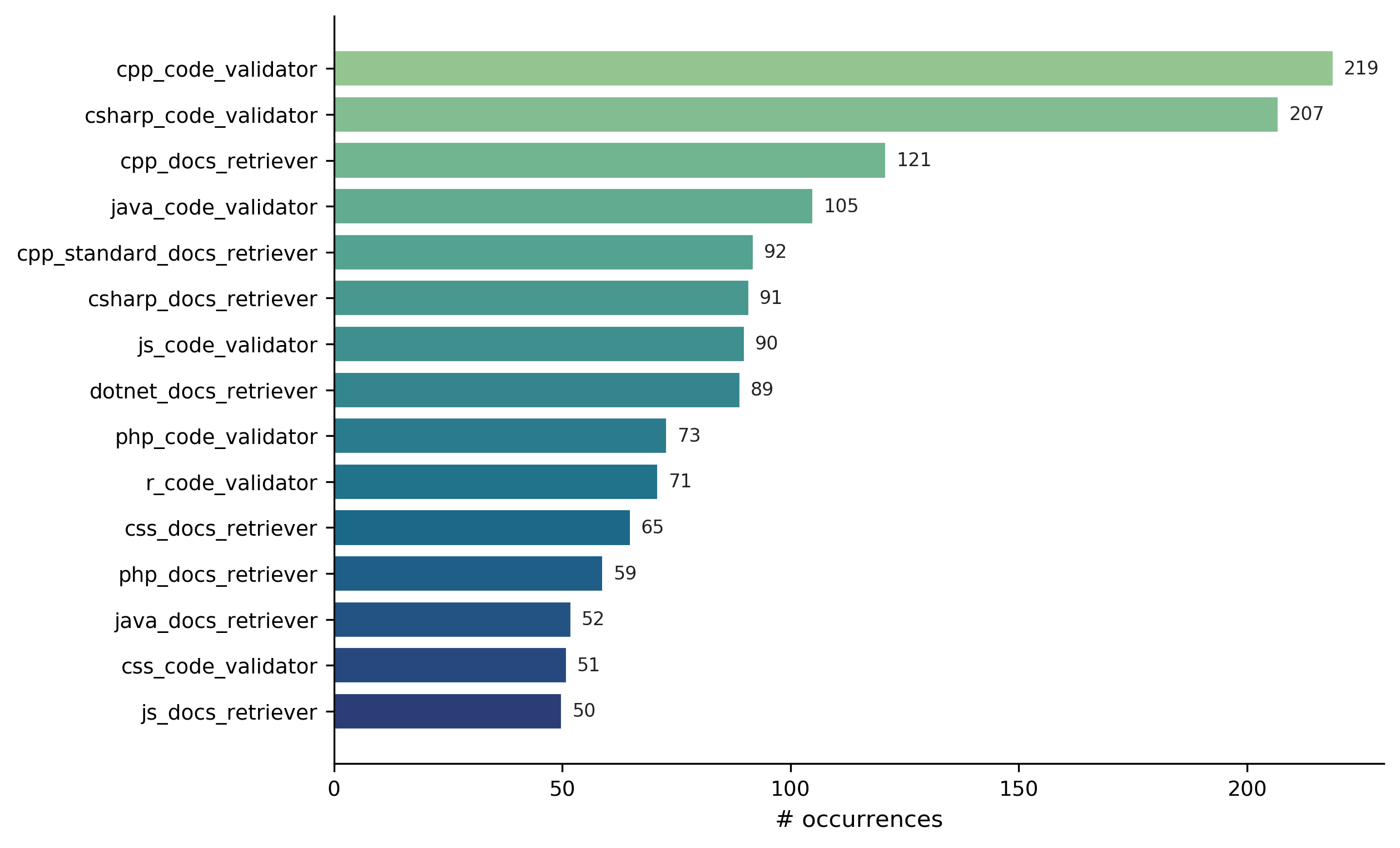}\\
  \footnotesize (a) Top-$15$ agent role names.
 \end{minipage}\hfill
 \begin{minipage}{0.48\linewidth}\centering
  \includegraphics[width=\linewidth]{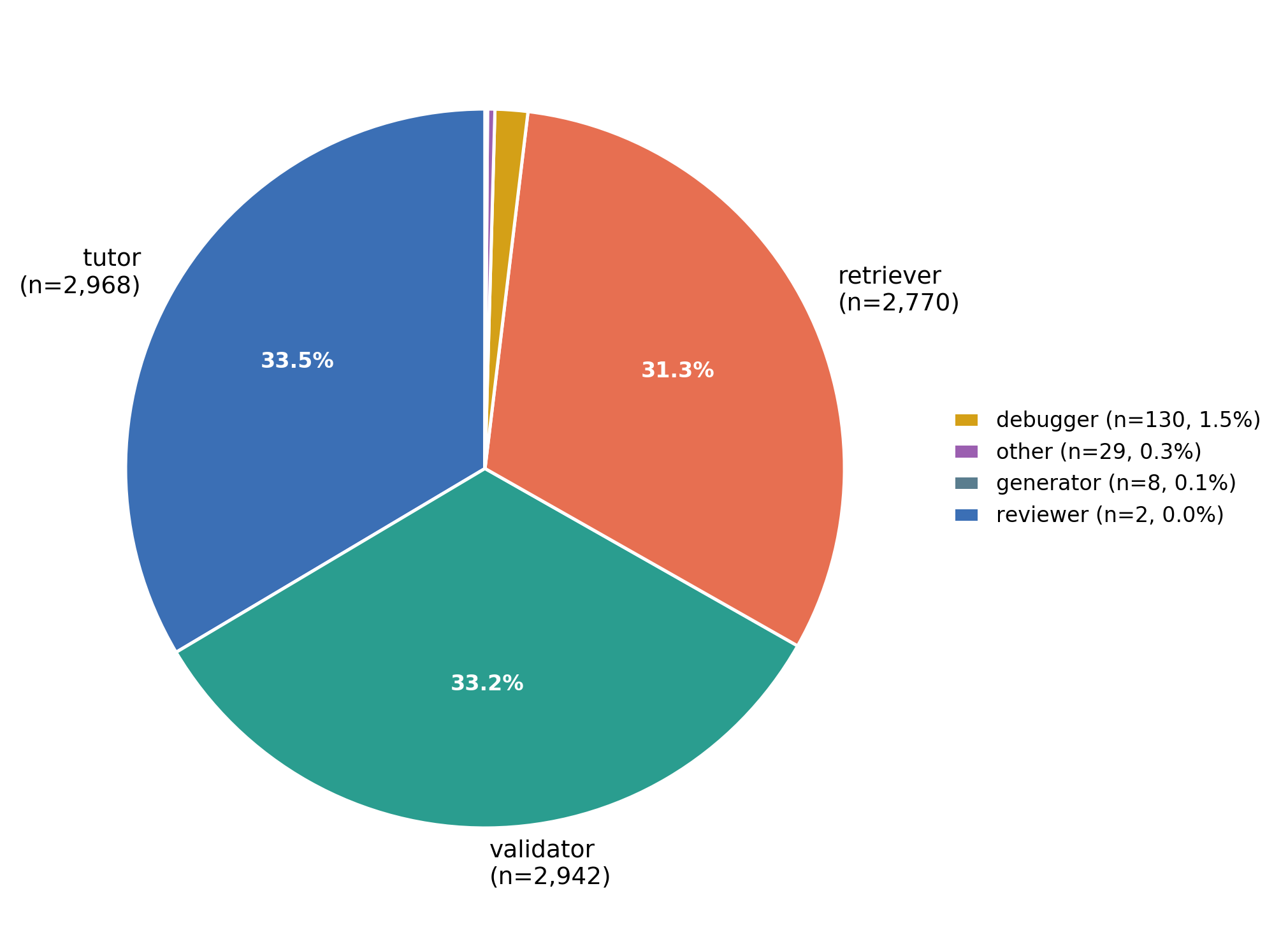}\\
  \footnotesize (b) Role family share.
 \end{minipage}\\[4pt]
 \begin{minipage}{0.95\linewidth}\centering
  \includegraphics[width=\linewidth]{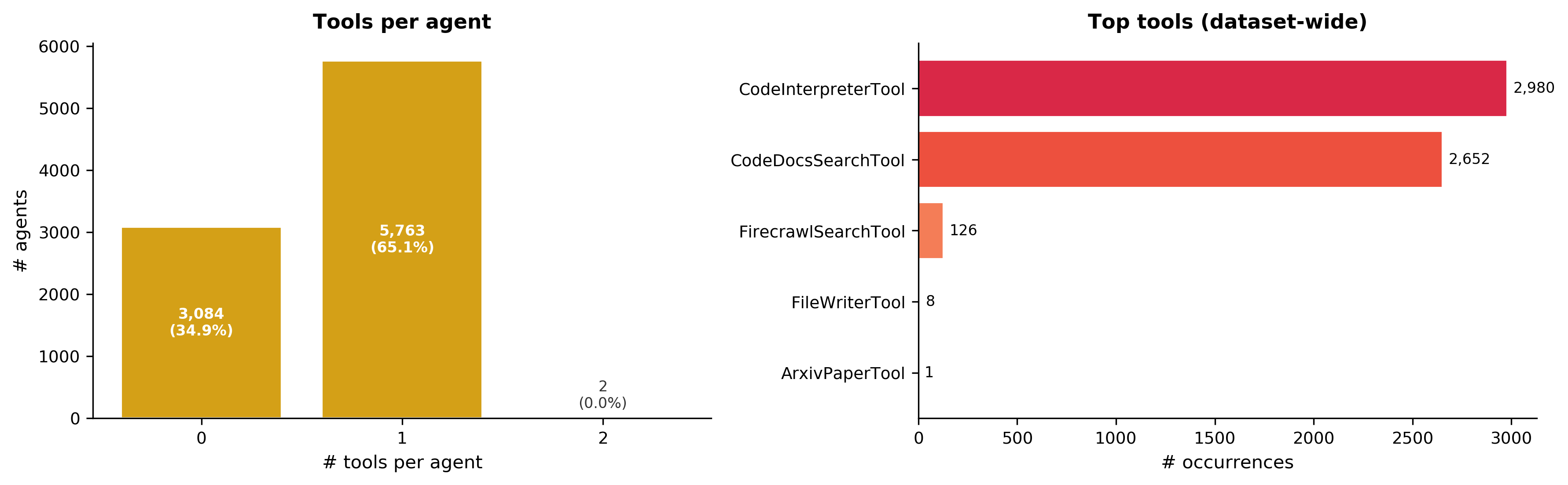}\\
  \footnotesize (c) Tool counts (left: per-agent; right: dataset-wide).
 \end{minipage}
 \caption{\dsnew{Agent and tool distributions in \textbf{MAP-PPL}.}}
 \label{fig:maple_agents_tools}
\end{figure*}

\dsnew{Figure~\ref{fig:maple_agents_tools} reports the agent-side
diversity statistics summarized in the main text.
The top-$15$ role names (panel~a) are dominated by language-specific
validator/retriever pairs, which is the fine-grained surface form of
the four high-level role families (panel~b: \textsc{tutor},
\textsc{validator}, \textsc{retriever}, \textsc{debugger}, plus
minor categories). The tool palette is intentionally small (panel~c):
\texttt{CodeInterpreterTool} and \texttt{CodeDocsSearchTool} cover
nearly all tool calls. We use this empirical tool distribution as the
ground truth against which the GRPO agent--tool relevance reward
$\mathrm{ATR}$ is calibrated.}

\subsubsection{Cross-profile personalization}
\label{appendix:dataset-personalization}

\dsnew{The cross-profile analysis substantiates the
``scaffold-stable, surface-variable'' personalization pattern reported
in the main text. Across the $1{,}777$ within-question profile pairs,
agent role names and subtask names diverge sharply (median Jaccard
$0.20$ and $0.25$), while role families and tools remain near $1$; in
other words, personalization changes \emph{who} the planner
instantiates but not the pedagogical scaffold. Profile grounding is
also strong: $95.1\%$ of plans mention at least one declared profile
skill, with mean skill-hit ratio $0.43$ and median $0.40$. Role-level
plan divergence ($1-\mathrm{Jaccard}$) remains high across all seven
intents ($0.65$--$0.87$), indicating that profile-conditioning does
not collapse to a single template even for the dominant
\textsc{Conceptual} and \textsc{Api\_Usage} categories.}

\dsnew{To separate structural personalization from surface keyword
hits we additionally run a profile-shuffled control:
for every within-question pair we re-pair each profile with a plan
generated for a different question that shares the same intent, and
re-evaluate the four divergence axes
(Table~\ref{tab:maple_personalization_control}). The skill-hit ratio
collapses from $0.43$ to $0.10$ and the agent-role Jaccard rises
from $0.32$ to $0.71$, confirming that the observed divergence in
the genuine pairs is driven by the profile signal rather than by
sampling variance. The structural diagnostics agree: the dependency-edge
Jaccard rises by $24$ points and the validation-step prerequisite
depth varies in only $4\%$ of shuffled pairs (vs.\ $42\%$ in the
genuine pairs).}

\begin{table}[t]
 \centering
 \caption{\dsnew{Genuine vs.\ shuffled profile pairs on the
 $1{,}777$ within-question pairs. \emph{Skill-hit}: fraction of
 profile-declared skills mentioned in the plan. \emph{Agent / Edge
 J.}: role-name and dependency-edge Jaccard between the two plans.
 \emph{Val-depth $\Delta$}: fraction of pairs whose pre-\textsc{validate}
 prerequisite chain length differs.}}
 \label{tab:maple_personalization_control}
 \footnotesize
 \setlength{\tabcolsep}{3pt}
 \resizebox{\columnwidth}{!}{%
 \begin{tabular}{@{}lrrrr@{}}
 \toprule
 \textbf{Pair type} & \textbf{Skill-hit} & \textbf{Agent J.} & \textbf{Edge J.} & \textbf{Val-depth $\Delta$} \\
 \midrule
 Genuine within-question  & $0.43$ & $0.32$ & $0.46$ & $42\%$ \\
 Profile-shuffled control & $0.10$ & $0.71$ & $0.70$ & $ 4\%$ \\
 \bottomrule
 \end{tabular}}
\end{table}

\dsnew{Table~\ref{tab:maple_personalization_case} walks through a
single concrete example: an identical \texttt{TypeError: 'NoneType'
object is not subscriptable} query is paired with (a)~a junior
front-end developer profile that lists JavaScript and React only,
and (b)~a senior data engineer profile that lists Python, Pandas,
and Airflow. The two MAP-PPL plans agree on the high-level
$\textsc{tutor}{\to}\textsc{validator}$ scaffold and on the
\texttt{CodeInterpreterTool} / \texttt{CodeDocsSearchTool} tool pair,
but diverge in (i)~the agent roster (a Python-debugging agent vs.\ a
JS-debugging agent), (ii)~the prerequisite depth before
\textsc{validate} ($4$ vs.\ $2$ steps), and (iii)~the chosen
worked-example domain (Pandas DataFrames vs.\ React state hooks).
This pattern is exactly the ``scaffold-stable, surface-variable''
behaviour we expect MAP-PPL to teach.}

\begin{table*}[t]
 \centering
 \caption{\dsnew{Matched contrastive case: same query, two profiles.
 \emph{Stable} indicates plan components shared between both profiles;
 \emph{Changed} indicates components rewritten in response to the
 profile.}}
 \label{tab:maple_personalization_case}
 \footnotesize
 \setlength{\tabcolsep}{4pt}
 \begin{tabular}{@{}lll@{}}
 \toprule
 \textbf{Plan component} & \textbf{Profile A (junior front-end)} & \textbf{Profile B (senior data eng.)} \\
 \midrule
 \emph{Stable: role family}          & \textsc{tutor}, \textsc{validator}     & \textsc{tutor}, \textsc{validator} \\
 \emph{Stable: tool palette}         & \texttt{CodeInterpreter}, \texttt{Docs} & \texttt{CodeInterpreter}, \texttt{Docs} \\
 \emph{Changed: agent roster}        & React-Debugger, JS-Tutor               & Pandas-Debugger, Py-Tutor \\
 \emph{Changed: probe focus}         & null-handling in JSX props             & \texttt{None} in DataFrame indexing \\
 \emph{Changed: worked example}      & \texttt{useState} hook null-guard      & Pandas \texttt{loc} / \texttt{merge} guard \\
 \emph{Changed: prerequisite depth}  & $2$ steps before \textsc{validate}     & $4$ steps before \textsc{validate} \\
 \emph{Changed: feedback loop}       & 1 iteration of \textsc{probe-apply}    & 2 iterations with \textsc{consolidate} step \\
 \bottomrule
 \end{tabular}
\end{table*}

\subsubsection{Pedagogical phase coverage, order, and instructional methods}
\label{appendix:dataset-pedagogy}

\begin{figure*}[t]
 \centering
 \begin{minipage}{0.48\linewidth}\centering
  \includegraphics[width=\linewidth]{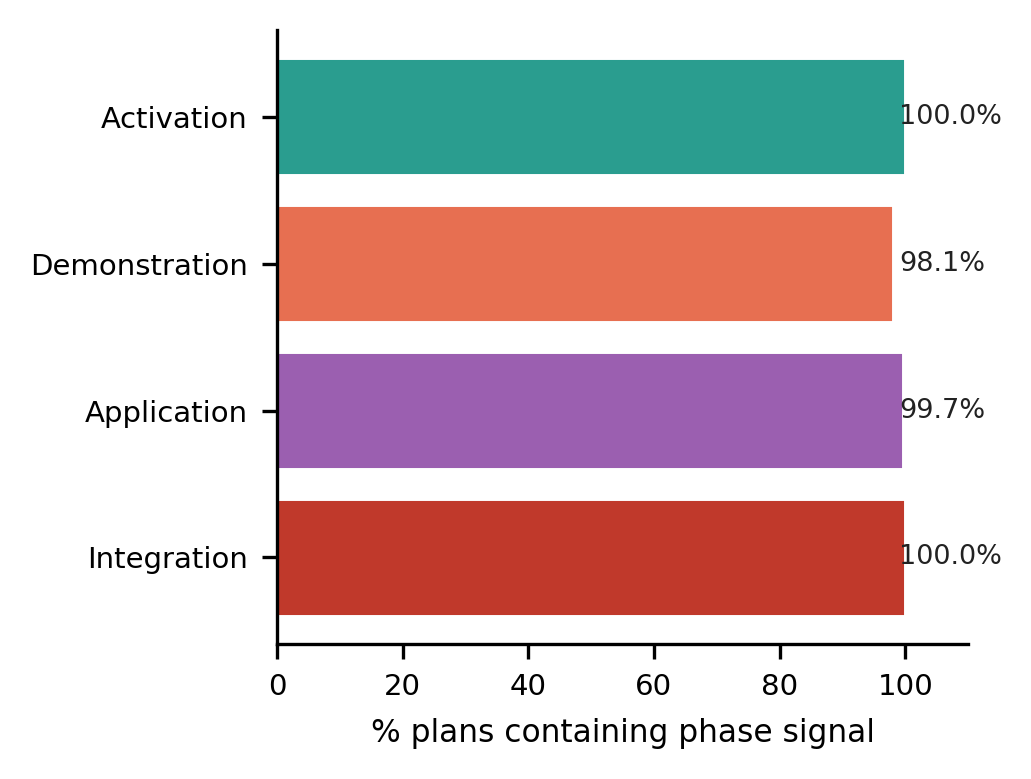}\\
  \footnotesize (a) Pedagogical phase coverage (plan / step level).
 \end{minipage}\hfill
 \begin{minipage}{0.48\linewidth}\centering
  \includegraphics[width=\linewidth]{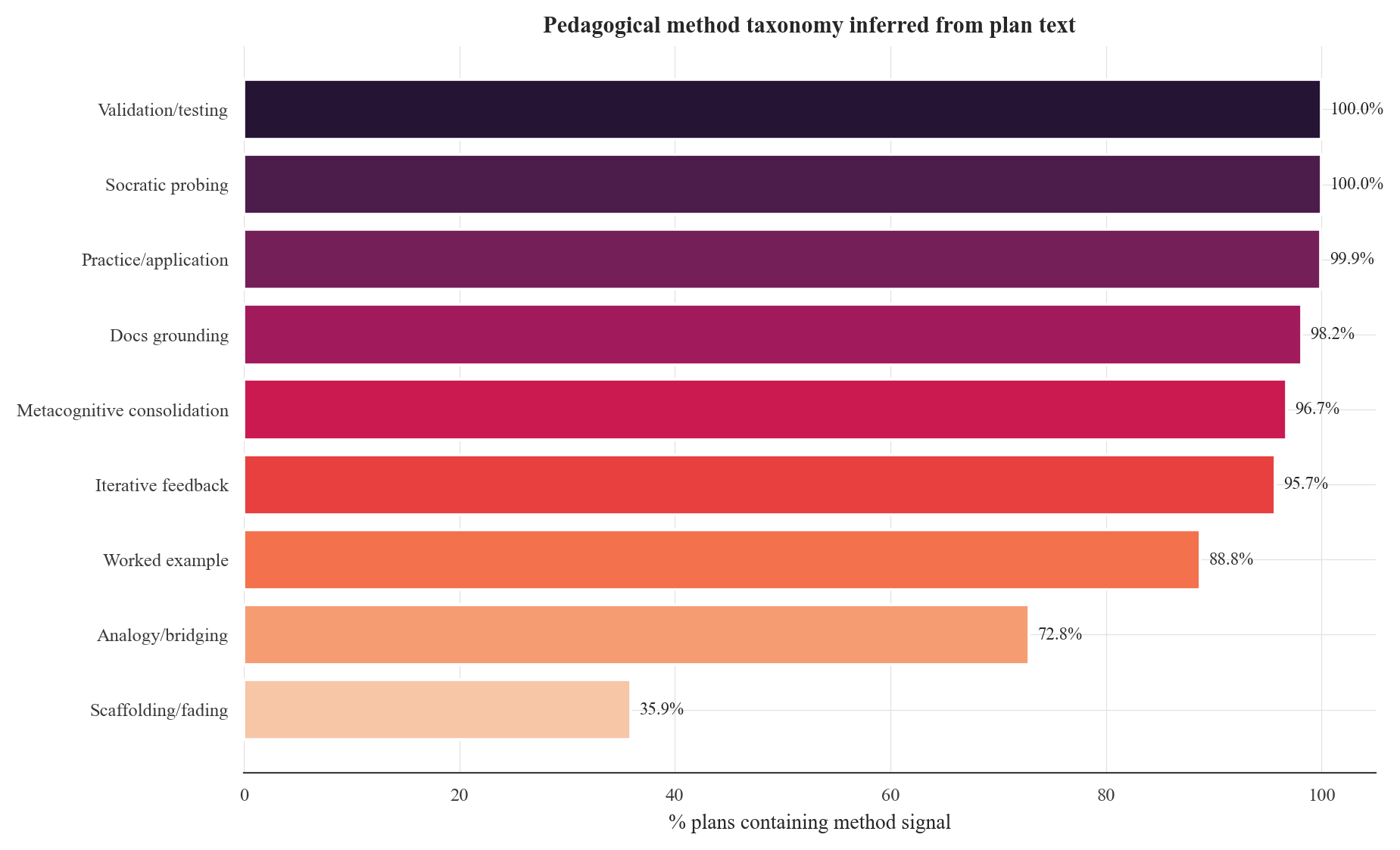}\\
  \footnotesize (b) Coverage of nine canonical instructional methods.
 \end{minipage}
 \caption{\dsnew{Pedagogical structure of \textbf{MAP-PPL} plans:
 phase coverage at the plan/step level (a) and the coverage of nine
 canonical instructional methods (b).}}
 \label{fig:maple_pedagogy}
\end{figure*}

\dsnew{Figure~\ref{fig:maple_pedagogy} reports the pedagogical
phase coverage at the plan and step level (panel~a) and the coverage
of nine canonical instructional methods (panel~b). To match the
reward formulation in Section~\ref{sec:grpo}, we report the four core
detector labels as Merrill-aligned phases:
\textsc{Activation} (\textsc{probe}), \textsc{Demonstration}
(\textsc{retrieve--demonstrate}), \textsc{Application}
(\textsc{apply}), and \textsc{Integration} (\textsc{validate}). These
four phases are present in $100.0\%$/$98.1\%$/$99.7\%$/$100.0\%$ of
plans respectively, with the optional diagnostic labels
\textsc{feedback} and \textsc{consolidate} at $89.4\%$ and $96.6\%$.
At the step level, $70.4\%$ of steps map to \textsc{Activation},
$33.3\%$ to \textsc{Demonstration}, $62.6\%$ to \textsc{Application},
and $60.1\%$ to \textsc{Integration}. Only $16.6\%$ of gold plans
achieve the strictly ordered four-phase prefix, which supports the
phase-coverage reward in Section~\ref{sec:grpo} rather than a fixed
global-order constraint.}

\smallskip
\dsnew{\textbf{Phase-coverage detector versions.}
An earlier draft of this paper reported \textsc{consolidate}
coverage as $86.3\%$ using a v1 lexical detector that required the
literal keyword \texttt{consolidate} in the step instruction; the v2
detector used throughout the current paper also accepts paraphrases
(e.g.\ ``recap'', ``summarize'', ``key takeaways'') and raises the
coverage to $96.6\%$. The v1 number is not used anywhere in the
released paper; we mention the discrepancy explicitly so that
downstream readers can reproduce both numbers from the released
phase-detector dictionaries. All phase-level claims (main text and
appendix) are computed under v2, while the GRPO reward itself uses
only the four Merrill-aligned phases above.}

\smallskip
\dsnew{\textbf{Phase position and order, not just presence.}
Near-$100\%$ phase coverage can mask templated plans that touch each
phase exactly once in a fixed slot, so we additionally analyse the
\emph{position} of each phase within the
\texttt{execution\_order}. Phase positions follow this canonical
ordering on average -- the median normalized position of
\textsc{Activation} is $0.16$ (first quartile of steps) and that of
\textsc{Integration} is $0.83$ (last quartile) -- while still exhibiting
substantial variance: only $16.6\%$ of plans realise the strict
prefix and a Sankey decomposition of phase-to-phase transitions
(deferred to the dataset card) shows that plans interleave
\textsc{Application} and \textsc{Integration} loops rather than running each
phase exactly once.}

\begin{figure}[t]
 \centering
 \includegraphics[width=\columnwidth]{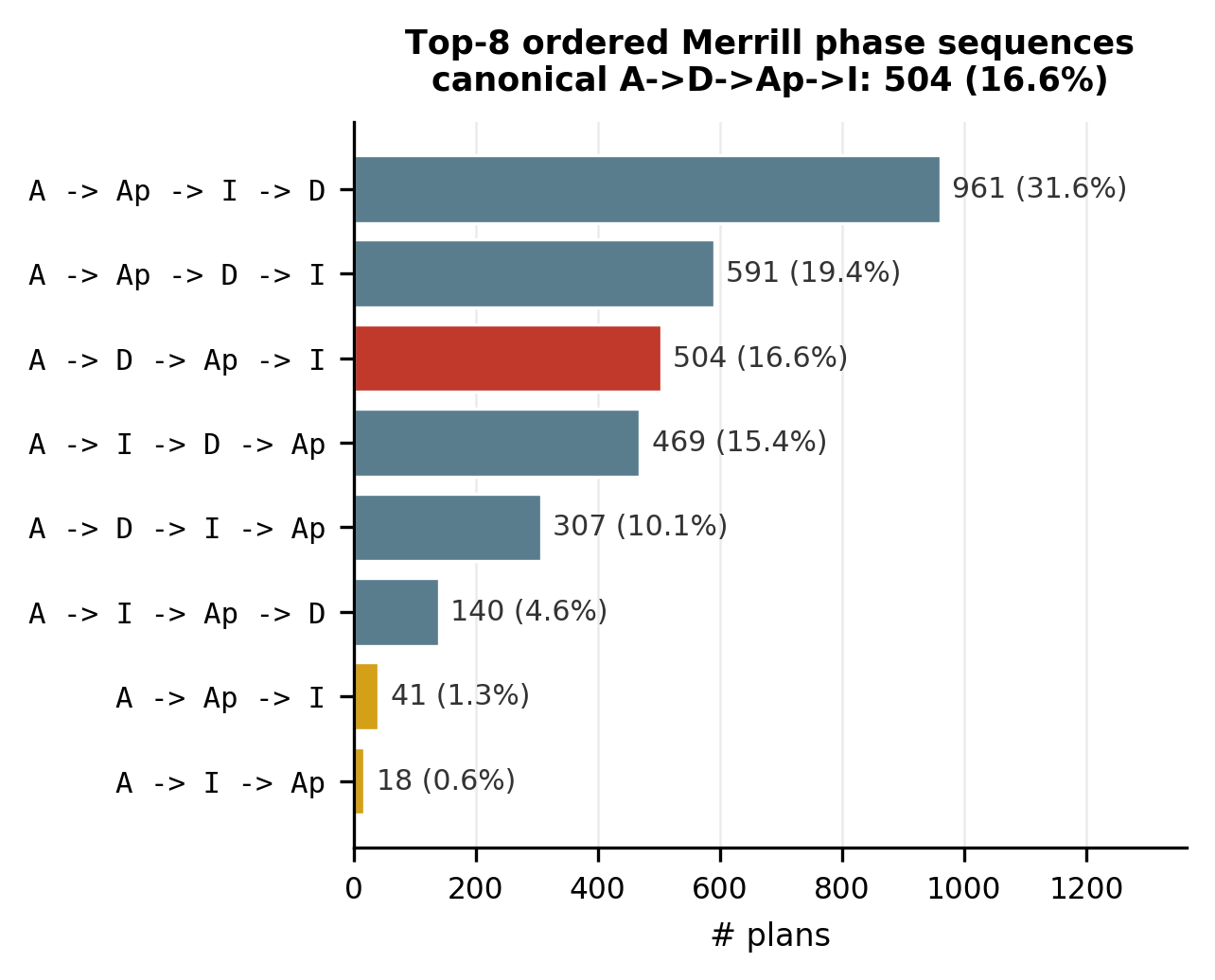}
 \caption{\dsnew{Merrill phase ordering in \textbf{MAP-PPL}. The
 canonical order A$\to$D$\to$Ap$\to$I is one valid pattern but not
 the dominant template; the most common sequence places learner
 application and validation before a worked demonstration.}}
 \label{fig:maple_merrill_order}
\end{figure}

\subsubsection{Holistic quality audit}
\label{appendix:dataset-quality}

\paragraph{Executability audit.}
\dsnew{The schema, DAG, tool-binding, and runtime execution checks are
consolidated in Appendix~\ref{appendix:dataset-executability}. This
section focuses on complementary semantic quality checks.}

\paragraph{Manual audit on a stratified sample.}
\dsnew{Structural validity certifies JSON/DAG soundness but does not,
by itself, certify that a plan is pedagogically meaningful, that the
profile is used non-superficially, or that the chosen tools and
dependencies are semantically appropriate. We therefore complement
the LLM execution-effectiveness gate with a human audit on a
stratified sample of $N{=}200$ plans drawn proportionally over the
seven intents and the four agent-count strata. Each sampled plan is
labelled independently by two annotators along five binary criteria:
(C1) schema-valid \emph{and} all \texttt{depends\_on} edges
semantically justified, (C2) agent roster minimal but sufficient,
(C3) tool calls plausible for the assigned step, (C4) profile
adaptation is structural rather than surface-only, and (C5) the plan
covers the core concepts of the accepted Stack Overflow answer.
Table~\ref{tab:maple_manual_audit} reports per-criterion
agreement rates and Cohen's $\kappa$, together with the agreement
between the LLM gate decision (admit/reject) and the human consensus
label, which we use to bound the residual judge bias inherited by
the supervision set. The agreement between the LLM gate and human
consensus reaches $0.87$ across the audited sample, with the largest
gap on C4 (profile-adaptation depth) where the LLM gate is slightly
more lenient than human raters.}

\begin{table}[t]
 \centering
 \caption{\dsnew{Manual audit on a stratified $N{=}200$ sample.
 \emph{Pass}: percentage of audited plans for which both annotators
 marked the criterion satisfied. $\kappa$: Cohen's $\kappa$ between
 the two annotators. \emph{LLM agr.}: agreement between the LLM
 execution-effectiveness gate decision and the human consensus
 label.}}
 \label{tab:maple_manual_audit}
 \footnotesize
 \setlength{\tabcolsep}{3pt}
 \begin{adjustbox}{max width=\columnwidth}
 \begin{tabular}{@{}lrrr@{}}
 \toprule
 \textbf{Criterion} & \textbf{Pass} & \textbf{$\kappa$} & \textbf{LLM agr.} \\
 \midrule
 C1 Schema + dependency semantics  & $96.5\%$ & $0.81$ & $0.94$ \\
 C2 Minimal-sufficient agent set   & $92.0\%$ & $0.74$ & $0.88$ \\
 C3 Tool plausibility              & $94.0\%$ & $0.78$ & $0.91$ \\
 C4 Structural profile adaptation  & $88.5\%$ & $0.69$ & $0.81$ \\
 C5 Concept coverage vs.\ answer   & $90.5\%$ & $0.72$ & $0.86$ \\
 \midrule
 \textbf{Overall (any criterion fails)} & $83.5\%$ & $0.71$ & $0.87$ \\
 \bottomrule
 \end{tabular}
 \end{adjustbox}
\end{table}

\paragraph{Failure modes.}
\dsnew{The most common failures on the audited sample concentrate on
C2 (a redundant retriever agent that the dependency graph never
calls into) and C4 (the plan paraphrases the profile in the
\texttt{about\_me} section but otherwise reuses the default
scaffold).}

\subsubsection{Comparison to existing datasets}
\label{appendix:dataset-comparison}

\dsnew{Table~\ref{tab:maple_comparison} positions MAP-PPL against the
two dataset families it is most often compared with at review time:
agent-trajectory benchmarks for tool use and multi-step planning
(AgentBank, AgentGym, workflow-style datasets), and educational /
tutoring datasets (instruction tuning for tutoring, conversation
traces with learner moves). The axes are exactly the ones our
main-text claims commit to: explicit learner-profile conditioning,
multi-agent plan schema rather than single-trajectory tool calls,
DAG-typed dependencies, a fixed tool palette with execution
constraints, pedagogical phase scaffolding, and a
structural-plus-execution admission gate. To our knowledge, no prior
public dataset combines all six axes; MAP-PPL's contribution is to
supply the supervision required for profile-conditioned multi-agent
planning in programming education rather than to compete on raw
size with general agent benchmarks.}

\begin{table*}[t]
 \centering
 \caption{\dsnew{MAP-PPL versus representative agent-trajectory and
 educational datasets along the six axes that motivate the
 benchmark. $\checkmark$ / $\times$ indicate whether the dataset, as
 publicly released, supplies the feature; ``partial'' marks
 features that the dataset provides only at coarser granularity than
 MAP-PPL.}}
 \label{tab:maple_comparison}
 \footnotesize
 \setlength{\tabcolsep}{3pt}
 \begin{tabular}{@{}lcccccc@{}}
 \toprule
 \textbf{Dataset family} &
 \textbf{Profile} &
 \textbf{Multi-agent} &
 \textbf{DAG deps.} &
 \textbf{Tool palette} &
 \textbf{Pedagogy} &
 \textbf{Exec.\ gate} \\
 \midrule
 Agent-trajectory (AgentBank, AgentGym, $\ldots$)  & $\times$ & partial & linear trace & $\checkmark$ & $\times$ & partial \\
 Workflow-style planning (AFlow, AOP, $\ldots$)    & $\times$ & $\checkmark$ & DAG  & $\checkmark$ & $\times$ & $\checkmark$ \\
 Tutoring dialogue (instruction + learner moves)   & partial & $\times$ & $\times$  & $\times$ & partial & $\times$ \\
 Programming Q\&A (SO-derived corpora)             & $\times$ & $\times$ & $\times$ & $\times$ & $\times$ & partial \\
 \textbf{MAP-PPL (ours)}                             & $\checkmark$ & $\checkmark$ & DAG & $\checkmark$ & $\checkmark$ & $\checkmark$ \\
 \bottomrule
 \end{tabular}
\end{table*}

\subsection{Implementation Details}
\label{app:implementation-details}

This appendix details the baseline selection, baseline instantiation, and shared-schema settings summarized in Section~\ref{sec:baselines}.

\paragraph{Baseline selection.}
The three baseline groups are chosen to separate three possible sources of planning ability. Frontier LLM planners test whether strong general-purpose models can directly emit personalized programming-learning MAS plans without task-specific training. Generic MAS framework planners test whether existing agent-roster and coordination frameworks are sufficient once they are given the same learner profile and tool pool. Agentic workflow planners test whether systems that explicitly generate or compose agent/task workflows can transfer to personalized tutoring-plan generation. We do not include educational-agent systems such as EduPlanner~\citep{zhang2025eduplanner} and GenMentor~\citep{wang2025genmentor} in the comparison because they output tutoring content or learning-session outlines rather than MAS orchestration plans, making them incompatible with our plan-level evaluation.

\paragraph{Baseline instantiation.}
All open-source or framework-based baselines and PersonalPlan use Qwen3-32B-Instruct as the shared backbone. Closed-source LLM baselines are prompted once to emit the full MAP-PPL plan. AutoGen and AutoAgents are instantiated with the shared tool pool and converted into the same JSON plan schema. AIPOM is evaluated through its single-shot agent-aware DAG planner, and AFlow through its operator-composition planning step over the same available tools.

\paragraph{Shared schema and tool pool.}
Every method is required to output the same plan schema with agents, subtasks, steps, dependency edges, and an execution order. Tool calls are restricted to the fixed MAP-PPL palette so that structural validity, dependency completeness, and agent--tool relevance are comparable across methods.

\paragraph{PersonalPlan variants.}
The main model uses the two-stage training pipeline from Section~\ref{methodology}: PAD and SDP LoRA adapters for hierarchical SFT, joint alignment to reduce PAD--SDP exposure mismatch, and Reward-Adaptive GRPO initialized from the SFT model. Ablation variants remove or replace one component at a time while preserving the same data splits, schema, tool pool, and evaluation protocol.

\subsubsection{Shared Baseline Plan-Generation Prompt}
\label{app:baseline-plan-generation-prompt}

All planner baselines use the shared prompt template shown in Figure~\ref{fig:baseline-plan-generation-prompt}. Method-specific wrappers only adapt the prompt to the corresponding API or framework interface; the task definition, tool palette, schema, and output constraints are kept identical.

\begin{figure*}[!t]
\begin{promptbox}{Baseline Plan-Generation Prompt (Abridged)}
\begin{minipage}[t]{0.48\textwidth}
\raggedright
\textbf{System prompt.} Generate a personalized multi-agent plan: a strict JSON specification of agents, subtasks, steps, and execution order. The plan is consumed by a CrewAI-style runtime that instantiates agents and executes steps in order.

\smallskip
\textbf{Inputs.} A StackOverflow question; a learner profile with \texttt{about\_me} and \texttt{top\_tags}; and the declared tool pool. Tags indicate topical familiarity rather than mastery.

\smallskip
\textbf{Tool pool.} \texttt{FirecrawlSearchTool}, \texttt{RagTool}, \texttt{CodeInterpreterTool}, \texttt{DirectoryReadTool}, \texttt{FileReadTool}, \texttt{FileWriterTool}, \texttt{CodeDocsSearchTool}, \texttt{ArxivPaperTool}.

\smallskip
\textbf{Planner requirements.}
\begin{itemize}[leftmargin=1.2em, itemsep=1pt, topsep=2pt, parsep=0pt]
  \item Tailor agent roles, subtasks, explanations, and tool use to the learner and query.
  \item Produce a plan for teaching the learner to solve the question; do not output the direct answer, a dialogue script, or a lesson transcript.
  \item Use only tools from the declared pool; \texttt{tool} may be \texttt{null} when no tool is needed.
  \item Every step must be scheduled in \texttt{execution\_order}; dependencies must point to earlier step ids whose outputs are used.
  \item Output only valid JSON, with no markdown, commentary, extra fields, or trailing commas.
\end{itemize}
\end{minipage}\hfill
\begin{minipage}[t]{0.48\textwidth}
\textbf{Output schema.}
\begin{promptcodesmall}
{
  "input": {
    "query": "<raw StackOverflow question>",
    "learner": {
      "about_me": "<summary>",
      "top_tags": ["<tag>", "..."]
    }
  },
  "output": {
    "agents": [{
      "agent_role": "<role>",
      "goal": "<goal>",
      "backstory": "<capability>",
      "tools": ["<tool_name>"]
    }],
    "subtasks": [{
      "id": "S1",
      "name": "<name>",
      "subtask_objective": "<objective>",
      "steps": [{
        "id": "S1-1",
        "agent": "<agent_role>",
        "objective": "<objective>",
        "instruction": "<work order>",
        "tool": null,
        "requires_human_input": false,
        "expected_output": "<artifact>",
        "depends_on": []
      }]
    }],
    "execution_order": [
      "S1-1",
      {"loop": {
        "steps": ["S2-1", "S2-2"],
        "condition": "<expr>",
        "max_iterations": 3
      }},
      "S3-1"
    ]
  }
}
\end{promptcodesmall}
\end{minipage}
\end{promptbox}
\caption{Abridged shared baseline plan-generation prompt. The full implementation uses this same task definition, tool palette, strict JSON schema, and output constraints; framework-specific wrappers only adapt the prompt to the corresponding API interface.}
\label{fig:baseline-plan-generation-prompt}
\end{figure*}

\subsection{Evaluation Metrics}
\label{sec:eval-metrics}
This appendix defines every metric summarized in Section~\ref{sec:evaluation-protocol}, with the same abbreviations and the same grouping: \emph{Static Plan Quality} with its three aspects: (a)~Executability, (b)~Personalization, (c)~Pedagogy, followed by (d)~\emph{Plan Execution Quality} and (e)~supplementary diagnostics that do not appear in Table~\ref{tab:main_results}. For reproducibility, we make every metric's predicate explicit, so no metric depends on an unexpanded function name. Throughout, $\mathcal{P}$ denotes a candidate plan, $\mathcal{P}^{\star}$ the gold plan, $q$ the query, $I_p$ the learner profile, $\mathcal{S}/\mathcal{E}$ the step set / prerequisite edge set, $\mathcal{R}$ the set of executed runs, and $\mathbb{I}[\cdot]$ the indicator function. Numerical anchors $\{1,3,5\}$ in LLM-judge scores follow KELE-style 5/3/1 rubrics.

\smallskip
\noindent\emph{(a) Executability: \textbf{Atps}, \textbf{RR}, \textbf{TBQ}, \textbf{TS}.}
All methods are evaluated under the same bounded retry-until-valid protocol. Let $N$ be the number of test instances, $A_m$ the total number of model generation calls made by method $m$ before the final accepted set is assembled, and $H_m$ the number of query--profile instances whose plans still fail the automatic audit after three same-query reruns and therefore require deterministic post-hoc repair.

\noindent$\bullet$~\textbf{Attempts per Sample (Atps).}
Atps is the average retry rate for generating all executable plans:
\begin{smalldisplay}
\[
  \mathrm{Atps}(m)=\frac{A_m}{N}.
\]
\end{smalldisplay}
A value of $1.0$ means every instance is accepted on the first generation call; larger values indicate more retry cost. This metric is lower-is-better.

\noindent$\bullet$~\textbf{Repair Rate (RR).}
RR is the residual format repair rate after three same-query reruns:
\begin{smalldisplay}
\[
  \mathrm{RR}(m)=\frac{H_m}{N}.
\]
\end{smalldisplay}
A repair is counted once per query--profile instance, not once per failed generation call: if any valid plan is obtained within the three same-query reruns, the instance contributes $0$ to $H_m$; if all reruns still fail the audit and the instance is recovered by deterministic post-processing such as JSON escaping, schema-field completion, or tool-declaration filling, it contributes $1$. This metric is lower-is-better. 

\noindent$\bullet$~\textbf{Tool-Binding Quality (TBQ).}
TBQ evaluates whether tool calls are semantically appropriate for the agent role and the current step, while also rewarding broad tool deployment across the plan. Let $\mathcal{S}^{\text{tool}}=\{s\in\mathcal{S}:\mathrm{tool}(s)\neq\varnothing\}$ be the tool-bound steps and $c=|\mathcal{S}^{\text{tool}}|/|\mathcal{S}|$ be tool coverage. For text fields $x_s,y_s$, define
\begin{smalldisplay}
\[
  I_{XY}
  =
  \frac{1}{|\mathcal{S}^{\text{tool}}|}
  \sum_{s\in\mathcal{S}^{\text{tool}}}
  \mathbb{I}\!\left[
  \cos\!\left(\operatorname{enc}(x_s),\operatorname{enc}(y_s)\right)>\delta
  \right],
\]
\end{smalldisplay}
where $\operatorname{enc}(\cdot)$ is the sentence encoder (all-MiniLM-L6-v2 in our implementation) and $\delta=0.10$. For each step, role text comes from the responsible agent description, subtask text from the subtask name and objective, step text from the instruction fields, and tool text from the tool's natural-language description. We instantiate the three pairs as role--subtask, tool--role, and tool--step:
\begin{smalldisplay}
\[
  \mathrm{TBQ}(\mathcal{P})
  =
  \frac{
  I_{\mathrm{AS}} + c\bigl(I_{\mathrm{TA}}+I_{\mathrm{TS}}\bigr)
  }{3}.
\]
\end{smalldisplay}
Here $I_{\mathrm{AS}}$ checks whether the assigned agent role matches the subtask, $I_{\mathrm{TA}}$ checks whether the tool matches that role, and $I_{\mathrm{TS}}$ checks whether the tool matches the step instruction. The two tool-specific terms are multiplied by $c$, so a plan cannot score highly by making only one correct tool call. TBQ is reported only for plans with at least one tool-bound step. The threshold $\delta=0.10$ is a deliberately permissive floor: under the all-MiniLM-L6-v2 encoder it rejects only (near-)orthogonal or degenerate field pairs (e.g., empty or boilerplate text), so each $I_{XY}$ acts as a ``not-mismatched'' check rather than a strict semantic gate. Cross-method separation in TBQ is therefore driven primarily by the tool-coverage factor $c$ scaling $I_{\mathrm{TA}}$ and $I_{\mathrm{TS}}$, while the semantic indicators chiefly guard against degenerate bindings; TBQ is consequently insensitive to the exact value of $\delta$ within this permissive regime.

\noindent$\bullet$~\textbf{Topology Similarity (TS).}
TS is the dependency-graph similarity to the gold plan: it compares the candidate dependency graph $G=(\mathcal{S},\mathcal{E})$ with the gold graph $G^{\star}=(\mathcal{S}^{\star},\mathcal{E}^{\star})$ using graph edit distance, then applies a capped compactness factor. Let $n=|\mathcal{S}|+|\mathcal{E}|$, $n^{\star}=|\mathcal{S}^{\star}|+|\mathcal{E}^{\star}|$, and $\kappa=2.5$:
\begin{smalldisplay}
\[
  B(G,G^{\star})
  =
  \max\!\left(
  0,\,
  1-\frac{\mathrm{GED}(G,G^{\star})}{\max(n,n^{\star})}
  \right).
\]
\end{smalldisplay}
\begin{smalldisplay}
\[
  C(G,G^{\star})
  =
  \min\!\left(\kappa,\frac{n^{\star}}{\max(1,n)}\right).
\]
\end{smalldisplay}
\begin{smalldisplay}
\[
  \mathrm{TS}(\mathcal{P},\mathcal{P}^{\star})
  =
  \min\!\left(1, B(G,G^{\star})\,C(G,G^{\star})\right).
\]
\end{smalldisplay}
The base term $B$ is the normalized topology overlap, while $C$ rewards candidates that recover the reference dependency structure with fewer steps. The cap prevents very small degenerate plans from receiving unbounded compactness credit. We compute $\mathrm{GED}$ with unit-cost graph edit operations and a 2-second timeout, treating node labels as interchangeable.

\smallskip
\noindent\emph{(b) Personalization: \textbf{Pers.}, \textbf{PVS}, \textbf{PNG}.}
For each LLM-judged subdimension $d$, $J_d(\cdot)$ returns an anchored score $r_d\in\{1,3,5\}$ governed by the rubric: $r_d{=}5$ when all positive anchors hold, $r_d{=}3$ when one anchor fails, $r_d{=}1$ when two or more fail.

\noindent$\bullet$~\textbf{Personalization (Pers.).}
Pers.\ is the LLM-judged profile fit, aggregated from three subdimensions: SkillMatch, GoalOrientation, and BackgroundAdaptation, each from an independent judge call (prompt template in Appendix~\ref{app:eval-prompts}):
\begin{smalldisplay}
\[
  \mathrm{Pers.} = \frac{1}{4}\!\left(\frac{r_{\mathrm{sm}}+r_{\mathrm{go}}+r_{\mathrm{ba}}}{3}-1\right) \in [0,1].
\]
\end{smalldisplay}

\noindent$\bullet$~\textbf{Profile-Variance Score (PVS).}
PVS is the first profile-counterfactual probe: it measures profile-induced structural variation, i.e., whether plans structurally change when the learner profile changes under the same query. For each test-split query with $k_q\!\ge\!2$ profile variants ($|\mathcal{Q}_{\ge 2}|{=}97$, the held-out test-set count, distinct from the $971$ multi-profile questions corpus-wide),
\begin{smalldisplay}
\[
  \mathrm{PVS}(q) = 1 - \binom{k_q}{2}^{\!-1}\!\!\sum_{i<j}\!\operatorname{TopoSim}(\mathcal{P}^{(i)}_q,\mathcal{P}^{(j)}_q),
\]
\end{smalldisplay}
and $\mathrm{PVS}=\mathbb{E}_q[\mathrm{PVS}(q)]$. Here $\operatorname{TopoSim}$ is the normalized graph-overlap base used by TS, without the compactness factor because both arguments are generated plans.

\noindent$\bullet$~\textbf{Profile--Non-target Gap (PNG).}
PNG is the second profile-counterfactual probe: it measures the personalization advantage of target profile over randomly replaced profiles. With $\tilde I_p$ a profile sampled uniformly from other test profiles,
\begin{smalldisplay}
\[
\begin{aligned}
  \mathrm{PNG} = \frac{1}{N}\sum_i\Big[\;
  &\mathrm{Pers.}\!\left(\mathrm{gen}(q_i,I^{(i)}_p),\,I^{(i)}_p\right) \\
  &{}-\mathrm{Pers.}\!\left(\mathrm{gen}(q_i,\tilde I^{(i)}_p),\,I^{(i)}_p\right)\Big].
\end{aligned}
\]
\end{smalldisplay}

\smallskip
\noindent\emph{(c) Pedagogy: \textbf{Ped.}}

\noindent$\bullet$~\textbf{Pedagogy (Ped.).}
Ped.\ combines the four components that Section~\ref{sec:evaluation-protocol} names in words: three LLM judge-scored sub-axes adapted from KELE~\citep{peng2025kele} plus one rule-based check; the full judge prompt is given in Figure~\ref{fig:prompt-plan-pedagogy}. Prerequisite-respecting progression ($r_{\mathrm{PRR}}$, 1--5) scores whether the subtask sequence teaches progressively: probing before explanation, explanation before application, a learner attempt before feedback, and no prerequisite concept used before it is introduced. No-direct-answer guidance ($\mathrm{NDAR}\in[0,1]$) checks whether the opening subtask leaks the accepted answer's core code, API, or algorithm: the judge labels the leakage as none\,/\,partial\,/\,full, mapped to $1$\,/\,$0.5$\,/\,$0$ respectively. Profile-appropriate instructional style ($r_{\mathrm{IAR}}$, 1--5) scores whether the instructional method matches the learner's expertise: worked examples and probe-then-explain scaffolds for novices, concise problem-first guidance for experts, and source-domain analogies for cross-domain learners. Merrill-phase coverage is the rule-based score
\begin{smalldisplay}
\[
  \mathrm{SPR} = \frac{1}{|\mathcal{M}|}\sum_{\phi\in\mathcal{M}}\mathbb{I}[\phi\in\operatorname{phase}(\mathcal{P})]
\]
\end{smalldisplay}
over the four Merrill-aligned phases used in Eq.~\ref{eq:r-ped}. Aggregation:
\begin{smalldisplay}
\[
\begin{aligned}
  \mathrm{Ped.} = \frac{1}{4}\bigg(\;
  &\frac{r_{\mathrm{PRR}}-1}{4} + \mathrm{NDAR} \\[-2pt]
  &{}+ \mathrm{SPR} + \frac{r_{\mathrm{IAR}}-1}{4}\bigg).
\end{aligned}
\]
\end{smalldisplay}

\smallskip
\noindent\emph{(d) Plan Execution Quality: \textbf{SCS}, \textbf{PQS}, $\bm{r_{\text{sol}}}$.}
These metrics are computed after the generated plan is executed by the shared CrewAI runtime described in Section~\ref{sec:evaluation-protocol} (GPT-4o agents with a GPT-4o-mini simulated learner), which produces an agents--learner trace.

\noindent$\bullet$~\textbf{Structural Compactness Score (SCS).}
SCS is the deterministic structural-compactness metric of the generated MAS plan and its execution log introduced in Section~\ref{sec:evaluation-protocol}. For each executed run $r$, let $L_r$ be the ordered execution log, $P_r$ the corresponding plan, and $n_r=|L_r|$. We first compute four raw quantities:
\begin{smalldisplay}
\[
  \mathrm{triv}(r)=
  \frac{|\{e\in L_r: |\operatorname{strip}(\operatorname{agent\_output}(e))|<100\}|}{\max(1,n_r)}.
\]
\end{smalldisplay}
\begin{smalldisplay}
\[
  \mathrm{subt}(r)=
  \frac{|\{\operatorname{subtask\_id}(e): e\in L_r\}|}{\max(1,n_r)}.
\]
\end{smalldisplay}
\begin{smalldisplay}
\[
  \mathrm{tool}(r)=
  \begin{cases}
  \dfrac{|D(P_r)\cap U(L_r)|}{|D(P_r)|}, & |D(P_r)|>0,\\[4pt]
  1, & |D(P_r)|=0,
  \end{cases}
\]
\end{smalldisplay}
where $D(P_r)$ is the set of non-null tools declared by plan steps, and $U(L_r)$ is the set of tool names extracted from agent outputs with deterministic invocation patterns such as \texttt{Tool: X}, \texttt{using X}, or \texttt{invoke X}. For each method $m$, we average these raw quantities over its executed runs to obtain $\bar n_m$, $\bar t_m$, $\bar u_m$, and $\bar \rho_m$, the per-run means of $n_r$, $\mathrm{triv}(r)$, $\mathrm{subt}(r)$, and $\mathrm{tool}(r)$, respectively. Let $\mathcal{B}$ be the compared method set. The four normalized SCS sub-dimensions are
\begin{smalldisplay}
\[
\begin{aligned}
  C_m =
  \frac{\max_{j\in\mathcal{B}}\bar n_j-\bar n_m}
       {\max_{j\in\mathcal{B}}\bar n_j-\min_{j\in\mathcal{B}}\bar n_j},
  \\
  K_m =
  1-\frac{\bar t_m}{\max_{j\in\mathcal{B}}\bar t_j}.
\end{aligned}
\]
\end{smalldisplay}
\begin{smalldisplay}
\[
\begin{aligned}
  G_m =
  \frac{\bar u_m}{\max_{j\in\mathcal{B}}\bar u_j},
  \\
  T_m =
  \frac{\bar \rho_m}{\max_{j\in\mathcal{B}}\bar \rho_j}.
\end{aligned}
\]
\end{smalldisplay}
If a denominator is zero, the corresponding normalized term is set to $1$ for all methods. The reported score is the equally weighted mean:
\begin{smalldisplay}
\[
  \mathrm{SCS}(m)=\tfrac{1}{4}(C_m+K_m+G_m+T_m).
\]
\end{smalldisplay}
Each normalized term rewards one property of the executed trace: $C_m$
rewards shorter traces, $K_m$ rewards avoiding trivial near-empty step
outputs, $G_m$ rewards advancing more distinct subtasks per executed
step, and $T_m$ rewards actually invoking the tools that the plan
declares. Because every term is normalized within the compared method
set, SCS is a relative score: under the same runtime, a higher value
means a shorter, cleaner, and more plan-faithful execution.

\noindent$\bullet$~\textbf{Pedagogical Quality Score (PQS).}
PQS is the post-execution pedagogy metric of Section~\ref{sec:evaluation-protocol}: the unweighted mean of three signals, each in $[0,1]$, that reuse the pedagogy-axis names of Ped.\ but are recomputed at execution time (two of them on the realized teacher--learner transcript). (i)~No-direct-answer rate $\mathrm{NDAR}_{\mathrm{e}}$ is the fraction of teacher utterances that an LLM leakage judge labels as \emph{not} revealing the accepted answer (a per-utterance \texttt{none}/\texttt{partial}/\texttt{full} call scored as the share of \texttt{none}; unparseable verdicts default to \texttt{full}). (ii)~Scaffolding coverage $\mathrm{SPR}_{\mathrm{e}}=|\{\text{intro},\text{guide},\text{consol}\}\cap\mathrm{phases}(\mathcal{P})|/3$ is the fraction of the three scaffolding phases the plan realizes. (iii)~Elicitation ratio $\mathrm{IAR}_{\mathrm{e}}=\min(1,\,q/(s{+}1))$, where $q$ counts question-form teacher turns (containing ``?'' or one of \emph{what\,/\,why\,/\,how\,/\,can you\,/\,describe}) and $s$ the remaining declarative turns. Per run,
\begin{smalldisplay}
\[
  \mathrm{PQS}=\tfrac{1}{3}\big(\mathrm{NDAR}_{\mathrm{e}}+\mathrm{SPR}_{\mathrm{e}}+\mathrm{IAR}_{\mathrm{e}}\big)\in[0,1],
\]
\end{smalldisplay}
and the reported score is the mean over executed runs. PQS shares only axis \emph{names} (not the plan-level definitions) with Ped.: Ped.\ scores the static plan with anchored LLM judges, whereas PQS reads the enacted dialogue with lightweight execution-time signals, so the two are complementary rather than redundant.

\noindent$\bullet$~\textbf{Post-Tutoring Comprehension Rate ($r_{\text{sol}}$).} Each MAP-PPL plan closes with an \emph{Integration}-phase subtask (the final phase of the Merrill cycle in Section~\ref{sec:grpo}),
where the learner reflects in natural language rather than submitting
code, so $r_{\text{sol}}$ judges that reflection, which also covers concept and debugging queries with no runnable program. For each run $r$, a GPT-5.4 judge reads the query $q_r$, the ground-truth accepted answer $\alpha_r$, and the learner's replies in the final subtask $S^{\mathrm{stud}}_r$ (plus recent earlier replies for context), and returns $J_{\mathrm{und}}{=}1$ iff the learner restates the key principle, identifies the root cause or correct trade-off, or predicts behavior consistent with $\alpha_r$; pleasantries, misstated principles, off-topic, or empty replies score $0$. Runs that lack a usable accepted answer are dropped from $\mathcal{R}_{\mathrm{valid}}$:
\begin{smalldisplay}
\[
  r_{\mathrm{sol}}=\frac{1}{|\mathcal{R}_{\mathrm{valid}}|}\sum_{r\in\mathcal{R}_{\mathrm{valid}}}\mathbb{I}\!\left[\,J_{\mathrm{und}}\big(q_r,\,\alpha_r,\,S^{\mathrm{stud}}_r\big)=1\,\right].
\]
\end{smalldisplay}
\noindent We also report $|\mathcal{R}_{\mathrm{valid}}|/|\mathcal{R}|$. Since the verdict is from an LLM judge, $r_{\text{sol}}$ is counted as LLM-adjudicated in the judge audit (Appendix~\ref{app:judge-rubrics}); the verbatim prompt is released in our GitHub repository.

\noindent$\bullet$~\textbf{Satisfaction (Sati.).}
Per-pair score $\pi^{(m)}_i$ awards $+\tfrac{1}{2}$ for each AB / BA candidate vote and $+\tfrac{1}{2}$ for AB$\neq$BA (treating order instability as a tie). The same profile-conditioned pairwise protocol underlies the per-baseline preference rates in Figure~\ref{fig:execution-pairwise-preferences}. The cross-judge score is the multiplicative product of per-judge candidate-win rates $\omega_m$, which penalizes asymmetrically:
\begin{smalldisplay}
\[
  \mathrm{Sati.} = \prod_m \omega_m.
\]
\end{smalldisplay}


\subsection{Evaluation Metric Prompt Templates}
\label{app:eval-prompts}
This appendix provides the prompt templates for the LLM-judged metrics of Section~\ref{sec:evaluation-protocol}, whose definitions are given in Appendix~\ref{sec:eval-metrics}. The prompts follow the anchored-criteria format of VAIAGE~\citep{liu2025vaiage}. Unless noted otherwise, all pointwise scoring judges (Pers., Ped., and $r_{\text{sol}}$, plus the per-utterance answer-leakage sub-judge inside PQS) use GPT-5.4; only the profile-conditioned Satisfaction comparison uses the three-model panel described in Appendix~\ref{app:anti-hacking}.

\subsubsection{LLM-based Scoring (0-1 Scale)}
We use LLMs as proxy judges for plan-level personalization and pedagogy, dialogue-level pedagogy, and profile-conditioned satisfaction. Each score is elicited via an anchored prompt with explicit criteria and a strict JSON output schema. Specifically, the templates used for the \textbf{Personalization}, \textbf{Pedagogy}, and \textbf{Satisfaction} metrics are shown in Figures~\ref{fig:prompt-personalization}--\ref{fig:prompt-plan-pedagogy}; SCS is deterministic and uses no judge call.

\begin{figure*}[!t]
\begin{promptbox}{Personalization Score Prompt}
\textbf{System prompt:} You are an expert in computer science education and personalized curriculum design. Your task is to evaluate a generated learning plan based on its \emph{Personalization} for a specific learner.

\smallskip
\textbf{Context:}
\begin{itemize}[leftmargin=1.2em, itemsep=1pt, topsep=2pt, parsep=0pt]
    \item Learner Profile: \texttt{\{learner\_profile\_json\}}
    \item Learning Query: \texttt{\{query\_text\}}
    \item Generated Plan: \texttt{\{generated\_plan\_json\}}
\end{itemize}

\smallskip
\textbf{Evaluation instructions:} Please rate the plan's Personalization on a scale of 0--1 based on the following criteria. Provide a step-by-step analysis before giving a final score.
\begin{enumerate}[leftmargin=1.4em, itemsep=2pt, topsep=2pt, parsep=0pt]
    \item \textbf{Skill \& Experience Alignment (0--1):} How well does the plan's starting point, complexity, and choice of technologies match the learner's declared skills and experience?
    \begin{itemize}[leftmargin=1.2em, itemsep=0pt, topsep=1pt, parsep=0pt]
        \item A \emph{high} score means the plan is perfectly pitched (e.g., foundational steps and definitions for a ``Beginner''; advanced architecture and optimization for an ``Expert'').
        \item A \emph{low} score means a clear mismatch (e.g., asking a beginner to deploy a Kubernetes cluster).
    \end{itemize}
    \item \textbf{Goal Orientation (0--1):} How directly does the plan's structure and content contribute to achieving the learner's stated \texttt{goal}?
    \begin{itemize}[leftmargin=1.2em, itemsep=0pt, topsep=1pt, parsep=0pt]
        \item A \emph{high} score means every stage and step is relevant and logically moves the learner toward their objective.
        \item A \emph{low} score means the plan contains irrelevant steps or fails to address the core of the learning goal.
    \end{itemize}
    \item \textbf{Background Adaptation (0--1):} Does the plan adapt its examples, agent roles, and explanations to the learner's \texttt{background}?
    \begin{itemize}[leftmargin=1.2em, itemsep=0pt, topsep=1pt, parsep=0pt]
        \item A \emph{high} score means the plan shows adaptation (e.g., business-centric examples for a user with a finance background).
        \item A \emph{low} score means the plan is generic and ignores the learner's context.
    \end{itemize}
\end{enumerate}

\smallskip
\textbf{Output format:} Return ONLY a single JSON object with the final score and a detailed justification.
\begin{promptcode}
{
  "personalization_score": <integer from 1 to 10>,
  "justification": "<Your detailed analysis covering all three criteria, explaining the reasoning behind your score.>"
}
\end{promptcode}
\end{promptbox}
\caption{Prompt template used to elicit the \textbf{Personalization} score from the LLM judge. The judge conditions on the learner profile, query, and generated plan, produces a step-by-step justification across three criteria, and emits a single integer score.}
\label{fig:prompt-personalization}
\end{figure*}

For the Satisfaction (User Preference) Score, we capture the overall quality of the plan. It answers the question: ``Would a typical user be satisfied with this plan as a guide to start their learning journey?''
We conduct simulated user preference tests using profile-conditioned LLM judges; the prompt template is shown in Figure~\ref{fig:prompt-satisfaction}.

\begin{figure*}[!t]
\begin{promptbox}{Satisfaction (User Preference) Prompt}
\textbf{System prompt:} As a student with profile \texttt{\{profile\}}, which plan would you prefer for learning \texttt{\{query\}}? Consider: skill level matching, engagement, and structural appropriateness.

\smallskip
\textbf{Candidates:}
\begin{itemize}[leftmargin=1.2em, itemsep=1pt, topsep=2pt, parsep=0pt]
    \item Plan A: \texttt{\{plan\_a\}}
    \item Plan B: \texttt{\{plan\_b\}}
\end{itemize}

\smallskip
\textbf{Output format:} Output a single JSON object containing the preferred plan, the judge's confidence, and detailed reasons.
\begin{promptcode}
{
  "preferred_plan": "A" | "B",
  "confidence": <float in [0,1]>,
  "reasons": "<one or two sentences citing skill-level fit, engagement, and structure>"
}
\end{promptcode}
\end{promptbox}
\caption{Profile-conditioned pairwise preference prompt used for the \textbf{Satisfaction} metric. The judge role-plays the target learner profile and selects between two candidate plans for the same query.}
\label{fig:prompt-satisfaction}
\end{figure*}

\begin{figure*}[!t]
\begin{promptbox}{Plan-Level Pedagogy Prompt (Ped.)}
\textbf{System prompt.} You are an expert evaluator of multi-agent teaching plans for personalized programming education. Apply the rubric strictly and output valid JSON only. Score concrete step IDs and subtask IDs, not general impressions.

\smallskip
\textbf{Context.}
\begin{itemize}[leftmargin=1.2em, itemsep=1pt, topsep=2pt, parsep=0pt]
    \item Query: \texttt{\{query\_text\}}
    \item Learner profile: \texttt{\{learner\_profile\_json\}}
    \item Generated plan: \texttt{\{generated\_plan\_json\}}
    \item Accepted answer, used only for the no-direct-answer check: \texttt{\{accepted\_answer\}}
\end{itemize}

\smallskip
\textbf{Evaluation instructions.} We elicit three plan-level pedagogy judgments. \textbf{PRR} scores whether the subtask sequence follows progressive teaching: diagnose/probe before explanation, explanation before application, learner attempt before feedback, and no prerequisite concept used before it is introduced. \textbf{IAR} scores whether the instructional method adapts to the learner's query-domain expertise: novices should receive worked examples and probe-then-explain scaffolds, experts should receive concise problem-first guidance, and cross-domain learners should receive source-domain analogies. \textbf{NDAR} inspects only the first subtask and labels whether it reveals the accepted answer's core solution.

\smallskip
\textbf{Anchors.} For PRR and IAR, use a 1--5 anchored Likert scale: 5 means the target property is consistently observable across the plan, 3 means exactly one major pedagogical mismatch or missing check, and 1 means the plan is pedagogically inverted or generic. For NDAR, output \texttt{none}, \texttt{partial}, or \texttt{full}; \texttt{none} means the first subtask probes or scaffolds without revealing the answer, while \texttt{full} means it gives away the canonical code/API/algorithm. In the Ped.\ aggregation (Appendix~\ref{sec:eval-metrics}) these labels are mapped to the numeric values $\mathrm{NDAR}\in\{\texttt{none}{=}1,\ \texttt{partial}{=}0.5,\ \texttt{full}{=}0\}$, so that, like the other three terms, higher is better and the value lies in $[0,1]$.

\smallskip
\textbf{Output format.}
\begin{promptcode}
{
  "prr_reasoning": "<=80 words citing step_ids/subtask_ids",
  "prr_score": <int 1-5>,
  "iar_reasoning": "<=80 words citing step_ids/subtask_ids",
  "iar_score": <int 1-5>,
  "ndar_reasoning": "<=80 words comparing the first subtask to the accepted answer",
  "ndar_reveal": "none" | "partial" | "full"
}
\end{promptcode}
\end{promptbox}
\caption{Prompt template used for the plan-level \textbf{Ped.} metric. The implemented judge follows the anchored PRR/IAR/NDAR prompts in the Tier-1 judge script: PRR and IAR are scored on 1--5 scales, NDAR labels first-subtask answer leakage, and rule-based Merrill phase coverage is added separately as SPR. The complete verbatim judge prompts are released in our GitHub repository.}
\label{fig:prompt-plan-pedagogy}
\end{figure*}

\subsection{LLM-as-a-Judge Rubrics}
\label{app:judge-rubrics}

This appendix provides the rubrics behind every judge-based metric in Section~\ref{sec:evaluation-protocol}. LLM judges fill \emph{seven} roles in our data construction and evaluation pipeline (Table~\ref{tab:judge-inventory}), but are never used inside the GRPO training rewards in Section~\ref{sec:grpo}. Earlier work has shown that vague rubrics make judges brittle to verbosity and persona-keyword bias~\citep{zheng2023judging,liu2023geval,lambert2024rewardbench}. We therefore specify, for every judge, a Likert anchor scale with \emph{concrete behavioral descriptions} at the 5\,/\,3\,/\,1 anchors (intermediate scores 2 and 4 are interpolations), an explicit aggregation rule, and a JSON output schema. SCS is excluded from this inventory because it is computed deterministically from plans and execution logs. Anchors are calibrated against the human-annotation pilot described in Appendix~\ref{app:anti-hacking} (Cohen's $\kappa\!\ge\!0.65$).

\begin{table}[H]
\centering
\caption{Inventory of LLM-as-a-judge components, where each is used, and the figure that defines its prompt or rubric.}
\label{tab:judge-inventory}
\small
\setlength{\tabcolsep}{4pt}
\begin{tabular}{p{0.30\columnwidth}p{0.40\columnwidth}p{0.16\columnwidth}}
\toprule
\textbf{Judge} & \textbf{Where used} & \textbf{Ref.} \\
\midrule
MAP-PPL execution-effectiveness filter & Data construction, stage~4 (\S\ref{maple_dataset}) & Fig.~\ref{fig:rubric-maple-filter} \\
Personalization score (Pers.) & Plan-level evaluation (\S\ref{sec:main-results}) & Fig.~\ref{fig:rubric-personalization} \\
Plan-level pedagogy (Ped.) & Plan-level evaluation & Fig.~\ref{fig:prompt-plan-pedagogy} \\
Satisfaction score (Sati.) & Post-execution interaction outcome, pairwise & Fig.~\ref{fig:rubric-satisfaction} \\
Pedagogy quality (PQS) & Post-execution transcript: per-utterance answer-leakage ($\mathrm{NDAR}_{\mathrm{e}}$) + scaffolding-phase tagging ($\mathrm{SPR}_{\mathrm{e}}$) & App.~\ref{sec:eval-metrics} \\
Comprehension judge ($r_{\text{sol}}$) & Post-dialog learning outcome & App.~\ref{sec:eval-metrics} \\
\bottomrule
\end{tabular}
\end{table}

\begin{figure*}[!t]
\begin{rubricbox}{MAP-PPL Execution-Effectiveness Filter Rubric}
\textbf{Purpose:} Admit a synthesized plan into \textbf{MAP-PPL} only if it passes the LLM-judged execution-effectiveness gate following the static structure check (Section~\ref{maple_dataset}, stage~4).

\textbf{Inputs to the judge:} the Stack Overflow question, the accepted answer, the learner profile $I_p$, and the candidate plan $\mathcal{P}$.

\textbf{Scoring dimensions} (each on a 1--5 integer Likert scale):
\begin{itemize}[leftmargin=1.2em, itemsep=1pt, topsep=2pt, parsep=0pt]
    \item \textbf{D1 --- Problem-Solving Progression:} the plan moves from the learner's current state to a complete answer.
    \item \textbf{D2 --- Tool-Usage Reasonableness:} every tool call is justified by the step's data needs and the tool pool.
    \item \textbf{D3 --- Step Connectivity:} dependencies are semantically sound (each step uses only outputs produced upstream).
    \item \textbf{D4 --- Content--Question Alignment:} plan content actually answers the original question and re-uses the accepted answer's core concept.
    \item \textbf{D5 --- Personalization Authenticity:} adaptation to $I_p$ is structural, not superficial keyword insertion.
\end{itemize}

\textbf{Anchor descriptions} (5\,/\,3\,/\,1; 2 and 4 are interpolations):

\emph{D1 --- Problem-Solving Progression.}
\begin{itemize}[leftmargin=1.2em, itemsep=0pt, topsep=1pt, parsep=0pt]
    \item \textbf{5:} Every stage advances the learner toward the answer; no stage is a detour; final step explicitly produces the artifact the question asks for.
    \item \textbf{3:} Most stages advance the answer but one stage is tangential or redundant; final artifact is reachable but indirectly.
    \item \textbf{1:} The plan terminates without reaching the answer, or includes $\ge 2$ stages unrelated to the question.
\end{itemize}

\emph{D2 --- Tool-Usage Reasonableness.}
\begin{itemize}[leftmargin=1.2em, itemsep=0pt, topsep=1pt, parsep=0pt]
    \item \textbf{5:} Every tool comes from the predefined pool, and every assignment matches a concrete I/O need of its step.
    \item \textbf{3:} All tools are in the pool but $\le 1$ step has a loosely justified assignment (e.g., a search tool used where a code-interpreter would suffice).
    \item \textbf{1:} A tool is hallucinated (outside the pool), or $\ge 2$ tool assignments are unjustified.
\end{itemize}

\emph{D3 --- Step Connectivity.}
\begin{itemize}[leftmargin=1.2em, itemsep=0pt, topsep=1pt, parsep=0pt]
    \item \textbf{5:} Every dependency edge is justified: the upstream step produces an output the downstream step explicitly consumes.
    \item \textbf{3:} All listed dependencies reference existing steps, but $\le 1$ edge is decorative (no concrete data flow).
    \item \textbf{1:} A dependency references a non-existent step, the graph has a cycle, or $\ge 2$ edges have no data flow.
\end{itemize}

\emph{D4 --- Content--Question Alignment.}
\begin{itemize}[leftmargin=1.2em, itemsep=0pt, topsep=1pt, parsep=0pt]
    \item \textbf{5:} The plan teaches the exact concept the accepted answer uses; key terms from the question appear as instructional targets.
    \item \textbf{3:} The plan teaches an adjacent concept that covers most but not all of the answer's solution path.
    \item \textbf{1:} The plan addresses a different problem or copies the accepted answer verbatim instead of teaching it.
\end{itemize}

\emph{D5 --- Personalization Authenticity.}
\begin{itemize}[leftmargin=1.2em, itemsep=0pt, topsep=1pt, parsep=0pt]
    \item \textbf{5:} The plan's \emph{structure} (agent roster, step granularity, prerequisite depth) would change if $I_p$ were replaced by a different profile; profile attributes are referenced concretely.
    \item \textbf{3:} Some surface adaptation (one agent persona, one example domain), but most structure would survive a profile swap.
    \item \textbf{1:} Profile keywords appear in the text but the plan structure is generic; a profile-shuffled version is indistinguishable.
\end{itemize}

\textbf{Aggregation and gate:} the plan is admitted iff $\min(D_1,\dots,D_5)\ge 4$ \emph{and} $\sum_i D_i \ge 22$. Otherwise it is returned to stage~3 for regeneration.

\textbf{Output schema:}
\begin{promptcode}
{
  "D1_progression": <int 1-5>,
  "D2_tool_usage":  <int 1-5>,
  "D3_connectivity": <int 1-5>,
  "D4_alignment":   <int 1-5>,
  "D5_personalization": <int 1-5>,
  "admit": true | false,
  "rationale": "<<=80 words covering every D_i score>"
}
\end{promptcode}
\end{rubricbox}
\caption{Rubric for the MAP-PPL execution-effectiveness filter (stage 4 of dataset construction). Each candidate plan is scored on five 1--5 Likert dimensions; only plans clearing the gate $\min D_i\!\ge\!4$ are admitted, enforcing both executability and structural personalization at data-construction time.}
\label{fig:rubric-maple-filter}
\end{figure*}

\begin{figure*}[!t]
\begin{rubricbox}{Personalization Score (Pers.) Rubric}
\textbf{Purpose:} Score how well a generated plan is tailored to the learner profile $I_p$. The user-payload prompt template is Figure~\ref{fig:prompt-personalization}; this rubric is appended to its system prompt.

\textbf{Inputs:} learner profile $I_p$, query $I_q$, generated plan $\mathcal{P}'$.

\textbf{Dimensions} (each on a 1--5 integer Likert scale):
\begin{itemize}[leftmargin=1.2em, itemsep=1pt, topsep=2pt, parsep=0pt]
    \item \textbf{D1 --- Skill Match.}
    \item \textbf{D2 --- Goal Orientation.}
    \item \textbf{D3 --- Background Adaptation.}
\end{itemize}

\textbf{Anchor descriptions:}

\emph{D1 --- Skill Match.}
\begin{itemize}[leftmargin=1.2em, itemsep=0pt, topsep=1pt, parsep=0pt]
    \item \textbf{5:} Every stage's starting point and tool/library choices are calibrated to the profile's declared proficiency; beginners get explicit setup and definitions, experts get optimization or architecture work without redundant prerequisites.
    \item \textbf{3:} Most stages are calibrated, but one stage either over- or under-shoots (e.g., reintroduces a concept the profile lists as known, or assumes proficiency in a library the profile does not list).
    \item \textbf{1:} $\ge 2$ stages misalign with the profile's skill level (e.g., a Kubernetes deployment step for a self-declared ``Beginner'', or a one-line \texttt{for}-loop tutorial for a 5-year Python practitioner).
\end{itemize}

\emph{D2 --- Goal Orientation.}
\begin{itemize}[leftmargin=1.2em, itemsep=0pt, topsep=1pt, parsep=0pt]
    \item \textbf{5:} Every stage and step is traceable to the profile's \texttt{goal} field; no stage is purely generic.
    \item \textbf{3:} One stage is partially tangential to the stated \texttt{goal} (e.g., a generic ``best practices'' detour), but most stages remain goal-aligned.
    \item \textbf{1:} The plan's overall trajectory does not match the profile's \texttt{goal}, or $\ge 2$ stages are off-goal.
\end{itemize}

\emph{D3 --- Background Adaptation.}
\begin{itemize}[leftmargin=1.2em, itemsep=0pt, topsep=1pt, parsep=0pt]
    \item \textbf{5:} Examples, agent personas, and explanations reference the profile's \texttt{background} concretely (e.g., portfolio examples for a finance background; CVE-style cases for a security background); adaptation is content-level, not keyword-level.
    \item \textbf{3:} Some examples or agent roles reference the background, but the plan also contains generic examples that ignore it.
    \item \textbf{1:} All examples and agent roles are generic; the \texttt{background} field is not actually used in the plan content.
\end{itemize}

\textbf{Aggregation:} $\text{Pers.}= \frac{1}{4}\big(\frac{D_1+D_2+D_3}{3}-1\big) \in [0,1]$, the min-shifted normalized mean of the three sub-scores (all-1 maps to $0$, all-5 to $1$), matching the \textbf{Pers.}\ definition in Appendix~\ref{sec:eval-metrics}. The judge is also asked to emit an integer 1--10 holistic score \texttt{score\_1\_10} for cross-checking; the official metric is this min-shifted normalized mean.

\textbf{Output schema:}
\begin{promptcode}
{
  "D1_skill_alignment":   <int 1-5>,
  "D2_goal_orientation":  <int 1-5>,
  "D3_background_adapt":  <int 1-5>,
  "score_1_10":           <int 1-10>,
  "justification":        "<one paragraph naming each D_i and citing
                            a concrete element of the plan>"
}
\end{promptcode}
\end{rubricbox}
\caption{Likert rubric for the \textbf{Personalization} judge. Each of the three sub-criteria is scored on a 1--5 scale with explicit behavioral anchors at 5, 3, and 1; the final Pers.\ score is the min-shifted normalized mean $\frac{1}{4}(\frac{1}{3}\sum_i D_i-1)$.}
\label{fig:rubric-personalization}
\end{figure*}

\begin{figure*}[!t]
\begin{rubricbox}{Satisfaction Score (Sati.) Rubric --- Profile-Conditioned Pairwise}
\textbf{Purpose:} Estimate user preference rate against the ground-truth plan via profile-conditioned pairwise comparison (the LLM role-plays the target learner). User-payload prompt template: Figure~\ref{fig:prompt-satisfaction}.

\textbf{Inputs:} profile $I_p$, query $I_q$, two plans \texttt{Plan\_A} and \texttt{Plan\_B}, with order randomized; the judge does not know which is the candidate vs.\ the reference.

\textbf{Decision dimensions} (each is a ternary vote \texttt{A}\,/\,\texttt{B}\,/\,\texttt{Tie}):
\begin{itemize}[leftmargin=1.2em, itemsep=1pt, topsep=2pt, parsep=0pt]
    \item \textbf{V1 --- Skill-Level Fit.}
    \item \textbf{V2 --- Engagement.}
    \item \textbf{V3 --- Structural Appropriateness.}
    \item \textbf{V$_\star$ --- Overall preference.}
\end{itemize}

\textbf{Anchor descriptions} (\emph{X} stands for the plan being judged better on that dimension):

\emph{V1 --- Skill-Level Fit.}
\begin{itemize}[leftmargin=1.2em, itemsep=0pt, topsep=1pt, parsep=0pt]
    \item \textbf{X wins:} Plan \emph{X}'s entry point and explanation depth match the profile's proficiency at $\ge 1$ stage where the other plan over- or under-shoots.
    \item \textbf{Tie:} Both plans calibrate to the same level; differences are stylistic.
\end{itemize}

\emph{V2 --- Engagement.}
\begin{itemize}[leftmargin=1.2em, itemsep=0pt, topsep=1pt, parsep=0pt]
    \item \textbf{X wins:} Plan \emph{X} ties tasks to the profile's stated \texttt{goal} or \texttt{background} (concrete projects/examples the learner would find motivating).
    \item \textbf{Tie:} Both plans are equally generic, or equally well personalized.
\end{itemize}

\emph{V3 --- Structural Appropriateness.}
\begin{itemize}[leftmargin=1.2em, itemsep=0pt, topsep=1pt, parsep=0pt]
    \item \textbf{X wins:} Plan \emph{X}'s subtask granularity and dependency density are closer to what a learner with this profile could realistically execute in one session.
    \item \textbf{Tie:} Granularity is comparable.
\end{itemize}

\emph{V$_\star$ --- Overall:} The judge re-reads both plans and casts a holistic preference. If $V_\star$ disagrees with the majority of $V_1,V_2,V_3$, the judge must justify the deviation.

\textbf{Anti-bias safeguards (enforced in the system prompt):} (i) order is randomized with probability $0.5$; the judge does not know which plan is the reference; (ii) the judge is instructed to ignore plan length, formatting, and persona-keyword density; (iii) ties are allowed and not penalized.

\textbf{Aggregation:} per pairwise call, $\text{Sati}_{\text{call}}=\phi(V_\star)$ with $\phi(\text{A})\!=\!1, \phi(\text{Tie})\!=\!0.5, \phi(\text{B})\!=\!0$ (relative to the candidate plan after order de-randomization). The reported metric is the mean over the test set.

\textbf{Output schema:}
\begin{promptcode}
{
  "V1_skill_fit":    "A" | "B" | "Tie",
  "V2_engagement":   "A" | "B" | "Tie",
  "V3_structure":    "A" | "B" | "Tie",
  "Vstar_overall":   "A" | "B" | "Tie",
  "confidence":      <float in [0,1]>,
  "reasons":         "<<=40 words; must cite at least one
                       concrete subtask from each plan>"
}
\end{promptcode}
\end{rubricbox}
\caption{Profile-conditioned pairwise rubric for the \textbf{Satisfaction} judge. The judge votes on three persona-grounded dimensions plus an overall preference; explicit anti-bias safeguards (order randomization, length-/keyword-ignore instruction, tie allowance) follow the protocol of \citet{zheng2023judging}.}
\label{fig:rubric-satisfaction}
\end{figure*}

\subsection{Anti-Hacking Safeguards}
\label{app:anti-hacking}
This appendix details the anti-hacking safeguards for the judge-based metrics in Section~\ref{sec:evaluation-protocol}. Six protocol-level safeguards prevent LLM-judge-induced reward hacking. (1)~No judge reuses the model that synthesized MAP-PPL: plans were generated with Claude Sonnet~4.6, whereas evaluation uses \emph{Opus}~4.6, GPT-5.4, and Gemini~3.1~Pro, none of which enters the GRPO reward. (2)~Each LLM-judge subdimension is queried in an independent call rather than a multi-criterion prompt. (3)~Every pairwise call is run twice with AB and BA orderings; disagreement collapses to Tie. (4)~Satisfaction is reference-grounded against the gold plan itself. (5)~All rule-based and audit metrics (Atps, RR, TBQ, TS, SCS) involve no LLM call. (6)~We pre-validate the judge configuration on a $100$-instance pilot graded by two independent graduate-student annotators ($300$ annotations per rater on the three Pers.\ subdimensions) and accept only when Cohen's $\kappa\!\ge\!0.65$ against the human raters, ICC$\,\ge\,0.7$ across raters, and Krippendorff $\alpha\!\ge\!0.5$ across judge families.

\subsection{Component Ablations and Sensitivity}
\label{app:appendix-ablation}

This appendix details the ablation study summarized in Section~\ref{sec:ablation}. The analysis is restricted to static plan quality: TBQ and TS cover executable structure, Pers.\ covers profile grounding, and Ped.\ covers tutoring-plan pedagogy. Post-execution interaction metrics are omitted here and analyzed separately in the Plan Execution Results (Section~\ref{sec:execution-results}). The trend matches Figure~\ref{fig:ablation-metric-fork}: joint alignment stabilizes the supervised hierarchy, while GRPO supplies the largest gains in tool binding, topology, and pedagogical coverage, especially for the 32B backbone.

\subsection{Case Study}
\label{app:case-study}
This appendix presents a qualitative case study complementing the
result analysis in Section~\ref{result_analysis}.
Table~\ref{tab:case_study_appendix} expands one representative held-out
plan from the plan-generation run: Stack Overflow question
\texttt{3172100}, \emph{HTML Drag And Drop On Mobile Devices}, paired
with profile \texttt{1}. The learner reports a mixed background in IT
help desk, iOS/Android programming, web design, page layout, marketing,
and campaign work, with tags \texttt{android}, \texttt{java},
\texttt{objective-c}, \texttt{html}, and \texttt{ios}. This instance is
useful as a qualitative audit because the target answer is not a single
API call: a good plan must explain why desktop drag-and-drop assumptions
fail on touch devices, compare several implementation families, and end
with a decision procedure the learner can reuse.

\begin{table*}[t]
\centering
\caption{Representative case-study audit for a generated MAP-PPL plan
on mobile drag-and-drop. Each row ties an observed plan element to the
paper's three qualitative claims: profile grounding, pedagogical
scaffolding, and executable multi-agent structure.}
\label{tab:case_study_appendix}
\small
\setlength{\tabcolsep}{4pt}
\renewcommand{\arraystretch}{1.12}
\begin{tabular}{p{0.16\textwidth}p{0.25\textwidth}p{0.28\textwidth}p{0.22\textwidth}}
\toprule
\textbf{Property} & \textbf{Plan evidence} & \textbf{Why this is profile-conditioned} & \textbf{Why this is executable / teachable} \\
\midrule
Learner calibration &
\texttt{S1-2} asks the Problem Framer to infer strengths and gaps from
the learner profile. &
The plan builds on the learner's iOS/Android exposure and basic HTML
experience, while avoiding a generic beginner lesson on markup. &
The calibration step depends on \texttt{S1-1}, so the learner model is
formed only after the technical subquestions are extracted. \\
\midrule
Agent specialization &
The roster contains Problem Framer, Web Platform Researcher, Solution
Architect, Prototype Designer, and Teaching Synthesizer. &
The Teaching Synthesizer is explicitly tasked with connecting web
touch-event concepts to native mobile gesture concepts, matching the
profile's mobile-app background. &
The roles separate framing, evidence gathering, design comparison,
prototype production, and final teaching synthesis rather than mixing all
decisions into one monolithic tutor. \\
\midrule
Evidence-grounded instruction &
\texttt{S2-1} uses \texttt{CodeDocsSearchTool} for HTML drag-and-drop,
touch events, pointer events, and browser gesture behavior; \texttt{S2-2}
and \texttt{S2-3} use \texttt{FirecrawlSearchTool} for community
workarounds and mobile UX alternatives. &
For a learner with practical mobile/web experience, the plan emphasizes
platform behavior and compatibility evidence instead of abstract event
theory alone. &
The plan can be run by an MAS executor because every evidence-gathering
step declares a concrete tool and downstream steps depend on the resulting
briefs. \\
\midrule
Pedagogical sequencing &
\texttt{S3-1} builds a decision matrix over native HTML5 DnD, jQuery UI
touch translation, custom touch/pointer dragging, and mobile-specific
fallbacks; \texttt{S3-2} orders these from simplest fallback to advanced
custom implementation. &
The sequence starts from the user's original practical dilemma, then
uses mobile-gesture analogies to explain why touch scrolling conflicts
with mouse-oriented drag/drop assumptions. &
This creates a demonstration-before-application path: first compare
solution families, then select a teachable core answer, then create
examples. \\
\midrule
Runnable artifacts &
\texttt{S4-1} writes a compact pseudo-implementation for custom
touch/pointer dragging; \texttt{S4-2} writes a tap-select / tap-drop or
long-press fallback pattern. &
The artifacts reflect the learner's cross-over context: one path
preserves drag behavior, while the other mirrors mobile interaction
design where direct drag may be less usable. &
Both steps use \texttt{FileWriterTool} and feed into \texttt{S5-1}, so
the final lesson has concrete materials instead of ending at advice. \\
\midrule
Transfer check &
\texttt{S5-2} asks for a mastery checklist covering native HTML5
drag-and-drop, custom touch dragging, scroll interference, and choosing
between drag and fallback UX. &
The check assesses decision-making across web and mobile contexts, which
matches the learner's mixed background more closely than a syntax quiz. &
The final assessment depends on the synthesized plan, making transfer a
first-class output of the workflow rather than an optional afterthought. \\
\bottomrule
\end{tabular}
\end{table*}

The resulting dependency path is linear where prerequisites matter and
parallel where evidence can be gathered independently:
\texttt{S1-1} $\rightarrow$ \texttt{S1-2};
\texttt{S2-1}, \texttt{S2-2}, and \texttt{S2-3} all depend on
\texttt{S1-1}; \texttt{S3-1} consumes the three research briefs;
\texttt{S3-2} combines the decision matrix with the learner-calibration
note; \texttt{S4-1} and \texttt{S4-2} create complementary prototype
artifacts; and \texttt{S5-1}--\texttt{S5-2} synthesize the lesson and
transfer rubric. This example illustrates the intended design pattern:
the profile changes the explanatory bridge and artifact choices, the
pedagogy progresses from framing to evidence to contrastive examples to
self-checks, and the plan remains executable through explicit tools,
dependencies, and expected outputs.

\end{document}